\documentclass[prd,preprint,tightenlines,floatfix,
showpacs,preprintnumbers,nofootinbib,eqsecnum,superscriptaddress]{revtex4}

\usepackage[T1]{fontenc}		

\usepackage{amsmath,amsfonts,amssymb,amstext,mathrsfs}
\usepackage{mathpazo}

\usepackage[dvips]{graphicx}
\usepackage{epsf,float}
\usepackage{revsymb}

\usepackage{dcolumn}
\usepackage{braket}
\usepackage{color,xcolor}
\usepackage{graphicx}
\usepackage{subfigure}
\usepackage{multirow}
\usepackage{tabularx}
\usepackage{pstricks}
\usepackage[section]{placeins}
\usepackage{booktabs}
\usepackage{array}

\usepackage{hyperref}

\newcommand{\Pom}{\mathbb{P}}
\newcommand{\Ode}{\mathbb{O}}
\newcommand{\Reg}{\mathbb{R}}

\newcommand{\bdPt}{\mbox{\boldmath $dP_{t}$}}
\newcommand{\bqta}{\mbox{\boldmath $q_{t,1}$}}
\newcommand{\bqtb}{\mbox{\boldmath $q_{t,2}$}}
\newcommand{\bpta}{\mbox{\boldmath $p_{t,1}$}}
\newcommand{\bptb}{\mbox{\boldmath $p_{t,2}$}}

\newcommand{\bpa}{\mbox{\boldmath $p_{a}$}}
\newcommand{\bpb}{\mbox{\boldmath $p_{b}$}}
\newcommand{\bpip}{\mbox{\boldmath $p_{3}$}}
\newcommand{\bpim}{\mbox{\boldmath $p_{4}$}}

\newcommand{\bhk}{\mbox{\boldmath $\hat{k}$}}

\newcommand{\bhpa}{\mbox{\boldmath $\hat{p}_{a}$}}
\newcommand{\bhpb}{\mbox{\boldmath $\hat{p}_{b}$}}

\newcommand{\bea}{\mbox{\boldmath $e_{1}$}}
\newcommand{\beb}{\mbox{\boldmath $e_{2}$}}
\newcommand{\bec}{\mbox{\boldmath $e_{3}$}}

\begin{document}

\title{Central exclusive diffractive production\\ of $\pi^{+}\pi^{-}$ continuum,
scalar and tensor resonances\\ in $p p$ and $p \bar p$ scattering
within tensor pomeron approach}

\author{Piotr Lebiedowicz}
 \email{Piotr.Lebiedowicz@ifj.edu.pl}
\affiliation{Institute of Nuclear Physics, Polish Academy of Sciences, Radzikowskiego 152, PL-31-342 Krak\'ow, Poland}

\author{Otto Nachtmann}
 \email{O.Nachtmann@thphys.uni-heidelberg.de}
\affiliation{Institut f\"ur Theoretische Physik, Universit\"at Heidelberg,
Philosophenweg 16, D-69120 Heidelberg, Germany}

\author{Antoni Szczurek
\footnote{Also at University of Rzesz\'ow, PL-35-959 Rzesz\'ow, Poland.}}
\email{Antoni.Szczurek@ifj.edu.pl}
\affiliation{Institute of Nuclear Physics, Polish Academy of Sciences, Radzikowskiego 152, PL-31-342 Krak\'ow, Poland}

\begin{abstract}
\noindent
We consider central exclusive diffractive dipion production
in the reactions $pp \to pp \pi^{+} \pi^{-}$ 
and $p\bar{p} \to p\bar{p} \pi^{+} \pi^{-}$ at high energies.
We include the dipion continuum, the dominant scalar $f_{0}(500)$, $f_{0}(980)$, 
and tensor $f_{2}(1270)$ resonances decaying into the $\pi^{+} \pi^{-}$ pairs.
The calculation is based on a tensor pomeron model 
and the amplitudes for the processes are formulated in terms of vertices
respecting the standard crossing and charge-conjugation relations of 
Quantum Field Theory.
The formulae for the dipion continuum and tensor meson production 
are given here for the first time.
The theoretical results are compared with existing 
STAR, CDF, CMS experimental data 
and predictions for planned or being carried out experiments 
(ALICE, ATLAS) are presented.
We show the influence of the experimental cuts on the integrated cross section
and on various differential distributions for outgoing particles.
Distributions in rapidities and transverse momenta of outgoing protons 
and pions as well as correlations in azimuthal angle between them are presented.
We find that the relative contribution of resonant $f_2(1270)$ and
dipion continuum strongly depends on the cut on proton transverse momenta 
or four-momentum transfer squared $t_{1,2}$
which may explain some controversial observations made 
by different ISR experiments in the past. 
The cuts may play then the role of a $\pi \pi$ resonance filter.
We suggest some experimental analyses to fix model parameters
related to the pomeron-pomeron-$f_{2}$ coupling.
\end{abstract}

\maketitle

\section{Introduction}
The exclusive reaction $p p \to p p \pi^+ \pi^-$ is one
of the reactions being extensively studied by several experimental groups
such as 
COMPASS \cite{Austregesilo:2013vky,Austregesilo:2014oxa,Adolph:2015tqa}, 
STAR \cite{Adamczyk:2014ofa,Sikora_LowX}, CDF \cite{Aaltonen:2015uva,Albrow_Project_new},
ALICE \cite{Schicker:2014aoa}, ATLAS \cite{Staszewski:2011bg},
and CMS \cite{Osterberg:2014mta,CMS:2015diy}. 
It is commonly believed that at high energies
the pomeron-pomeron fusion is the dominant mechanism of the exclusive 
two-pion production. In the past two of us have formulated a simple
Regge inspired model of the two-pion continuum 
mediated by the double pomeron/reggeon exchanges 
with parameters fixed from phenomenological 
analyses of $NN$ and $\pi N$ scattering \cite{Lebiedowicz:2009pj}
\footnote{For another related work see \cite{Lebiedowicz:2011nb}
where the exclusive reaction $pp \to pp \pi^{+}\pi^{-}$
constitutes an irreducible background to the scalar $\chi_{c0}$ meson production.
These model studies were extended also to the $pp \to pp K^{+}K^{-}$ \cite{Lebiedowicz:2011tp}
and the $pp \to nn \pi^{+}\pi^{+}$ \cite{Lebiedowicz:2010yb} processes.}.
The number of free model parameters is then limited to a parameter
of form factor describing off-shellness of the exchanged pion.
The largest uncertainties in the model are due
to the unknown off-shell pion form factor and the absorption corrections
discussed recently in \cite{Lebiedowicz:2015eka}.
Although this model gives correct order of magnitude cross sections
it is not able to describe details of differential distributions,
in particular the distribution in dipion invariant mass where we
observe a rich pattern of structures. 
Clearly such an approach
does not include resonance contributions which interfere with the continuum. 
It is found that the pattern of visible structures depends on experiment
but as we rather advocate on the cuts used in a particular experiment
(usually these cuts are different for different experiments).

It was known for a long time that the commonly used vector pomeron has problems
from a field theory point of view.
Taken literally it gives opposite signs for $pp$ and $\bar{p}p$ total cross sections.
A way out of this dilemma was already shown in \cite{Nachtmann:1991ua}
where the pomeron was described as a coherent superposition of exchanges
with spin 2 + 4 + 6 + ... . 
This same idea is realised in a very practical way
in the tensor-pomeron model formulated in \cite{Ewerz:2013kda}.
In this model pomeron exchange can effectively be treated as the exchange of a rank-2 tensor.
The corresponding couplings of the tensorial object to proton and pion were worked out.
In Ref.~\cite{Lebiedowicz:2013ika} the model was applied to 
the production of several scalar and pseudoscalar mesons in the reaction $p p \to p p M$. 
A good description 
of the experimental distributions \cite{Barberis:1998ax} was achieved 
at relatively low energy where, however, reggeon exchanges still play a very important role
\footnote{The role of secondary reggeons in central pseudoscalar meson production 
was discussed also in Ref.~\cite{Kochelev:2000wm}.}.
The resonant ($\rho^0 \to \pi^{+}\pi^{-}$) and non-resonant (Drell-S\"oding)
photon-pomeron/reggeon $\pi^{+} \pi^{-}$ production in $pp$ collisions
was studied in \cite{Lebiedowicz:2014bea}.
In \cite{Bolz:2014mya} an extensive study of the photoproduction reaction
$\gamma p \to \pi^{+} \pi^{-} p$ in the framework
of the tensor-pomeron model was presented.

In most of the experimental preliminary spectra
of the $p p \to p p \pi^+ \pi^-$ reaction at higher energies a peak at
$M_{\pi \pi} \sim$ 1270 MeV is observed. One can expect that the peak
is related to the production of the well known tensor isoscalar
meson $f_2(1270)$ which decays with high probability into the $\pi^+ \pi^-$ channel. 
In principle, contributions from the $f_0(1370)$, $f_0(1500)$ and $f_0(1710)$
mesons are not excluded. 
The $f_0(1500)$ and $f_0(1710)$ mesons are often considered 
as potential candidates for scalar states with dominant glueball content
and it is expected that 
in pomeron-pomeron fusion the glueball production could be
prominently enhanced
due to the gluonic nature of the pomeron \cite{Szczurek:2009yk,Ochs:2013gi}.

For a study of the resonance production
observed in the $\pi^{+} \pi^{-}$ and $K^{+}K^{-}$ mass spectra
in the fixed target experiments at low energies
see Refs.~\cite{Armstrong:1991ch,Barberis:1999cq,Barberis:1999an,Kirk:2000ws,Kirk:2014nwa}. 
There is evidence from the analysis of the decay modes of the scalar states observed,
that the lightest scalar glueball manifests itself through 
the mixing with nearby $q \bar{q}$ states \cite{Kirk:2014nwa}
and that the difference in the transverse momentum vectors between the two exchange particles 
($dP_{t}$)
can be used to select out known $q \bar{q}$ states from non-$q \bar{q}$ candidates.
\footnote{It has been observed in Ref.~\cite{Barberis:1996iq}
that all the undisputed $q\bar{q}$ states 
(i.e. $\eta$, $\eta'$, $f_{1}(1285)$ etc.)
are suppressed as $dP_{t} \to 0$, whereas the glueball candidates, 
e.g. $f_{0}(1500)$, survive.
As can be seen there $\rho^{0}(770)$, $f_{2}(1270)$ and $f_{2}'(1525)$ 
have larger $dP_{t}$ and their cross sections peak at $\phi_{pp} = \pi$,
i.e. the outgoing protons are on opposite sides of the beam,
in contrast to the 'enigmatic' $f_{0}(980)$, $f_{0}(1500)$ and $f_{0}(1710)$ states.
}
Also the four-momentum transfer squared $|t|$
from one of the proton vertices for the resonances was determined.
For the tensor $f_{2}(1270)$ and $f_{2}'(1525)$ states
their fractional distributions show non-single exponential behaviour; 
see Fig.~5 of \cite{Barberis:1999cq}.
It has been observed in \cite{Armstrong:1991ch} that 
the $\rho(770)$, $\phi(1020)$, $f_{2}(1270)$ and $f'_{2}(1525)$ resonances
are visible more efficiently in the high $t = |t_{1} + t_{2}|$ region ($t > 0.3$~GeV$^{2}$)
and that at low $t$ their signals are suppressed.

In the ISR experiments, see \cite{Breakstone:1986xd,Breakstone:1989ty,Breakstone:1990at},
the $\pi^{+} \pi^{-}$ invariant mass distribution shows an enhancement 
in the low-mass ($S$-wave) region
and a very significant resonance structure.
A clear $f_{2}(1270)$ signal has been observed at $\sqrt{s} = 62$~GeV~\cite{Breakstone:1986xd}
and a cross section $\sigma(pp \to pp f_{2}, f_{2} \to \pi^{+} \pi^{-})$
of $(8 \pm 1 \pm 3)$ $\mu$b was determined,
where the four-momentum transfer squared is $|t| \geqslant 0.08$~GeV$^{2}$,
the scattered protons have $x_{F,p} \geqslant 0.9$,
and the pion c.m.~system rapidity is limited to the region $|{\rm y}_{\pi}| \geqslant 1.5$.
However, this behaviour is rather different from that observed in Ref.~\cite{Akesson:1985rn}.
In our opinion this is due to the different kinematic coverage of these two ISR experiments.
The experiment \cite{Akesson:1985rn} has been performed at $\sqrt{s} = 63$~GeV.
Compared to \cite{Breakstone:1986xd} their analysis covered
smaller four-momentum transfer squared
($0.01 \lesssim |t| \lesssim 0.06$~GeV$^{2}$), $x_{F,p} \geqslant 0.95$,
and the central rapidity region was more restricted.
Moreover, their $D$-wave cross section shows an enhancement between $1.2 - 1.5$~GeV
and the authors of \cite{Akesson:1985rn} argued that 
the $f_{2}(1270)$ alone does not explain the behaviour of the data
and additional states are needed, e.g. a scalar at around 1400~MeV 
and tensor at $m=1480 \pm 50$~MeV with $\Gamma=150 \pm 50$~MeV.
In the paper \cite{Breakstone:1990at} the cross section of central 
$\pi^{+} \pi^{-}$ production shows an enhancement in both the $S$ and $D$-waves 
near the mass of the $f_{2}(1270)$ and $f_{0}(1400)$.
The $D$-wave mass spectrum was described with the $f_{2}(1270)$ resonant state
and a broad background term.
A cross section for exclusive $f_{2}(1270)$ meson production of $5.0 \pm 0.7$ $\mu$b 
(not including a systematic error, estimated to be 1.5 $\mu$b) was obtained.

On the theoretical side, the production of the tensor meson $f_2$
was not considered so far in the literature, except in Ref.~\cite{Petrov:2004hh}.
We note that in a recent work \cite{Fiore:2015lnz} 
the authors also consider the resonance production through 
the pomeron-pomeron fusion at the LHC
but ignoring the spin effects in the pomeron-pomeron-meson vertices.
In the present paper we consider both, production of the two-pion
continuum and of the $f_0(500)$, the $f_0(980)$, and the $f_2(1270)$ resonances
in the $\pi^+ \pi^-$ channel, consistently within the tensor pomeron model.
This model allows also to calculate interference effects.
We consider the tensor-tensor-tensor coupling in a Lagrangian
formalism and present a list of possible couplings.
The specificities of the different couplings are discussed,
also in the context of experimental results.
We discuss a first qualitative attempt to ``reproduce'' the experimentally 
observed behaviours of the two-pion spectra obtained
in the $p p \to p p \pi^+ \pi^-$ reaction and discuss consequences
of experimental cuts on the observed spectra.
The calculations presented in section~\ref{sec:section_4} 
were done with a FORTRAN code using the VEGAS routine \cite{Lepage:1980dq}.
\section{Exclusive two-pion production}
\label{sec:excl}

We study central exclusive production of $\pi^+ \pi^-$ 
in proton-proton collisions at high energies
\begin{eqnarray}
p(p_{a},\lambda_{a}) + p(p_{b},\lambda_{b}) \to
p(p_{1},\lambda_{1}) + \pi^{+}(p_{3}) + \pi^{-}(p_{4}) + p(p_{2},\lambda_{2}) \,,
\label{2to4_reaction}
\end{eqnarray}
where $p_{a,b}$, $p_{1,2}$ and $\lambda_{a,b}$, 
$\lambda_{1,2} \in \lbrace +1/2, -1/2 \rbrace$ 
denote the four-momenta and helicities of the protons, 
and $p_{3,4}$ denote the four-momenta of the charged pions, respectively.

\begin{figure}
\includegraphics[width=6.cm]{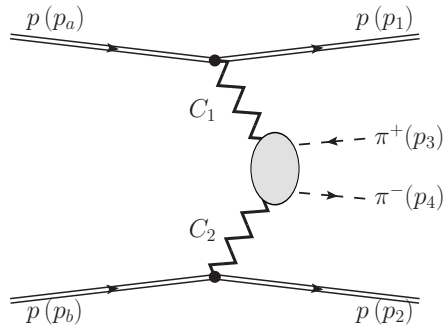}   
  \caption{\label{fig:generic_diagram}
  \small Generic ``Born level'' diagram for central exclusive $\pi^+ \pi^-$ production 
  in proton-proton collisions.
}
\end{figure}

The full amplitude of $\pi^{+} \pi^{-}$ production 
is a sum of continuum amplitude 
and the amplitudes through the $s$-channel resonances:
%
\begin{equation}
\begin{split}
{\cal M}_{pp \to pp \pi^{+} \pi^{-}} =
{\cal M}^{\pi \pi{\rm-continuum}}_{pp \to pp \pi^{+} \pi^{-}} + 
{\cal M}^{\pi \pi{\rm-resonances}}_{pp \to pp \pi^{+} \pi^{-}}\,.
\end{split}
\label{amplitude_pomTpomT}
\end{equation}
%
This amplitude for central exclusive $\pi^+ \pi^-$ production
is believed to be given by the fusion of two exchange objects.
The generic ``Born level'' diagram
is shown in Fig.~\ref{fig:generic_diagram},
where we label the exchange objects by their charge conjugation numbers
$C_{1}$, $C_{2} \in \lbrace +1, -1 \rbrace$.
At high energies the exchange objects to be considered are
the photon $\gamma$, the pomeron $\Pom$, the odderon $\Ode$, 
and the reggeons $\Reg = f_{2 \Reg}, a_{2 \Reg}, \omega_{\Reg}, \rho_{\Reg}$.
Their charge conjugation numbers and $G$ parities 
are listed in Table~\ref{tab:table1}.

\begin{table}
\caption{
Charge conjugation and $G$-parity quantum numbers of exchange objects
for resonance and continuum production.}
\label{tab:table1}
\begin{tabular}{|c|c|c|}
\hline
Exchange object  & $C$ & $G$  \\
\hline
$\Pom$ & 1 & 1 \\
$f_{2 \Reg}$ & 1 & 1 \\
$a_{2 \Reg}$ & 1 & -1 \\
\hline
$\gamma$ & -1 & \\
\hline
$\Ode$ & -1 & -1 \\
$\omega_{\Reg}$ & -1 & -1 \\
$\rho_{\Reg}$ & -1 & 1 \\
\hline
\end{tabular}
\end{table}

In calculating the amplitude (\ref{amplitude_pomTpomT}) from the diagram
Fig.~\ref{fig:generic_diagram}
a sum over all combinations of exchanges,
$(C_{1},C_{2})$ = $(1,1)$, $(-1,-1)$, $(1,-1)$, $(-1,1)$,
has to be taken:
\begin{eqnarray}
{\cal M}_{pp \to pp \pi^{+} \pi^{-}} =
{\cal M}^{(1,1)} +
{\cal M}^{(-1,-1)} +
{\cal M}^{(1,-1)} +
{\cal M}^{(-1,1)}\,.
\label{generic_amplitude}
\end{eqnarray}
Note that the $(1,1)$ and $(-1,-1)$ contributions
will produce a $\pi^{+}\pi^{-}$ state with charge conjugation $C = +1$.
The $(1,-1)$ and $(-1,1)$ contributions will produce a 
$\pi^{+}\pi^{-}$ state with $C = -1$.
This implies that the ${\cal M}^{(C_{1},C_{2})}$ amplitudes
have the following properties under exchange of the $\pi^{+}$ and $\pi^{-}$
momenta, keeping all other kinematic variables,
indicated by the dots, fixed:
\begin{equation}
\begin{split}
& {\cal M}^{(1,1)}(..., p_{3}, p_{4}) = {\cal M}^{(1,1)}(..., p_{4}, p_{3})\,,\\
& {\cal M}^{(-1,-1)}(..., p_{3}, p_{4}) = {\cal M}^{(-1,-1)}(..., p_{4}, p_{3})\,,\\
& {\cal M}^{(1,-1)}(..., p_{3}, p_{4}) = -{\cal M}^{(1,-1)}(..., p_{4}, p_{3})\,,\\
& {\cal M}^{(-1,1)}(..., p_{3}, p_{4}) = -{\cal M}^{(-1,1)}(..., p_{4}, p_{3})\,.
\end{split}
\label{amps_under_charge_conj}
\end{equation}

The interference of the amplitudes ${\cal M}^{(1,1)}+{\cal M}^{(-1,-1)}$
with ${\cal M}^{(1,-1)}+{\cal M}^{(-1,1)}$
will lead to asymmetries under exchange of the $\pi^{+}$ and $\pi^{-}$ momenta.
Such asymmetries can, for instance, 
be studied in the rest frame of the $\pi^{+}\pi^{-}$ pair
using a convenient coordinate system like the Collins-Soper frame \cite{Collins:1977iv}.
For a discussion of various reference frames 
see for instance \cite{Byckling_Kajantie,Bolz:2014mya}.
Asymmetries may be quite interesting from an experimental
point of view since they could allow to measure small contributions
in the amplitude which would be hard to detect otherwise.
Some details related for the asymmetries are given in Appendix~\ref{section:Asymmetries}.

\section{Two-pion continuum production}
\label{sec:continuum}

\begin{figure}
\includegraphics[width=6.cm]{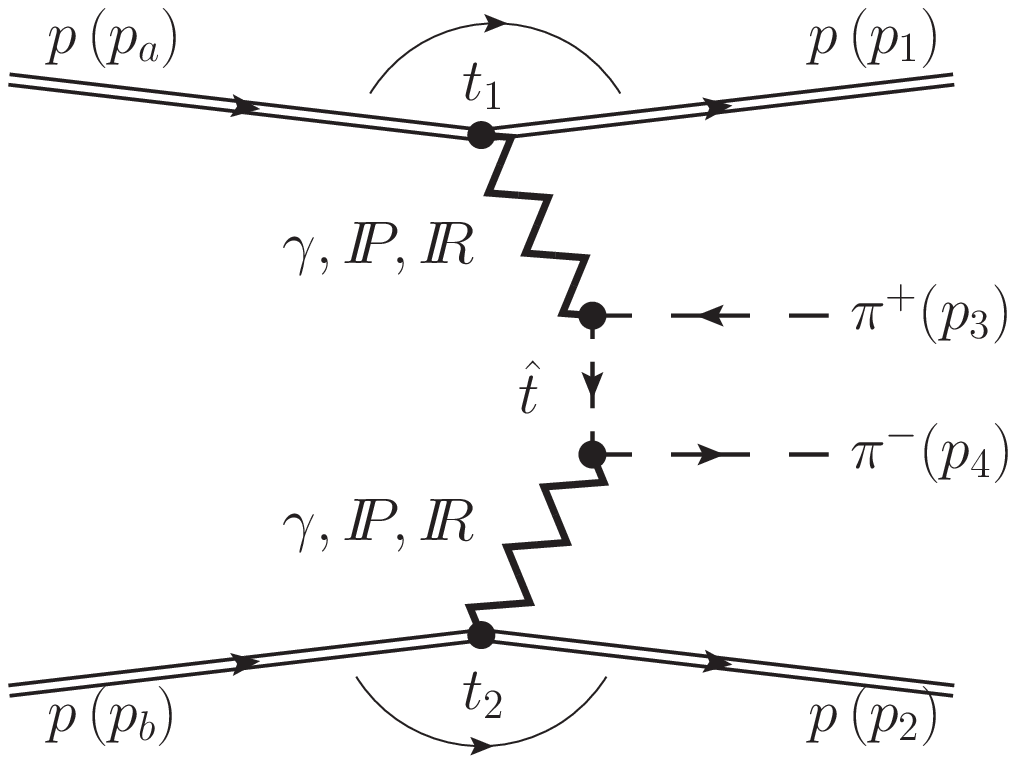}
\includegraphics[width=6.cm]{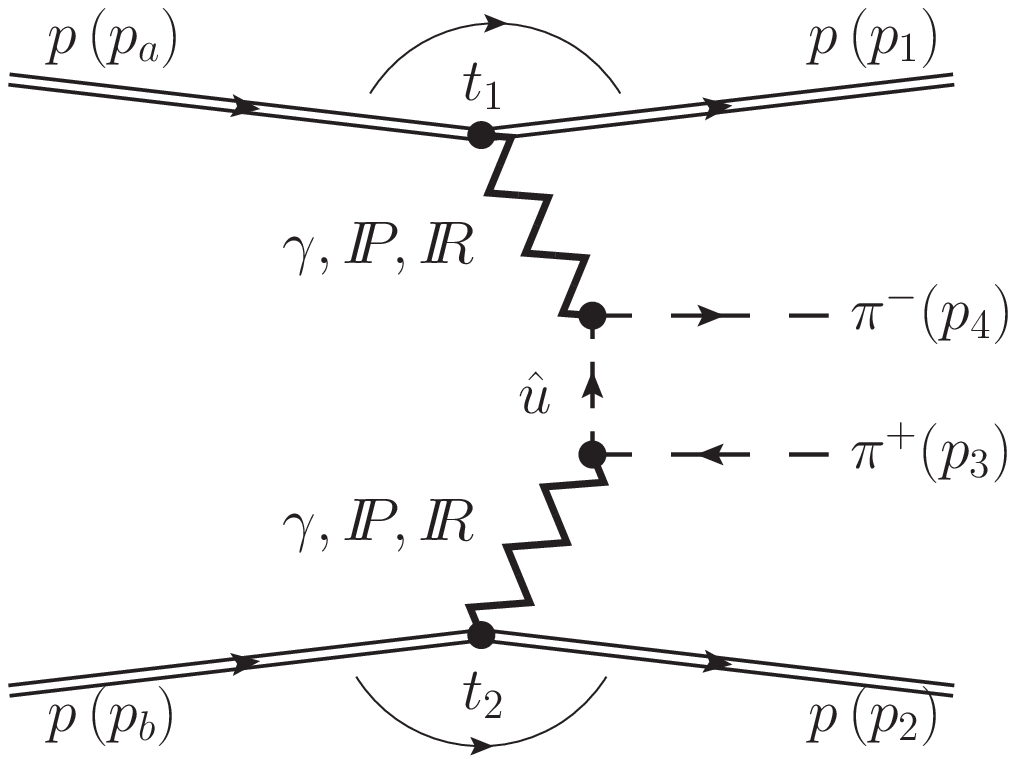}    
  \caption{\label{fig:diagrams_Born}
  \small The Born diagrams for double-pomeron/reggeon and photon mediated
  central exclusive continuum $\pi^+ \pi^-$ production in proton-proton collisions.
}
\end{figure}

The generic diagrams for exclusive two-pion continuum production
are shown in Fig.~\ref{fig:diagrams_Born}.
Taking into account the $G$ parity of -1
for the pions we get the following combinations $(C_{1},C_{2})$
of exchanges which can contribute
\footnote{Note that $G$ parity invariance forbids the vertices $a_{2 \Reg} \pi \pi$,
$\omega_{\Reg} \pi \pi$, $\Ode \pi \pi$, see Table~\ref{tab:table1}.
Thus, the exchanges of $a_{2 \Reg}$, $\omega_{R}$, $\Ode$ cannot
contribute to the dipion continuum.}
\begin{eqnarray}
&&(C_{1},C_{2}) = (1,1): \quad (\Pom + f_{2 \Reg}, \Pom + f_{2 \Reg})\,;
\label{c1c2_1}
\\
&&(C_{1},C_{2}) = (-1,-1): \quad (\rho_{\Reg} + \gamma, \rho_{\Reg} + \gamma)\,;
\label{c1c2_2}
\\
&&(C_{1},C_{2}) = (1,-1): \quad (\Pom + f_{2 \Reg}, \rho_{\Reg} + \gamma)\,;
\label{c1c2_3}
\\
&&(C_{1},C_{2}) = (-1,1): \quad (\rho_{\Reg} + \gamma, \Pom + f_{2 \Reg})\,.
\label{c1c2_4}
\end{eqnarray}
Note that for the cases involving the photon $\gamma$ in (\ref{c1c2_2}) to (\ref{c1c2_4})
one also has to take into account the diagrams involving
the corresponding contact terms; see \cite{Bolz:2014mya}.

From the above list of exchange contributions we have already treated
$(\Pom + f_{2 \Reg}, \gamma)$ and $(\gamma, \Pom + f_{2 \Reg})$ in \cite{Lebiedowicz:2014bea}.
At high energies the contributions involving $\rho_{\Reg}$ exchanges
are expected to be small since $\rho_{\Reg}$ is secondary reggeon and its coupling
to the proton is small; see e.g. (3.61), (3.62) of \cite{Ewerz:2013kda}.
The $(\gamma,\gamma)$ contribution is higher order in $\alpha_{em}$.
Thus we are left with the contribution
$(\Pom + f_{2 \Reg}, \Pom + f_{2 \Reg})$, (\ref{c1c2_1}), 
which we shall treat now.

The amplitude for the corresponding diagrams in Fig.~\ref{fig:diagrams_Born}
can be written as the following sum:
\footnote{We emphasize, that not only the leading pomeron exchanges contribute
to the dipion system with the isospin $I = 0$ and $C = +1$
but also the $\Pom f_{2 \Reg}$, $f_{2 \Reg} \Pom$, $f_{2 \Reg} f_{2 \Reg}$, 
$\rho_{\Reg} \rho_{\Reg}$ exchanges
and due to their non-negligible interference effects 
with the leading $\Pom \Pom$ term the subleading $f_{2 \Reg}$ exchanges 
must be included explicitly in our calculations.}
%
\begin{eqnarray}
{\cal M}^{\pi \pi{\rm-continuum}}_{pp \to pp \pi^{+} \pi^{-}} &=&
{\cal M}^{(\Pom \Pom \to \pi^{+}\pi^{-})} +
{\cal M}^{(\Pom f_{2 \Reg} \to \pi^{+}\pi^{-})} +
{\cal M}^{(f_{2 \Reg} \Pom \to \pi^{+}\pi^{-})} +
{\cal M}^{(f_{2 \Reg} f_{2 \Reg} \to \pi^{+}\pi^{-})}\,. \nonumber \\
\label{2to4_reaction_pp}
\end{eqnarray}
%
The $\Pom\Pom$-exchange amplitude can be written as
\begin{eqnarray}
{\cal M}^{(\Pom \Pom\to \pi^{+}\pi^{-})} =
{\cal M}^{({\rm \hat{t}})}_{\lambda_{a} \lambda_{b} \to \lambda_{1} \lambda_{2} \pi^{+}\pi^{-}}+
{\cal M}^{({\rm \hat{u}})}_{\lambda_{a} \lambda_{b} \to \lambda_{1} \lambda_{2} \pi^{+}\pi^{-}}
\,,
\label{pompom_amp}
\end{eqnarray}
where
\begin{equation}
\begin{split}
& {\cal M}^{({\rm \hat{t}})}_{\lambda_{a} \lambda_{b} \to \lambda_{1} \lambda_{2} \pi^{+}\pi^{-}} 
= \\
& \quad (-i)
\bar{u}(p_{1}, \lambda_{1}) 
i\Gamma^{(\Pom pp)}_{\mu_{1} \nu_{1}}(p_{1},p_{a}) 
u(p_{a}, \lambda_{a})\,
i\Delta^{(\Pom)\, \mu_{1} \nu_{1}, \alpha_{1} \beta_{1}}(s_{13},t_{1}) \,
i\Gamma^{(\Pom \pi \pi)}_{\alpha_{1} \beta_{1}}(p_{t},-p_{3}) \,
i\Delta^{(\pi)}(p_{t}) \\
& \quad \times  i\Gamma^{(\Pom \pi \pi)}_{\alpha_{2} \beta_{2}}(p_{4},p_{t})\,
i\Delta^{(\Pom)\, \alpha_{2} \beta_{2}, \mu_{2} \nu_{2}}(s_{24},t_{2}) \,
\bar{u}(p_{2}, \lambda_{2}) 
i\Gamma^{(\Pom pp)}_{\mu_{2} \nu_{2}}(p_{2},p_{b}) 
u(p_{b}, \lambda_{b}) \,,
\end{split}
\label{amplitude_t}
\end{equation}
\begin{equation} 
\begin{split}
& {\cal M}^{({\rm \hat{u}})}_{\lambda_{a} \lambda_{b} \to \lambda_{1} \lambda_{2} \pi^{+}\pi^{-}} 
= \\ 
& \quad (-i)\,
\bar{u}(p_{1}, \lambda_{1}) 
i\Gamma^{(\Pom pp)}_{\mu_{1} \nu_{1}}(p_{1},p_{a}) 
u(p_{a}, \lambda_{a}) \,
i\Delta^{(\Pom)\, \mu_{1} \nu_{1}, \alpha_{1} \beta_{1}}(s_{14},t_{1}) \,
i\Gamma^{(\Pom \pi \pi)}_{\alpha_{1} \beta_{1}}(p_{4},p_{u}) 
\,i\Delta^{(\pi)}(p_{u})  \\
& \quad \times  
i\Gamma^{(\Pom \pi \pi)}_{\alpha_{2} \beta_{2}}(p_{u},-p_{3})\, 
i\Delta^{(\Pom)\, \alpha_{2} \beta_{2}, \mu_{2} \nu_{2}}(s_{23},t_{2}) \,
\bar{u}(p_{2}, \lambda_{2}) 
i\Gamma^{(\Pom pp)}_{\mu_{2} \nu_{2}}(p_{2},p_{b}) 
u(p_{b}, \lambda_{b}) \,,
\end{split}
\label{amplitude_u}
\end{equation}
where $p_{t} = p_{a} - p_{1} - p_{3}$ and 
$p_{u} = p_{4} - p_{a} + p_{1}$, $s_{ij} = (p_{i} + p_{j})^{2}$.
The kinematic variables for reaction (\ref{2to4_reaction}) are
\begin{eqnarray}
&&s = (p_{a} + p_{b})^{2} = (p_{1} + p_{2} + p_{3} + p_{4})^{2}, 
\quad s_{34} = M_{\pi \pi}^{2} = (p_{3} + p_{4})^{2},\nonumber \\
&&t_1 = q_{1}^{2}, \quad t_2 = q_{2}^{2}, \quad q_1 = p_{a} - p_{1}, \quad q_2 = p_{b} - p_{2} \,.
\label{2to4_kinematic}
\end{eqnarray}
%
Here $\Delta^{(\Pom)}$ and $\Gamma^{(\Pom pp)}$ 
denote the effective propagator and proton vertex function, respectively, for the tensorial pomeron.
For the explicit expressions, see Sec.~3 of \cite{Ewerz:2013kda}.
The normal pion propagator is $i\Delta^{(\pi)}(k) = i/(k^{2}-m_{\pi}^{2})$.
In a similar way the $\Pom f_{2 \Reg}$, $f_{2 \Reg} \Pom$ 
and $f_{2 \Reg} f_{2 \Reg}$ amplitudes can be written.

The propagator of the tensor-pomeron exchange is written as
(see Eq.~(3.10) of \cite{Ewerz:2013kda}):
\begin{eqnarray}
i \Delta^{(\Pom)}_{\mu \nu, \kappa \lambda}(s,t) = 
\frac{1}{4s} \left( g_{\mu \kappa} g_{\nu \lambda} 
                  + g_{\mu \lambda} g_{\nu \kappa}
                  - \frac{1}{2} g_{\mu \nu} g_{\kappa \lambda} \right)
(-i s \alpha'_{\Pom})^{\alpha_{\Pom}(t)-1}
\label{pomeron_propagator}
\end{eqnarray}
and fulfils the following relations
\begin{equation} 
\begin{split}
&\Delta^{(\Pom)}_{\mu \nu, \kappa \lambda}(s,t) = 
\Delta^{(\Pom)}_{\nu \mu, \kappa \lambda}(s,t) =
\Delta^{(\Pom)}_{\mu \nu, \lambda \kappa}(s,t) =
\Delta^{(\Pom)}_{\kappa \lambda, \mu \nu}(s,t) \,, \\
&g^{\mu \nu} \Delta^{(\Pom)}_{\mu \nu, \kappa \lambda}(s,t) = 0, \quad 
g^{\kappa \lambda} \Delta^{(\Pom)}_{\mu \nu, \kappa \lambda}(s,t) = 0 \,.
\end{split}
\label{pomeron_propagator_aux}
\end{equation}
For the $f_{2 \Reg}$ reggeon exchange a similar form of the effective propagator 
and the $f_{2 \Reg}pp$ and $f_{2 \Reg} \pi \pi$ effective vertices is assumed, 
see (3.12) and (3.49), (3.53) of \cite{Ewerz:2013kda}.
Here the pomeron and reggeon trajectories $\alpha_{i}(t)$, where $i = \Pom, \Reg$, 
are assumed to be of standard linear forms, see e.g. \cite{Donnachie:2002en},
\begin{eqnarray}
&&\alpha_{\Pom}(t) = \alpha_{\Pom}(0)+\alpha'_{\Pom}\,t, \quad 
\alpha_{\Pom}(0) = 1.0808,\quad \alpha'_{\Pom} = 0.25 \; \mathrm{GeV}^{-2}\,,
\\
&&\alpha_{\Reg}(t) = \alpha_{\Reg}(0)+\alpha'_{\Reg}\,t, \quad 
\alpha_{\Reg}(0) = 0.5475,\quad \alpha'_{\Reg} = 0.9 \; \mathrm{GeV}^{-2}\,.
\label{trajectory}
\end{eqnarray}

The corresponding coupling of tensor pomeron to protons (antiprotons)
including a vertex form-factor, 
is written as (see Eq.~(3.43) of \cite{Ewerz:2013kda}):
\begin{eqnarray}
&&i\Gamma_{\mu \nu}^{(\Pom pp)}(p',p)= 
i\Gamma_{\mu \nu}^{(\Pom \bar{p} \bar{p})}(p',p)
\nonumber\\
&& \qquad =-i 3 \beta_{\Pom NN} F_{1}\bigl((p'-p)^2\bigr)
\left\lbrace 
\frac{1}{2} 
\left[ \gamma_{\mu}(p'+p)_{\nu} 
     + \gamma_{\nu}(p'+p)_{\mu} \right]
- \frac{1}{4} g_{\mu \nu} ( p\!\!\!/' + p\!\!\!/ )
\right\rbrace , \qquad \;\;\;
\label{vertex_pomNN}
\end{eqnarray}
where 
$\beta_{\Pom NN} = 1.87$~GeV$^{-1}$.
Starting with the $\Pom \pi \pi$ coupling Lagrangian, 
see Eq.~(7.3) of \cite{Ewerz:2013kda},
we have the following $\Pom \pi \pi$ vertex (see Eq.~(3.45) of \cite{Ewerz:2013kda})
%
\begin{eqnarray}
i\Gamma_{\mu \nu}^{(\Pom \pi \pi)}(k',k)=
-i 2 \beta_{\Pom \pi \pi} 
\left[ (k'+k)_{\mu}(k'+k)_{\nu} - \frac{1}{4} g_{\mu \nu} (k' + k)^{2} \right] \, F_{M}((k'-k)^2)\,,
\label{vertex_pompipi}
\end{eqnarray}
where $\beta_{\Pom \pi \pi} = 1.76$~GeV$^{-1}$
gives proper phenomenological normalization.
The form factors 
taking into account that the hadrons are extended objects 
(see Section~3.2 of \cite{Donnachie:2002en})
are chosen as 
%
%
%
\begin{eqnarray}
F_{1}(t)= \frac{4 m_{p}^{2}-2.79\,t}{(4 m_{p}^{2}-t)(1-t/m_{D}^{2})^{2}}\,, \qquad
F_{M}(t)=
\frac{1}{1-t/\Lambda_{0}^{2}}\,,
\label{Fpion}
\end{eqnarray}
where $m_{p}$ is the proton mass and $m_{D}^{2} = 0.71$~GeV$^{2}$
is the dipole mass squared and $\Lambda_{0}^{2} = 0.5$~GeV$^{2}$; 
see Eq.~(3.34) of \cite{Ewerz:2013kda}.
Alternatively, instead of the product of the form factors
$F_{1}(t) F_{M}(t)$ where the two factors 
are attached to the relevant vertices (see Eqs.~(\ref{vertex_pomNN}) and (\ref{vertex_pompipi}))
we can take single form factors $F^{(\Pom/\Reg)}_{\pi N}(t)$ in the exponential form
where the pomeron/reggeon slope parameters have been estimated 
from a fit to the $\pi p$ elastic scattering data,
see Eq.~(2.11) and Fig.~3 of \cite{Lebiedowicz:2015eka}.

The amplitudes (\ref{amplitude_t}) and (\ref{amplitude_u})
must be ``corrected'' for the off-shellness of the intermediate pions.
%
%
%
%
The form of the off-shell pion form factor is unknown in particular 
at higher values of $p_{t}^{2}$ or $p_{u}^{2}$.
The form factors are normalized to unity at
the on-shell point $\hat{F}_{\pi}(m_{\pi}^{2}) = 1$
and parametrised here in two ways:
%
\begin{eqnarray} 
&&\hat{F}_{\pi}(k^{2})=
\exp\left(\frac{k^{2}-m_{\pi}^{2}}{\Lambda^{2}_{off,E}}\right) \,,
\label{off-shell_form_factors_exp} \\
&&\hat{F}_{\pi}(k^{2})=
\dfrac{\Lambda^{2}_{off,M} - m_{\pi}^{2}}{\Lambda^{2}_{off,M} - k^{2}} \,,
\label{off-shell_form_factors_mon} 
\end{eqnarray}
where $\Lambda^{2}_{off,E}$ or $\Lambda^{2}_{off,M}$ could be
adjusted to experimental data.
It was shown in Fig.~9 of \cite{Lebiedowicz:2015eka} that
the monopole form (\ref{off-shell_form_factors_mon})
is supported by the preliminary CDF results \cite{Albrow_Project_new}
particularly at higher values of two-pion invariant mass, $M_{\pi\pi} > 1.5$~GeV.
Thus, in the numerical calculations below,
see section~\ref{sec:section_4},
we used the monopole form of the off-shell pion form factors.

In the high-energy small-angle approximation we have
%
\begin{equation}
\begin{split}
\bar{u}(p', \lambda') \,
\gamma_{\mu}(p'+p)_{\nu} \,
u(p, \lambda) 
\rightarrow (p' + p)_{\mu} (p' + p)_{\nu}\, 
\delta_{\lambda' \lambda}\,
\end{split}
\label{amplitude_approx}
\end{equation}
and we can write the leading terms of the amplitudes for the $pp \to pp \pi^{+} \pi^{-}$ process as
\begin{equation}
\begin{split}
& {\cal M}^{({\rm \hat{t}})}_{\lambda_{a} \lambda_{b} \to \lambda_{1} \lambda_{2} \pi^{+}\pi^{-}} 
\simeq 
3 \beta_{\Pom NN}  \, 2 (p_1 + p_a)_{\mu_{1}} (p_1 + p_a)_{\nu_{1}}\, 
\delta_{\lambda_{1} \lambda_{a}} \,F_{1}(t_{1}) F_{M}(t_{1}) \\
& \quad \times  2 \beta_{\Pom \pi \pi} \, (p_{t}-p_{3})^{\mu_{1}}(p_{t}-p_{3})^{\nu_{1}} \,
\frac{1}{4 s_{13}} (- i s_{13} \alpha'_{\Pom})^{\alpha_{\Pom}(t_{1})-1}
\frac{[\hat{F}_{\pi}(p_{t}^{2})]^{2}}{p_{t}^{2}-m_{\pi}^{2}}  \\
& \quad \times  
2 \beta_{\Pom \pi \pi} \, (p_{4}+p_{t})^{\mu_{2}}(p_{4}+p_{t})^{\nu_{2}} \,
\frac{1}{4 s_{24}} (- i s_{24} \alpha'_{\Pom})^{\alpha_{\Pom}(t_{2})-1} \\
& \quad \times  
3 \beta_{\Pom NN}  \, 2 (p_2 + p_b)_{\mu_{2}} (p_2 + p_b)_{\nu_{2}}\, 
\delta_{\lambda_{2} \lambda_{b}}\, F_{1}(t_{2}) F_{M}(t_{2}) \,,
\end{split}
\label{amplitude_t_approx}
\end{equation}
\begin{equation}
\begin{split}
& {\cal M}^{({\rm \hat{u}})}_{\lambda_{a} \lambda_{b} \to \lambda_{1} \lambda_{2} \pi^{+}\pi^{-}} 
\simeq 
3 \beta_{\Pom NN}  \, 2 (p_1 + p_a)_{\mu_{1}} (p_1 + p_a)_{\nu_{1}}\, 
\delta_{\lambda_{1} \lambda_{a}}\, F_{1}(t_{1}) F_{M}(t_{1}) \\
& \quad \times  2 \beta_{\Pom \pi \pi} \, (p_{4}+p_{u})^{\mu_{1}}(p_{4}+p_{u})^{\nu_{1}} \,
\frac{1}{4 s_{14}} (- i s_{14} \alpha'_{\Pom})^{\alpha_{\Pom}(t_{1})-1}
\frac{[\hat{F}_{\pi}(p_{u}^{2})]^{2}}{p_{u}^{2}-m_{\pi}^{2}} \\
& \quad \times  
2 \beta_{\Pom \pi \pi} \, (p_{u}-p_{3})^{\mu_{2}}(p_{u}-p_{3})^{\nu_{2}} \,
\frac{1}{4 s_{23}} (- i s_{23} \alpha'_{\Pom})^{\alpha_{\Pom}(t_{2})-1}  \\
& \quad \times  
3 \beta_{\Pom NN}  \, 2 (p_2 + p_b)_{\mu_{2}} (p_2 + p_b)_{\nu_{2}}\, 
\delta_{\lambda_{2} \lambda_{b}}\, F_{1}(t_{2}) F_{M}(t_{2}) \,.
\end{split}
\label{amplitude_u_approx}
\end{equation}
%

Now, we consider the vector pomeron exchange model.
We have the following ansatz for the $\Pom_{V} \pi^{-} \pi^{-}$ vertex
omitting the form factors ($M_{0} \equiv 1$~GeV)
\begin{eqnarray}
i\Gamma_{\mu}^{(\Pom_{V} \pi^{-} \pi^{-})}(k',k)=
-i 2 \beta_{\Pom \pi \pi} M_{0} (k'+k)_{\mu}\,.
\label{vertex_pompipi_vec_Eq1}
\end{eqnarray}
%
From isospin and charge-conjugation invariance we should have
\begin{eqnarray}
i\Gamma_{\mu}^{(\Pom_{V} \pi^{+} \pi^{+})}(k',k)=
i\Gamma_{\mu}^{(\Pom_{V} \pi^{-} \pi^{-})}(k',k) \,.
\label{vertex_pompipi_vec_Eq2}
\end{eqnarray}
But the crossing relations require
\begin{eqnarray}
i\Gamma_{\mu}^{(\Pom_{V} \pi^{+} \pi^{+})}(k',k)=
i\Gamma_{\mu}^{(\Pom_{V} \pi^{-} \pi^{-})}(-k,-k') \,.
\label{vertex_pompipi_vec_Eq3}
\end{eqnarray}
And from (\ref{vertex_pompipi_vec_Eq1}) we get
\begin{eqnarray}
i\Gamma_{\mu}^{(\Pom_{V} \pi^{-} \pi^{-})}(-k,-k')=
-i\Gamma_{\mu}^{(\Pom_{V} \pi^{-} \pi^{-})}(k',k) \,.
\label{vertex_pompipi_vec_Eq4}
\end{eqnarray}
Clearly, (\ref{vertex_pompipi_vec_Eq2}) 
and (\ref{vertex_pompipi_vec_Eq3}) plus (\ref{vertex_pompipi_vec_Eq4})
would lead to $i\Gamma_{\mu}^{(\Pom_{V} \pi \pi)}(k',k) \equiv 0$.
This is another manifestation that 
the $\Pom_{V} \pi \pi$ coupling for a vector pomeron has basic problems;
see also the discussion in Sec.~6.1 of \cite{Ewerz:2013kda}.

\section{Dipion resonant production}


In this section we consider the production
of $s$-channel resonances which decay to $\pi^+ \pi^-$
\begin{eqnarray}
p + p \to p + ({\rm resonance} \to \pi^+ \pi^-) + p \,.
\label{2to4_reaction_res}
\end{eqnarray}
The resonances which should be taken into account here and their
production modes via $(C_{1},C_{2})$ fusion are listed in Table~\ref{tab:table2}.
\begin{table}
\caption{
Resonances and $(C_{1},C_{2})$ production modes.}
\label{tab:table2}
\begin{tabular}{|c|c|c|}
\hline
$I^{G}J^{PC}$  & resonance & production $(C_{1},C_{2})$\\
\hline
$0^{+}0^{++}$ & $f_{0}(500)$ & $(\Pom + f_{2 \Reg}, \Pom + f_{2 \Reg})$, $(a_{2 \Reg}, a_{2 \Reg})$,\\
 & $f_{0}(980)$ & $(\Ode + \omega_{\Reg} + \gamma, \Ode + \omega_{\Reg} + \gamma)$, $(\rho_{\Reg}, \rho_{\Reg})$,\\
 & $f_{0}(1370)$ & $(\gamma, \rho_{\Reg})$, $(\rho_{\Reg}, \gamma)$\\
 & $f_{0}(1500)$ & \\ 
 & $f_{0}(1710)$ & \\
\hline
$1^{+}1^{--}$ & $\rho(770)$ & $(\gamma + \rho_{\Reg}, \Pom + f_{2 \Reg})$, $(\Pom + f_{2 \Reg}, \gamma + \rho_{\Reg})$,\\
 & $\rho(1450)$ & $(\Ode + \omega_{\Reg}, a_{2 \Reg})$, $(a_{2 \Reg},\Ode + \omega_{\Reg})$\\
 & $\rho(1700)$ & \\
\hline
$0^{+}2^{++}$ & $f_{2}(1270)$ & $(\Pom + f_{2 \Reg}, \Pom + f_{2 \Reg})$, $(a_{2 \Reg}, a_{2 \Reg})$,\\
 & $f_{2}'(1525)$ & $(\Ode + \omega_{\Reg} + \gamma, \Ode + \omega_{\Reg} + \gamma)$, $(\rho_{\Reg}, \rho_{\Reg})$,\\
 & $f_{2}(1950)$ & $(\gamma, \rho_{\Reg})$, $(\rho_{\Reg}, \gamma)$\\
\hline
$1^{+}3^{--}$ & $\rho_{3}(1690)$ & $(\gamma + \rho_{\Reg}, \Pom + f_{2 \Reg})$, $(\Pom + f_{2 \Reg}, \gamma + \rho_{\Reg})$,\\
 &  & $(\Ode + \omega_{\Reg}, a_{2 \Reg})$, $(a_{2 \Reg},\Ode + \omega_{\Reg})$\\
\hline
$0^{+}4^{++}$ & $f_{4}(2050)$ & $(\Pom + f_{2 \Reg}, \Pom + f_{2 \Reg})$, $(a_{2 \Reg}, a_{2 \Reg})$,\\
 &  & $(\Ode + \omega_{\Reg} + \gamma, \Ode + \omega_{\Reg} + \gamma)$, $(\rho_{\Reg}, \rho_{\Reg})$,\\
 &  & $(\gamma, \rho_{\Reg})$, $(\rho_{\Reg}, \gamma)$\\
\hline
\end{tabular}
\end{table}

\begin{figure}
\includegraphics[width=8.cm]{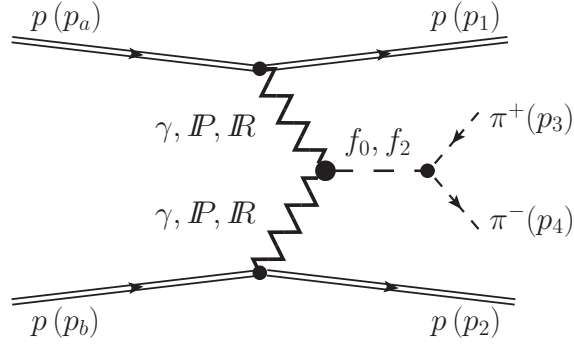}      
  \caption{\label{fig:resonant_Born}
  \small The Born diagram for double-pomeron/reggeon and photon mediated
  central exclusive $I^{G}J^{PC} = 0^{+}0^{++}$ and $0^{+}2^{++}$ resonances production 
  and their subsequent decays into $\pi^+ \pi^-$ in proton-proton collisions.
}
\end{figure}

The production of $\rho(770)$ and $\rho(1450)$ was already treated in \cite{Lebiedowicz:2014bea}.
Here we shall discuss the production of the $f_{0}$ and $f_{2}$ resonances;
see Fig.~\ref{fig:resonant_Born}.
We shall concentrate on the contributions from 
$(C_{1},C_{2})$ = $(\Pom + f_{2 \Reg}, \Pom + f_{2 \Reg})$.
We can justify this  as follows.
The contributions involving the odderon (if it exists at all),
the $a_{2 \Reg}$ and the $\rho_{\Reg}$ should be small
due to small couplings of these objects to the proton.
The secondary reggeons $a_{2 \Reg}$, $\omega_{\Reg}$, $\rho_{\Reg}$
should give small contributions at high energies.
We shall neglect contributions involving the photon $\gamma$
in the following. These are expected to become important
only for very small values of $|t_{1}|$ and/or $|t_{2}|$.
Thus we are left with $(\Pom + f_{2 \Reg}, \Pom + f_{2 \Reg})$
where $(\Pom,\Pom)$ fusion is the leading term and
$(\Pom,f_{2 \Reg})$ plus $(f_{2 \Reg},\Pom)$ is the first
non-leading term due to reggeons at high energies.

The amplitude for exclusive resonant $\pi^+ \pi^-$ production,
given by the diagram shown in Fig.~\ref{fig:resonant_Born},
can be written as
\begin{equation}
\begin{split}
{\cal M}^{\pi \pi{\rm-resonances}}_{pp \to pp \pi^{+} \pi^{-}}
= {\cal M}^{(\Pom \Pom \to f_{0} \to \pi^{+}\pi^{-})}_{\lambda_{a} \lambda_{b} \to \lambda_{1} \lambda_{2} \pi^{+}\pi^{-}}
 + {\cal M}^{(\Pom \Pom \to f_{2} \to \pi^{+}\pi^{-})}_{\lambda_{a} \lambda_{b} \to \lambda_{1} \lambda_{2} \pi^{+}\pi^{-}}\,.
\end{split}
\label{amplitude_f0f2_pomTpomT}
\end{equation}
%
\subsection{$I^{G}J^{PC} = 0^{+}0^{++}$}
For a scalar meson, $J^{PC} = 0^{++}$, the amplitude for $\Pom \Pom$ fusion can be written as
\begin{equation}
\begin{split}
{\cal M}^{(\Pom \Pom \to f_{0} \to \pi^{+}\pi^{-})}_{\lambda_{a} \lambda_{b} \to \lambda_{1} \lambda_{2} \pi^{+}\pi^{-}} 
= & (-i)\,
\bar{u}(p_{1}, \lambda_{1}) 
i\Gamma^{(\Pom pp)}_{\mu_{1} \nu_{1}}(p_{1},p_{a}) 
u(p_{a}, \lambda_{a})\;
i\Delta^{(\Pom)\, \mu_{1} \nu_{1}, \alpha_{1} \beta_{1}}(s_{1},t_{1}) \\
& \times 
i\Gamma^{(\Pom \Pom f_{0})}_{\alpha_{1} \beta_{1},\alpha_{2} \beta_{2}}(q_{1},q_{2}) \;
i\Delta^{(f_{0})}(p_{34})\;
i\Gamma^{(f_{0} \pi \pi)}(p_{34})\\
& \times 
i\Delta^{(\Pom)\, \alpha_{2} \beta_{2}, \mu_{2} \nu_{2}}(s_{2},t_{2}) \;
\bar{u}(p_{2}, \lambda_{2}) 
i\Gamma^{(\Pom pp)}_{\mu_{2} \nu_{2}}(p_{2},p_{b}) 
u(p_{b}, \lambda_{b}) \,,
\end{split}
\label{amplitude_f0_pomTpomT}
\end{equation}
where
$s_{1} = (p_{a} + q_{2})^{2} = (p_{1} + p_{34})^{2}$,
$s_{2} = (p_{b} + q_{1})^{2} = (p_{2} + p_{34})^{2}$, and
$p_{34} = p_{3} + p_{4}$.
The effective Lagrangians and the vertices for $\Pom \Pom$ fusion into the $f_{0}$ meson 
are discussed in Appendix~A of \cite{Lebiedowicz:2013ika}.
As was shown there the tensorial $\Pom \Pom f_{0}$ vertex 
corresponds to the sum of two lowest values of $(l,S)$, 
that is $(l,S) = (0,0)$ and $(2,2)$
with the corresponding coupling parameters 
$g_{\Pom \Pom M}'$ and $g_{\Pom \Pom M}''$, respectively.
The vertex including a form factor reads then as follows ($p_{34} = q_{1} + q_{2}$)
\begin{eqnarray}
i\Gamma_{\mu \nu,\kappa \lambda}^{(\Pom \Pom f_{0})} (q_{1},q_{2}) =
\left( i\Gamma_{\mu \nu,\kappa \lambda}'^{(\Pom \Pom f_{0})}\mid_{bare} +
       i\Gamma_{\mu \nu,\kappa \lambda}''^{(\Pom \Pom f_{0})} (q_{1}, q_{2})\mid_{bare} \right)
\tilde{F}^{(\Pom \Pom f_{0})}(q_{1}^{2},q_{2}^{2},p_{34}^{2}) \,;
\label{vertex_pompomS}
\end{eqnarray}
see (A.21) of \cite{Lebiedowicz:2013ika}.
Unfortunately, the pomeron-pomeron-meson form factor
is not well known as it is due to
nonperturbative effects related to the internal structure of the respective meson.
In practical calculations we take the factorized form
for the $\Pom \Pom f_{0}$ form factor
%
\begin{eqnarray}
\tilde{F}^{(\Pom \Pom f_{0})}(q_{1}^{2},q_{2}^{2},p_{34}^{2}) = 
F_{M}(q_{1}^{2}) F_{M}(q_{2}^{2}) F^{(\Pom \Pom f_{0})}(p_{34}^{2})\,
\label{Fpompommeson}
\end{eqnarray}
normalised to
$\tilde{F}^{(\Pom \Pom f_{0})}(0,0,m_{f_{0}}^{2}) = 1$.
%
We will further set
\begin{eqnarray}
F^{(\Pom \Pom f_{0})}(p_{34}^{2}) = 
\exp{ \left( \frac{-(p_{34}^{2}-m_{f_{0}}^{2})^{2}}{\Lambda_{f_{0}}^{4}} \right)}\,,
\quad \Lambda_{f_{0}} = 1\;{\rm GeV}\,.
\label{Fpompommeson_ff}
\end{eqnarray}
%

The scalar-meson propagator is taken as
\begin{eqnarray}
i\Delta^{(f_{0})}(p_{34}) = \dfrac{i}{p_{34}^{2}-m_{f_{0}}^2+i m_{f_{0}} \Gamma_{f_{0}}(p_{34}^{2})}\,,
\label{prop_scalar}
\end{eqnarray}
where the running (energy-dependent) width is parametrized as
\begin{eqnarray}
\Gamma_{f_{0}}(p_{34}^{2}) = \Gamma_{f_{0}}
\left( \frac{p_{34}^{2} - 4 m_{\pi}^2}{m_{f_{0}}^2 - 4 m_{\pi}^2} \right)^{1/2}
\theta(p_{34}^{2}-4 m_{\pi}^2)\,.
\end{eqnarray}
For the $f_{0} \pi \pi$ vertex we have ($M_{0} \equiv 1$~GeV)
\begin{eqnarray}
i\Gamma^{(f_{0} \pi \pi)}(p_{34}) = i g_{f_{0} \pi \pi} M_{0}\, F^{(f_{0} \pi \pi)}(p_{34}^{2})\,,
\end{eqnarray}
where $g_{f_{0} \pi \pi}$ is related to the partial decay width of the $f_{0}$ meson
(for an 'on-shell' $f_{0}$ state $p_{34}^{2} = m_{f_{0}}^{2}$)
\begin{eqnarray}
\Gamma(f_{0} \to \pi \pi) = 3 \Gamma(f_{0} \to \pi^{0} \pi^{0}) 
= \frac{3}{2} \Gamma(f_{0} \to \pi^{+} \pi^{-}) = \frac{3}{2} \frac{M_{0}^{2}}{16 \pi m_{f_{0}}}
|g_{f_{0} \pi \pi}|^{2} \left( 1-\frac{4 m_{\pi}^{2}}{m_{f_{0}}^{2}} \right)^{1/2}.
\nonumber\\
\label{gamma_f0pipi}
\end{eqnarray}
%
%
We also assume that $F^{(f_{0} \pi \pi)}(p_{34}^{2})$ = $F^{(\Pom \Pom f_{0})}(p_{34}^{2})$,
see Eq.~(\ref{Fpompommeson_ff}).
%

In the high-energy small-angle approximation we can write, setting $p_{34}^{2} = s_{34}$,
\begin{eqnarray}
&&{\cal M}^{(\Pom \Pom \to f_{0}\to \pi^{+}\pi^{-})}_{\lambda_{a} \lambda_{b} \to \lambda_{1} \lambda_{2} \pi^{+}\pi^{-}} 
\simeq 3 \beta_{\Pom NN}  \, 2(p_1 + p_a)^{\mu_{1}} (p_1 + p_a)^{\nu_{1}}\, 
\delta_{\lambda_{1} \lambda_{a}}\, F_1(t_1)  \;
\frac{1}{4 s_{1}} (- i s_{1} \alpha'_{\Pom})^{\alpha_{\Pom}(t_{1})-1} \nonumber \\ 
&& \quad \quad
\times 
\left.\Bigl[
g_{\Pom \Pom f_{0}}' M_{0} 
\Bigl( g_{\mu_{1} \mu_{2}} g_{\nu_{1} \nu_{2}} + 
  g_{\mu_{1} \nu_{2}} g_{\nu_{1} \mu_{2}} -
  \frac{1}{2} g_{\mu_{1} \nu_{1}} g_{\mu_{2} \nu_{2}} \Bigr) 
\right.
 \nonumber \\
&& \quad \quad\left. \quad
+ \frac{g_{\Pom \Pom f_{0}}''}{2 M_{0}} \Bigr( 
q_{1 \mu_{2}} q_{2 \mu_{1}} g_{\nu_{1} \nu_{2}} + 
q_{1 \mu_{2}} q_{2 \nu_{1}} g_{\mu_{1} \nu_{2}} +
q_{1 \nu_{2}} q_{2 \mu_{1}} g_{\nu_{1} \mu_{2}} + 
q_{1 \nu_{2}} q_{2 \nu_{1}} g_{\mu_{1} \mu_{2}} \right.
 \nonumber \\
&& \quad \quad\left. \quad 
-2 (q_{1}q_{2}) (g_{\mu_{1} \mu_{2}} g_{\nu_{1} \nu_{2}} 
              + g_{\nu_{1} \mu_{2}} g_{\mu_{1} \nu_{2}} ) \Bigr)
\right.\Bigr]  \nonumber \\
&& \quad \quad\times \dfrac{g_{f_{0} \pi \pi} M_{0}}{s_{34}-m_{f_{0}}^2+i m_{f_{0}} \Gamma_{f_{0}}(s_{34})}
\tilde{F}^{(\Pom \Pom f_{0})}(t_{1},t_{2},s_{34}) F^{(f_{0} \pi \pi)}(s_{34})\nonumber \\
&& \quad \quad\times 
\frac{1}{4 s_{2}} (- i s_{2} \alpha'_{\Pom})^{\alpha_{\Pom}(t_{2})-1}\,
3 \beta_{\Pom NN}  \, 2 (p_2 + p_b)^{\mu_{2}} (p_2 + p_b)^{\nu_{2}}\, 
\delta_{\lambda_{2} \lambda_{b}}\, F_1(t_2) \,.
\label{amplitude_f0_pomTpomT_approx}
\end{eqnarray}

From \cite{Agashe:2014kda} we have for the mass 
and width of the $f_{0}(500)$ and $f_{0}(980)$ mesons
\begin{eqnarray}
&& m_{f_{0}(500)} = 400 - 550 \;{\rm MeV}\,, 
\quad \Gamma_{f_{0}(500)} = 400 - 700 \;{\rm MeV}\,,\\
& &m_{f_{0}(980)} = 990 \pm 20 \;{\rm MeV}\,, 
\quad \Gamma_{f_{0}(980)} = 40 - 100 \;{\rm MeV}\,.
\end{eqnarray}
We get, assuming $\Gamma(f_{0} \to \pi\pi)/\Gamma_{f_{0}} = 100 \%$, 
$m_{f_{0}(500)} = 600$~MeV, $\Gamma_{f_{0}(500)} = 500$~MeV, 
$m_{f_{0}(980)} = 980$~MeV, $\Gamma_{f_{0}(980)} = 70$~MeV, 
and assuming $g_{f_{0} \pi \pi} > 0$
\begin{eqnarray}
g_{f_{0}(500) \pi \pi} = 3.37\,, \quad g_{f_{0}(980) \pi \pi} = 1.55\,.
\label{g_f0pipi}
\end{eqnarray}
%

\subsection{$I^{G}J^{PC} = 0^{+}2^{++}$}
The production of a tensor meson as $f_{2} \equiv f_{2}(1270)$ 
is more complicated to treat.
The amplitude for the $\pi \pi$ production through 
the $s$-channel $f_{2}$-meson exchange 
can be written as
\begin{equation}
\begin{split}
{\cal M}^{(\Pom \Pom \to f_{2}\to \pi^{+}\pi^{-})}_{\lambda_{a} \lambda_{b} \to \lambda_{1} \lambda_{2} \pi^{+}\pi^{-}} 
= & (-i)\,
\bar{u}(p_{1}, \lambda_{1}) 
i\Gamma^{(\Pom pp)}_{\mu_{1} \nu_{1}}(p_{1},p_{a}) 
u(p_{a}, \lambda_{a})\;
i\Delta^{(\Pom)\, \mu_{1} \nu_{1}, \alpha_{1} \beta_{1}}(s_{1},t_{1}) \\
& \times 
i\Gamma^{(\Pom \Pom f_{2})}_{\alpha_{1} \beta_{1},\alpha_{2} \beta_{2}, \rho \sigma}(q_{1},q_{2}) \;
i\Delta^{(f_{2})\,\rho \sigma, \alpha \beta}(p_{34})\;
i\Gamma^{(f_{2} \pi \pi)}_{\alpha \beta}(p_{3},p_{4})\\
& \times 
i\Delta^{(\Pom)\, \alpha_{2} \beta_{2}, \mu_{2} \nu_{2}}(s_{2},t_{2}) \;
\bar{u}(p_{2}, \lambda_{2}) 
i\Gamma^{(\Pom pp)}_{\mu_{2} \nu_{2}}(p_{2},p_{b}) 
u(p_{b}, \lambda_{b}) \,.
\end{split}
\label{amplitude_f2_pomTpomT}
\end{equation}

The pomeron-pomeron-$f_{2}$ coupling is the most complicated element of our amplitudes.
We have considered all possible tensorial structures for the coupling
(see Appendix~\ref{section:Tensorial_Couplings}).
Then, the $\Pom \Pom f_{2}$ vertex can be written as
\begin{eqnarray}
i\Gamma_{\mu \nu,\kappa \lambda,\rho \sigma}^{(\Pom \Pom f_{2})} (q_{1},q_{2}) =
\left( i\Gamma_{\mu \nu,\kappa \lambda,\rho \sigma}^{(\Pom \Pom f_{2})(1)} \mid_{bare}
+ \sum_{j=2}^{7}i\Gamma_{\mu \nu,\kappa \lambda,\rho \sigma}^{(\Pom \Pom f_{2})(j)}(q_{1},q_{2}) \mid_{bare} 
\right)
\tilde{F}^{(\Pom \Pom f_{2})}(q_{1}^{2},q_{2}^{2},p_{34}^{2}) \,.\nonumber\\
\label{vertex_pompomT}
\end{eqnarray}
Here $p_{34} = q_{1} + q_{2}$ and $\tilde{F}^{(\Pom \Pom f_{2})}$
is a form factor for which we make a factorised ansatz
\begin{eqnarray}
\tilde{F}^{(\Pom \Pom f_{2})}(q_{1}^{2},q_{2}^{2},p_{34}^{2}) = 
F_{M}(q_{1}^{2}) F_{M}(q_{2}^{2}) F^{(\Pom \Pom f_{2})}(p_{34}^{2})\,.
\label{Fpompommeson_tensor}
\end{eqnarray}
A possible choice for the 
$i\Gamma_{\mu \nu,\kappa \lambda,\rho \sigma}^{(\Pom \Pom f_{2})(j)}\mid_{bare}$
terms $j = 1, ..., 7$ is given in Appendix~\ref{section:Tensorial_Couplings}.
We are taking here the same form factor for each vertex with index $j$ ($j = 1, ..., 7$).
In principle, we could take a different form factor for each vertex.

Here, for qualitative calculations only, one may use 
the tensor-meson propagator with the simple Breit-Wigner form
\begin{eqnarray}
i\Delta_{\mu \nu, \kappa \lambda}^{(f_{2})}(p_{34})&=&
\frac{i}{p_{34}^{2}-m_{f_{2}}^2+i m_{f_{2}} \Gamma_{f_{2}}}
\left[ 
\frac{1}{2} 
( \hat{g}_{\mu \kappa} \hat{g}_{\nu \lambda}  + \hat{g}_{\mu \lambda} \hat{g}_{\nu \kappa} )
-\frac{1}{3} 
\hat{g}_{\mu \nu} \hat{g}_{\kappa \lambda}
\right] \,, 
\label{prop_f2}
\end{eqnarray}
where $\hat{g}_{\mu \nu} = -g_{\mu \nu} + p_{34 \mu} p_{34 \nu} / p_{34}^2$.
In (\ref{prop_f2}) $\Gamma_{f_{2}}$ is the total decay width of the $f_{2}(1270)$ resonance
and $m_{f_{2}}$ its mass.
The propagator (\ref{prop_f2}) fulfils then the following relations
\begin{eqnarray}
&&\Delta_{\mu \nu, \kappa \lambda}^{(f_{2})}(p_{34})=
\Delta_{\nu \mu, \kappa \lambda}^{(f_{2})}(p_{34})=
\Delta_{\mu \nu, \lambda \kappa}^{(f_{2})}(p_{34})=
\Delta_{\kappa \lambda, \mu \nu}^{(f_{2})}(p_{34})\,,
\label{eqn1}\\
&&g^{\mu \nu} \Delta_{\mu \nu, \kappa \lambda}^{(f_{2})}(p_{34}) = 0, \quad
  g^{\kappa \lambda} \Delta_{\mu \nu, \kappa \lambda}^{(f_{2})}(p_{34}) = 0\,.
\label{eqn2}
\end{eqnarray}

The $f_{2} \pi \pi$ vertex is written as (see Sec.~5.1 and Eqs.~(3.37), (3.38) of \cite{Ewerz:2013kda})
\begin{eqnarray}
i\Gamma_{\mu \nu}^{(f_{2} \pi \pi)}(p_{3},p_{4})=
-i \,\frac{g_{f_{2} \pi \pi}}{2 M_{0}} \,
\left[ (p_{3}-p_{4})_{\mu} (p_{3}-p_{4})_{\nu}
- \frac{1}{4} g_{\mu \nu} (p_{3}-p_{4})^{2} \right] \, F^{(f_{2} \pi \pi)}(p_{34}^{2})\,,
\nonumber \\
\label{vertex_f2pipi_N}
\end{eqnarray}
where $g_{f_{2} \pi \pi} = 9.26$ 
was obtained from the corresponding partial decay width, 
see (5.6) - (5.9) of \cite{Ewerz:2013kda}.
We assume that
\begin{eqnarray}
F^{(f_{2} \pi \pi)}(p_{34}^{2}) = F^{(\Pom \Pom f_{2})}(p_{34}^{2}) = 
\exp{ \left( \frac{-(p_{34}^{2}-m_{f_{2}}^{2})^{2}}{\Lambda_{f_{2}}^{4}} \right)}\,,
\quad \Lambda_{f_{2}} = 1\;{\rm GeV}\,.
\label{Fpompommeson_ff_tensor}
\end{eqnarray}

In the high-energy small-angle approximation we can write, setting $p_{34}^{2} = s_{34}$,
\begin{equation}
\begin{split}
&{\cal M}^{(\Pom \Pom \to f_{2}\to \pi^{+}\pi^{-})}_{\lambda_{a} \lambda_{b} \to \lambda_{1} \lambda_{2} \pi^{+}\pi^{-}} 
\simeq -\,     3 \beta_{\Pom NN}  \, 2 (p_1 + p_a)^{\mu_{1}} (p_1 + p_a)^{\nu_{1}}\, 
\delta_{\lambda_{1} \lambda_{a}}\, F_1(t_1) \, \frac{1}{4 s_{1}} (- i s_{1} \alpha'_{\Pom})^{\alpha_{\Pom}(t_{1})-1}
\\
& \qquad \times  
\Gamma_{\mu_{1} \nu_{1},\mu_{2} \nu_{2}, \rho \sigma}^{(\Pom \Pom f_{2})} (q_{1},q_{2})\,
\Delta^{(f_{2})\,\rho \sigma, \alpha \beta}(p_{34})\;
\frac{g_{f_{2} \pi \pi}}{2 M_{0}} \,
(p_{3}-p_{4})_{\alpha} (p_{3}-p_{4})_{\beta} \, F^{(f_{2} \pi \pi)}(s_{34}) 
\\
& \qquad \times  \frac{1}{4 s_{2}} (- i s_{2} \alpha'_{\Pom})^{\alpha_{\Pom}(t_{2})-1}\,
3 \beta_{\Pom NN} \, 2 (p_2 + p_b)^{\mu_{2}} (p_2 + p_b)^{\nu_{2}}\, 
\delta_{\lambda_{2} \lambda_{b}}\, F_1(t_2) \,.
\end{split}
\label{amplitude_f2_pomTpomT_approx}
\end{equation}
This general form is, however, not easy to be used as we do not
know the normalisation of each of the seven $\Pom \Pom f_{2}$ couplings.
In principle the parameters could be fitted to experimental data.
However, we are not yet ready to perform such an analysis at present.
Instead we will consider properties of each of the individual terms separately.

The production of the $f_{2}$ via $\Pom f_{2 \Reg}$, $f_{2 \Reg} \Pom$,
and $f_{2 \Reg} f_{2 \Reg}$ fusion can be treated 
in a completely analogous way to the $\Pom \Pom$ fusion.
But this would introduce further unknown parameters.
Therefore, we neglect in our present study the above terms which,
anyway, are non-leading at high energies.

\section{Preliminary results for present and future experiments}
\label{sec:section_4}

\begin{figure}[!ht]
\includegraphics[width = 0.45\textwidth]{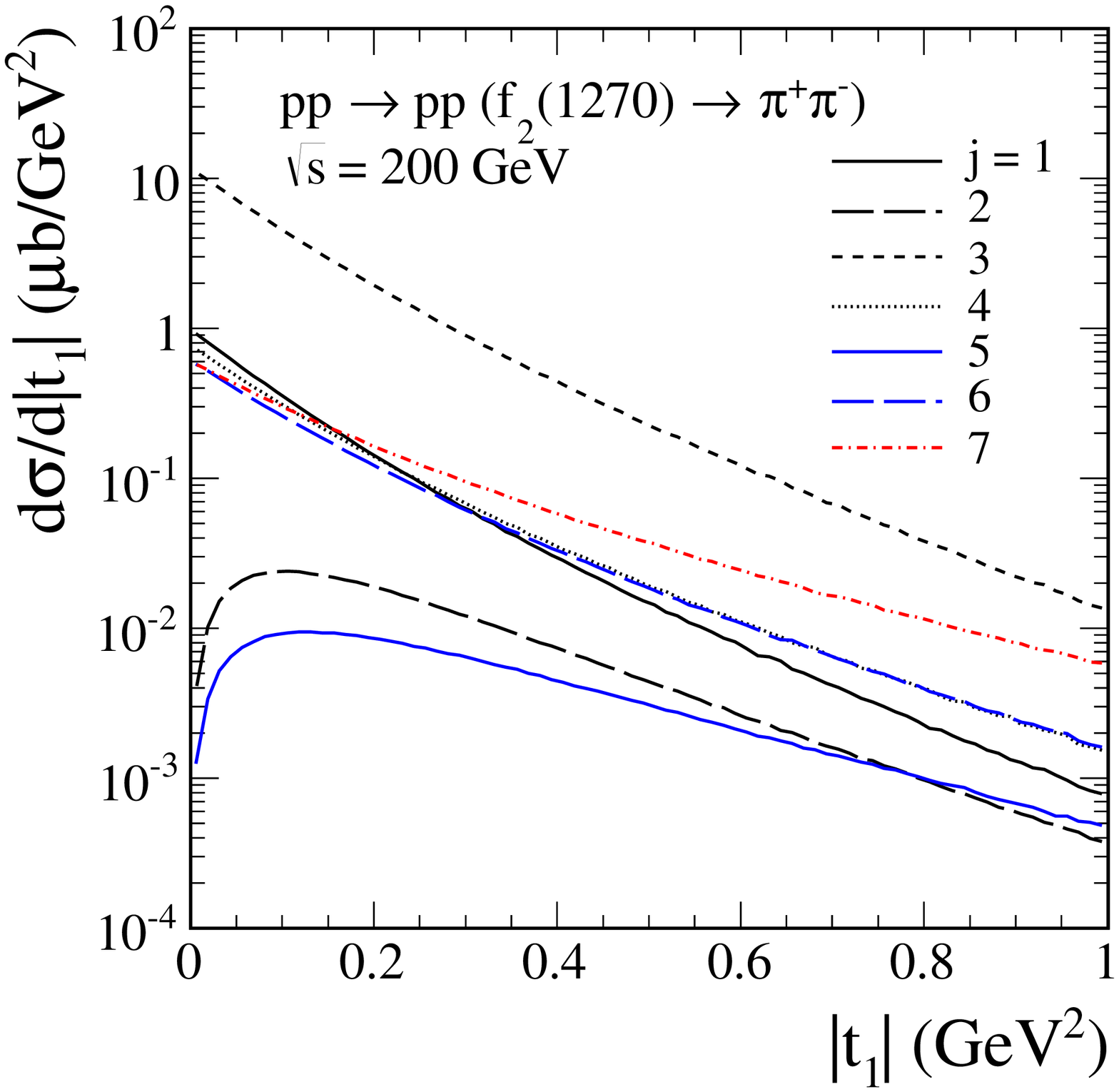}
\includegraphics[width = 0.45\textwidth]{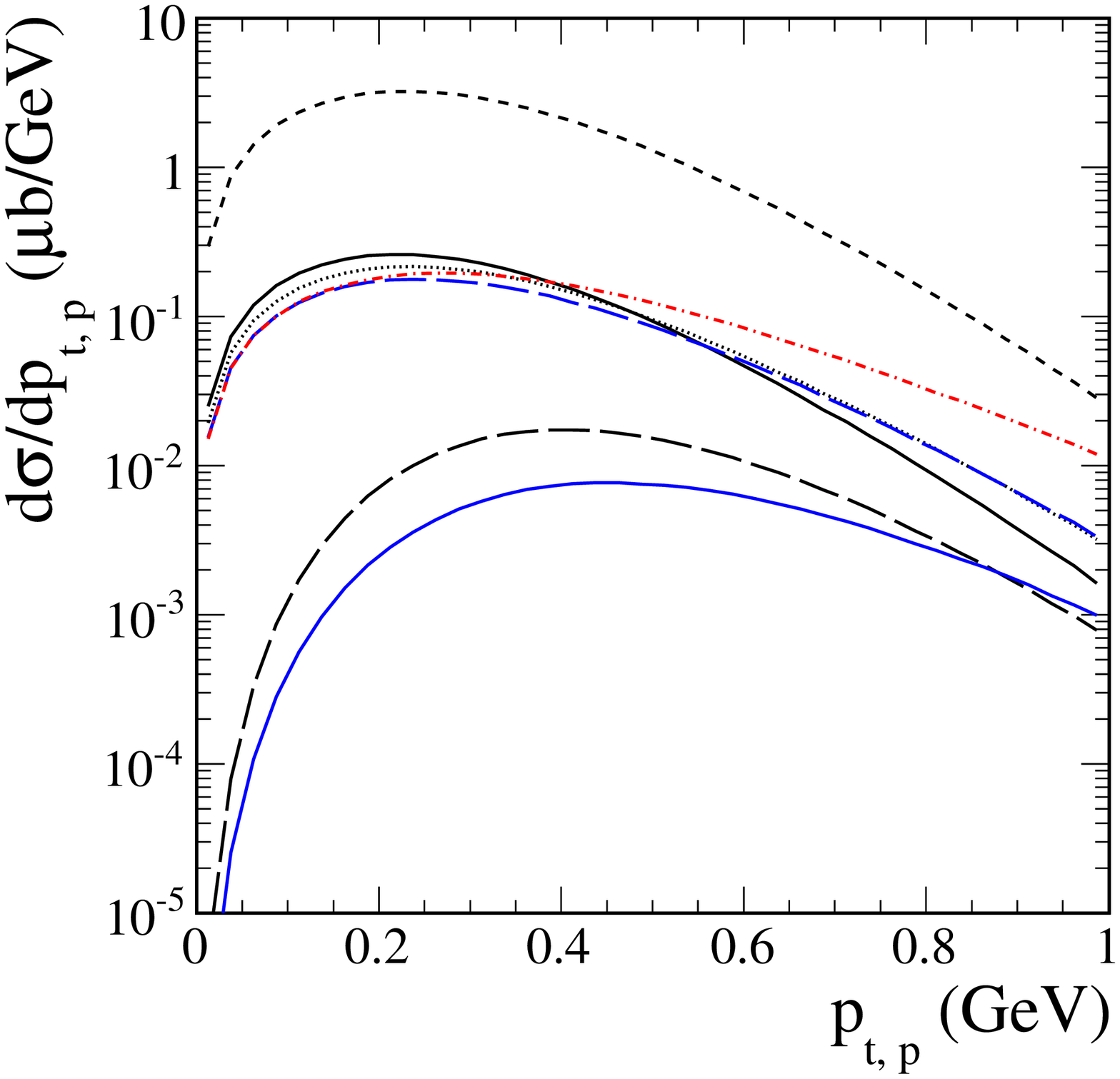}
\includegraphics[width = 0.45\textwidth]{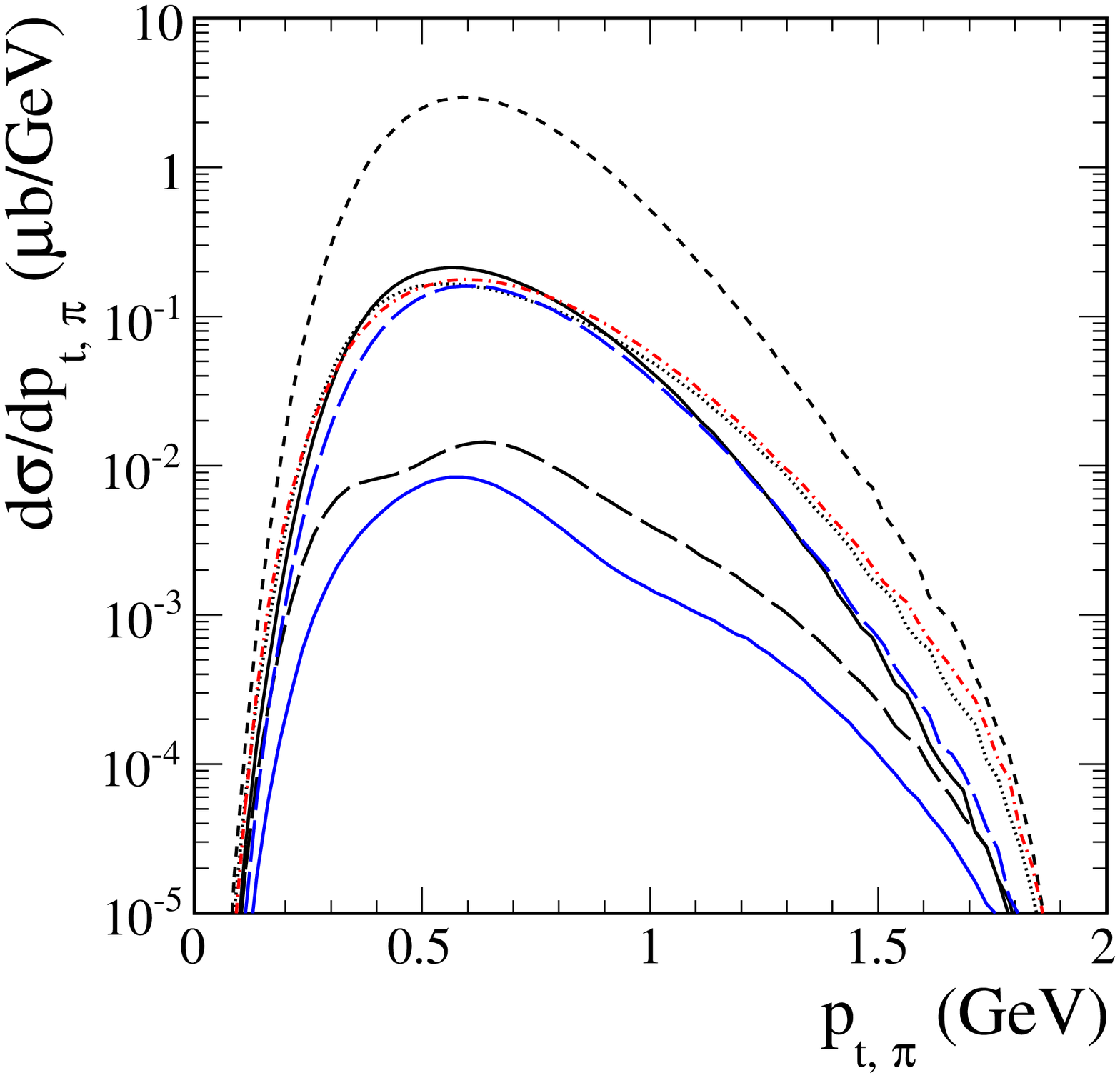}
\includegraphics[width = 0.45\textwidth]{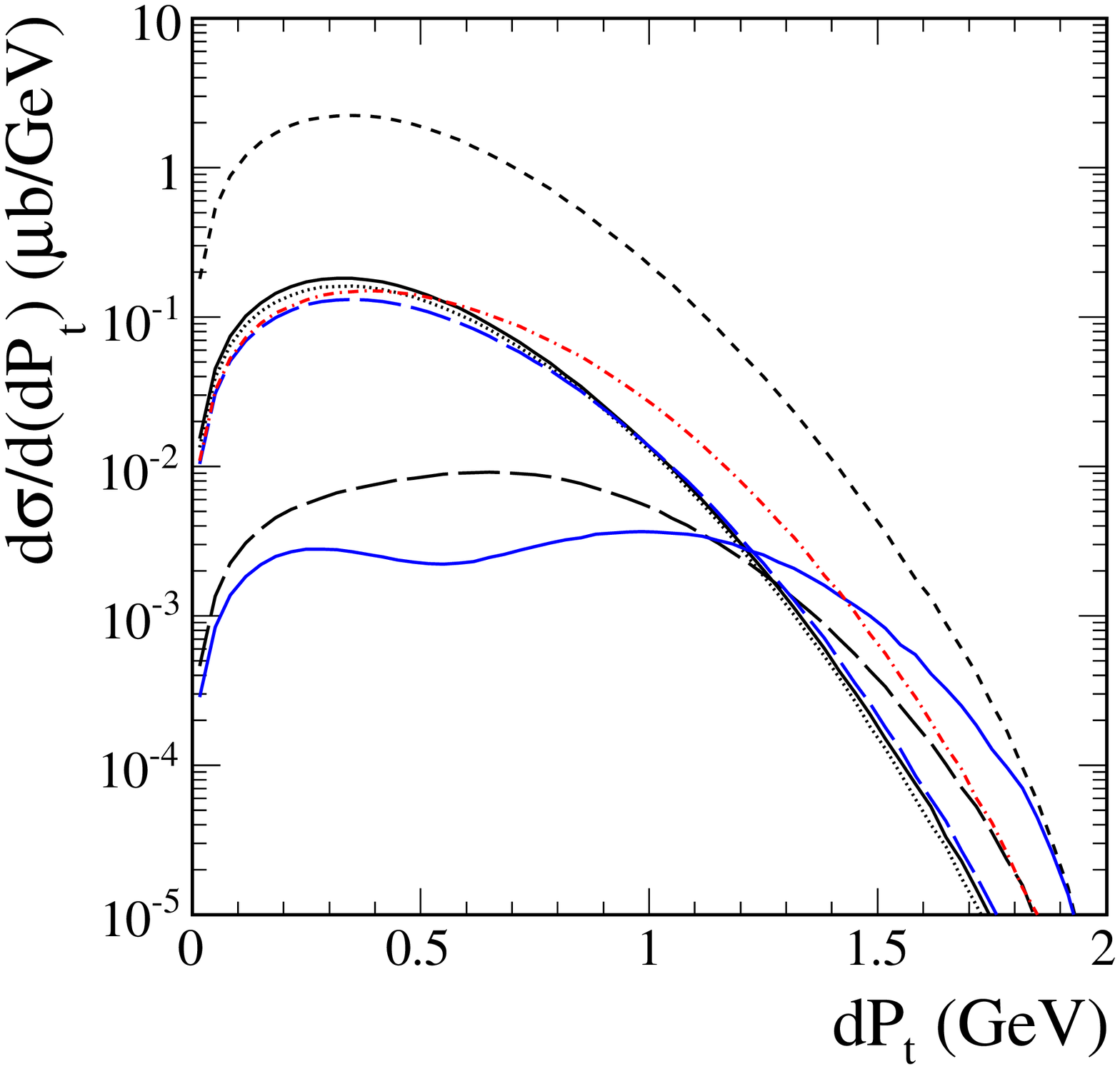}
  \caption{\label{fig:dsig_dt1}
  \small
The differential cross sections for the central exclusive
production of the $f_{2}(1270)$ meson by the fusion of two tensor pomerons 
at $\sqrt{s} = 200$~GeV and $|\eta_{\pi}|< 1$.
We show the individual contributions of the different couplings:
$j = 1$ (the black solid line), $j = 2$ (the black long-dashed line),
$j = 3$ (the black dashed line), $j = 4$ (the black dotted line), 
$j = 5$ (the blue solid line), $j = 6$ (the blue long-dashed line), 
and $j = 7$ (the red dot-dashed line).
For illustration the results have been obtained 
with coupling constants $g^{(j)}_{\Pom \Pom f_{2}} = 1.0$.
No absorption effects were included here.
}
\end{figure}

In this section we show some preliminary results 
of our calculations including the two-pion continuum, 
the $\rho(770)$, $f_0(500)$, $f_0(980)$ and $f_2(1270)$ resonances
which are known to decay into two pions \cite{Agashe:2014kda}.
We start from a discussion of some dependences
for the central exclusive production of the $f_{2}(1270)$ meson
at $\sqrt{s} = 200$~GeV and $|\eta_{\pi}|< 1$.
In Fig.~\ref{fig:dsig_dt1} we present different differential observables
in transferred four-momentum squared $t_{1}$ or $t_{2}$
between the initial and final protons,
in proton $p_{t,p}$ and pion $p_{t,\pi}$ transverse momenta 
as well as in the so-called ``glueball-filter variable'' 
defined by the difference of the transverse momentum vectors
$dP_{t} = |\bdPt|$ with $\bdPt = \bqta - \bqtb = \bptb - \bpta$.
We show results for the individual $j$ coupling terms 
(see Appendix~\ref{section:Tensorial_Couplings}).
The predictions differ considerably which could be checked experimentally.
We find that only in two cases ($j=2$ and 5)
the cross section $d\sigma/d|t|$ vanishes when $|t| \to 0$. 
Another possibility could be that two different amplitudes interfere such as to cancel exactly
for $|t|$ going to zero but cancel no longer for larger $|t|$,
but this seems to be rather improbable.

\begin{figure}[!ht]
\includegraphics[width = 0.45\textwidth]{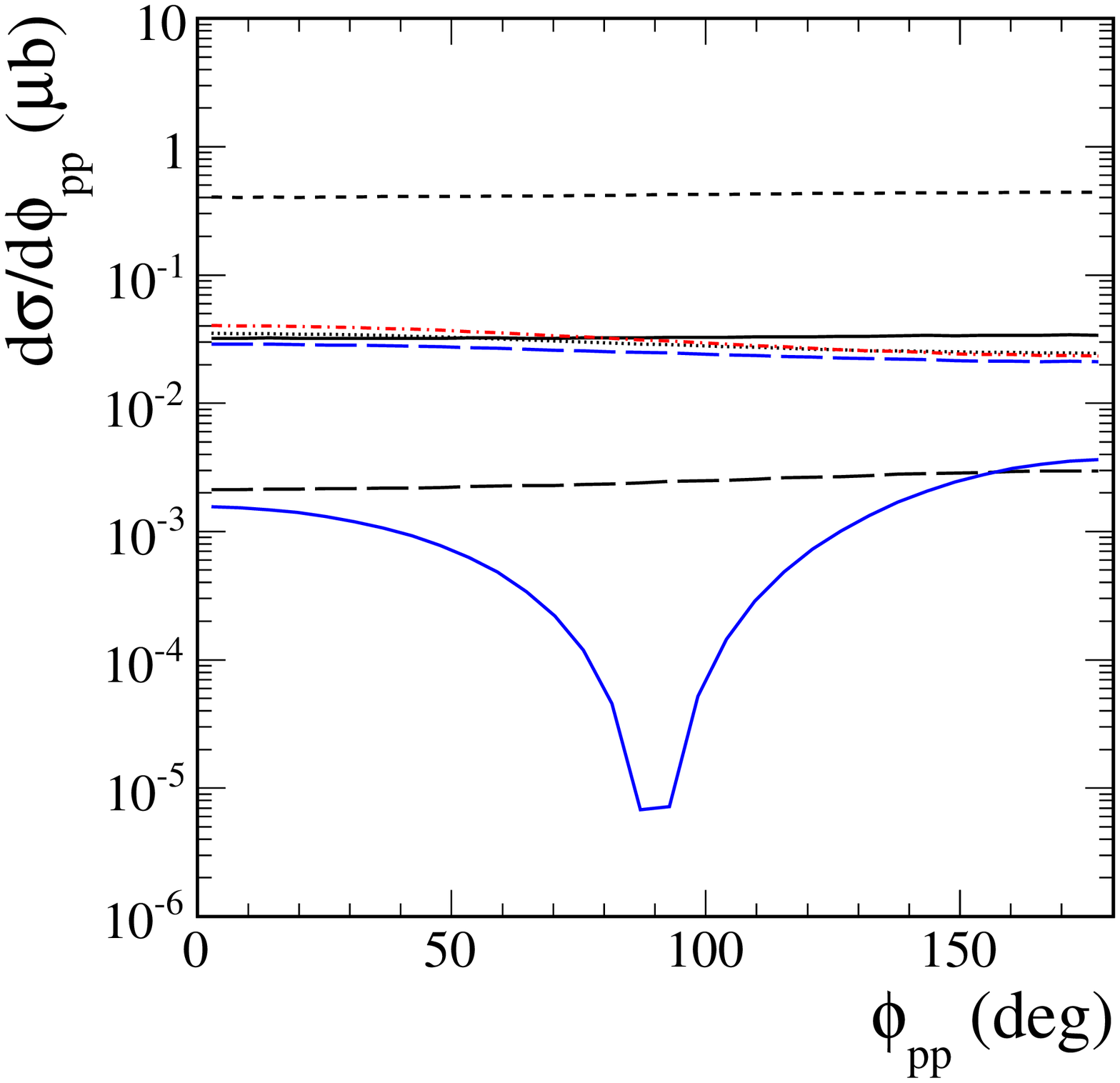}
\includegraphics[width = 0.45\textwidth]{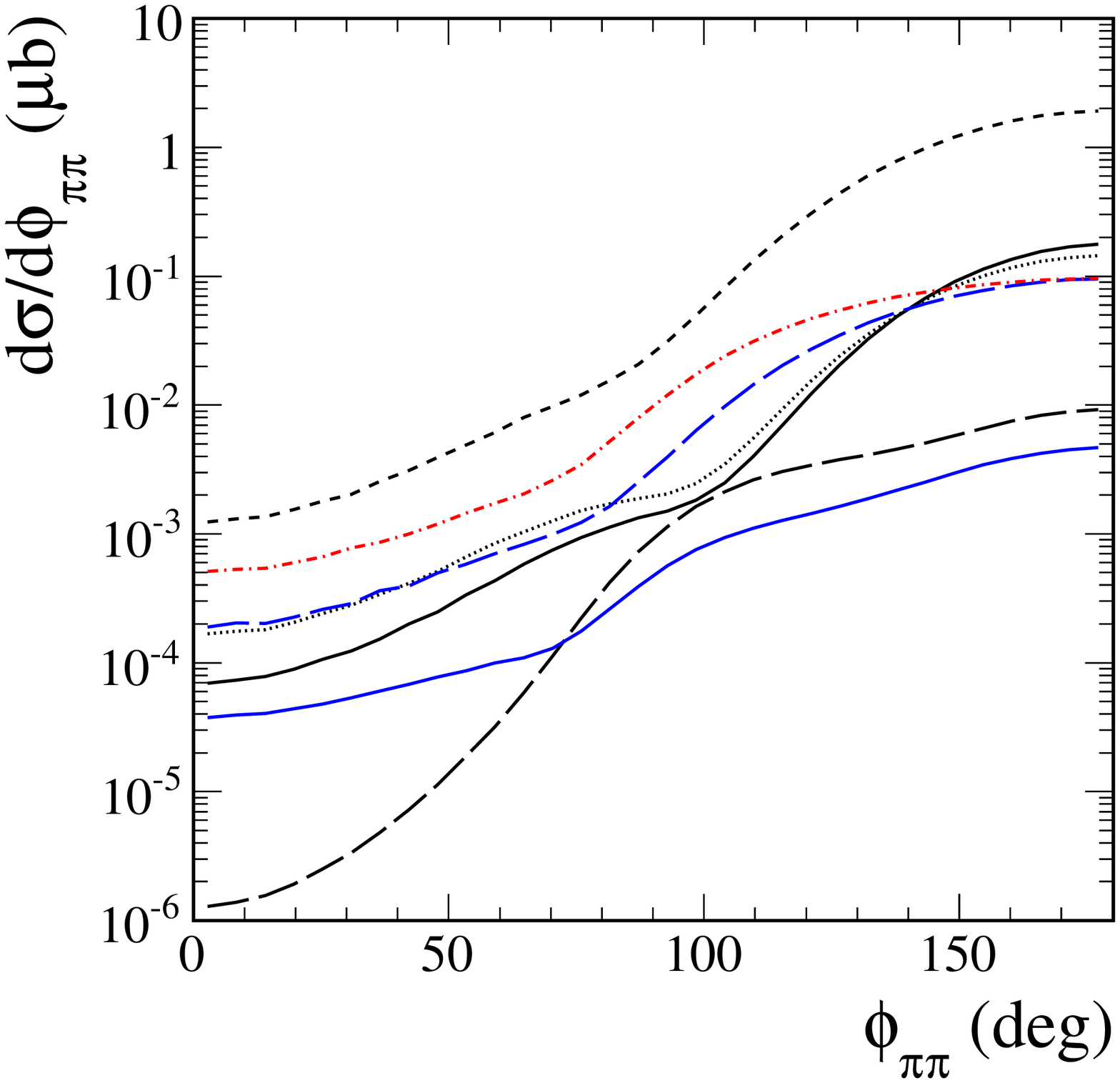}
  \caption{\label{fig:dsig_dphi}
  \small
The distributions in azimuthal angle between the outgoing protons 
(left panel) and between the outgoing pions (right panel)
for $\sqrt{s} = 200$~GeV and $|\eta_{\pi}|< 1$.
The meaning of the lines is the same as in Fig.~\ref{fig:dsig_dt1}.
No absorption effects were included here.
}
\end{figure}

The distributions in azimuthal angle between the outgoing protons, 
$\phi_{pp}$, and outgoing pions, $\phi_{\pi \pi}$,
for the central exclusive production of the $f_{2}(1270)$ meson
at $\sqrt{s} = 200$~GeV and $|\eta_{\pi}|< 1$ are shown in
Fig.~\ref{fig:dsig_dphi} separately for different couplings.
Only one of the seven couplings ($j=5$) gives a minimum at $\phi_{pp} = \pi/2$.
The shapes of the distributions in $\phi_{\pi \pi}$ are rather similar.

\begin{figure}[!ht]
\includegraphics[width = 0.45\textwidth]{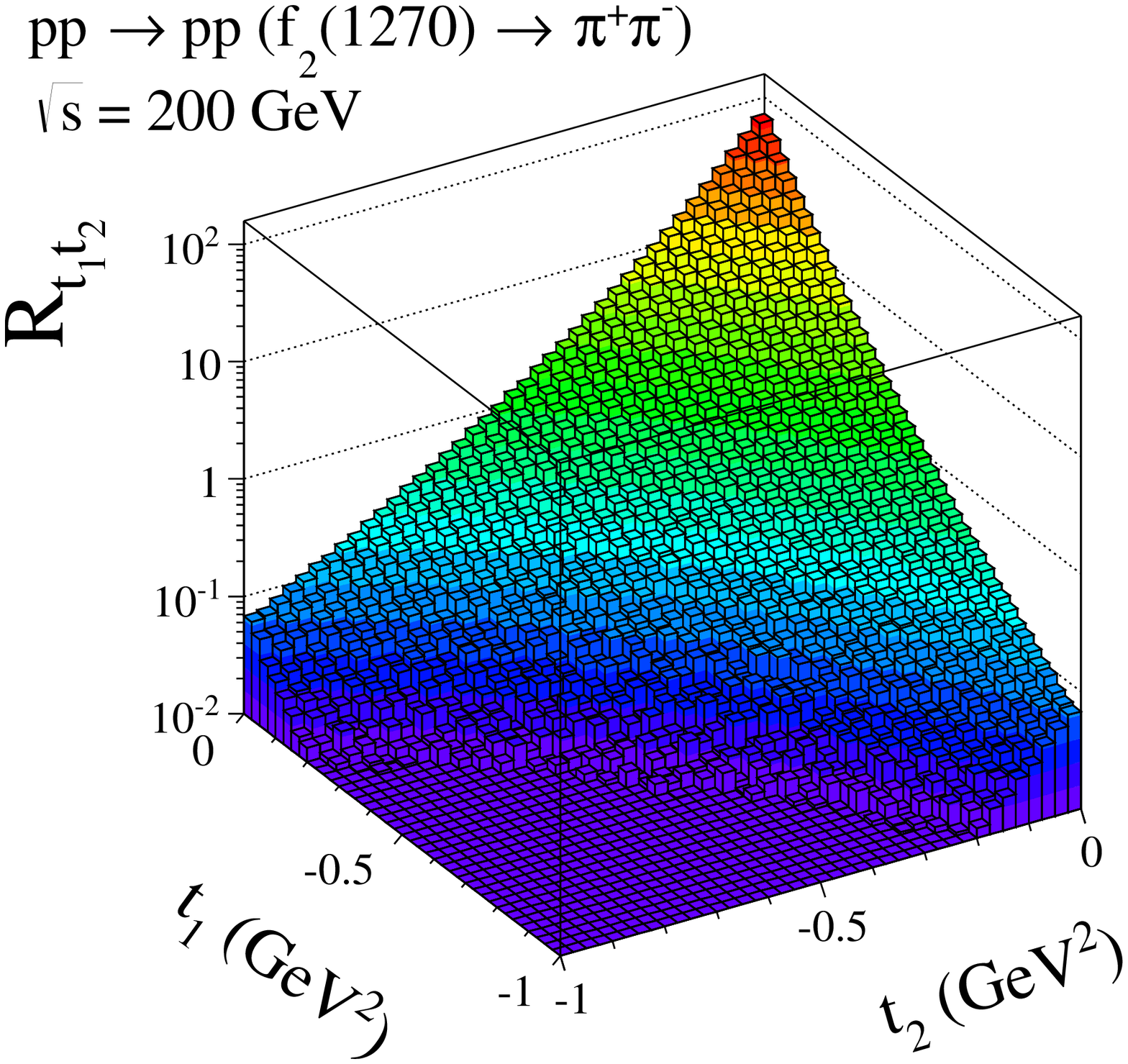}
\includegraphics[width = 0.45\textwidth]{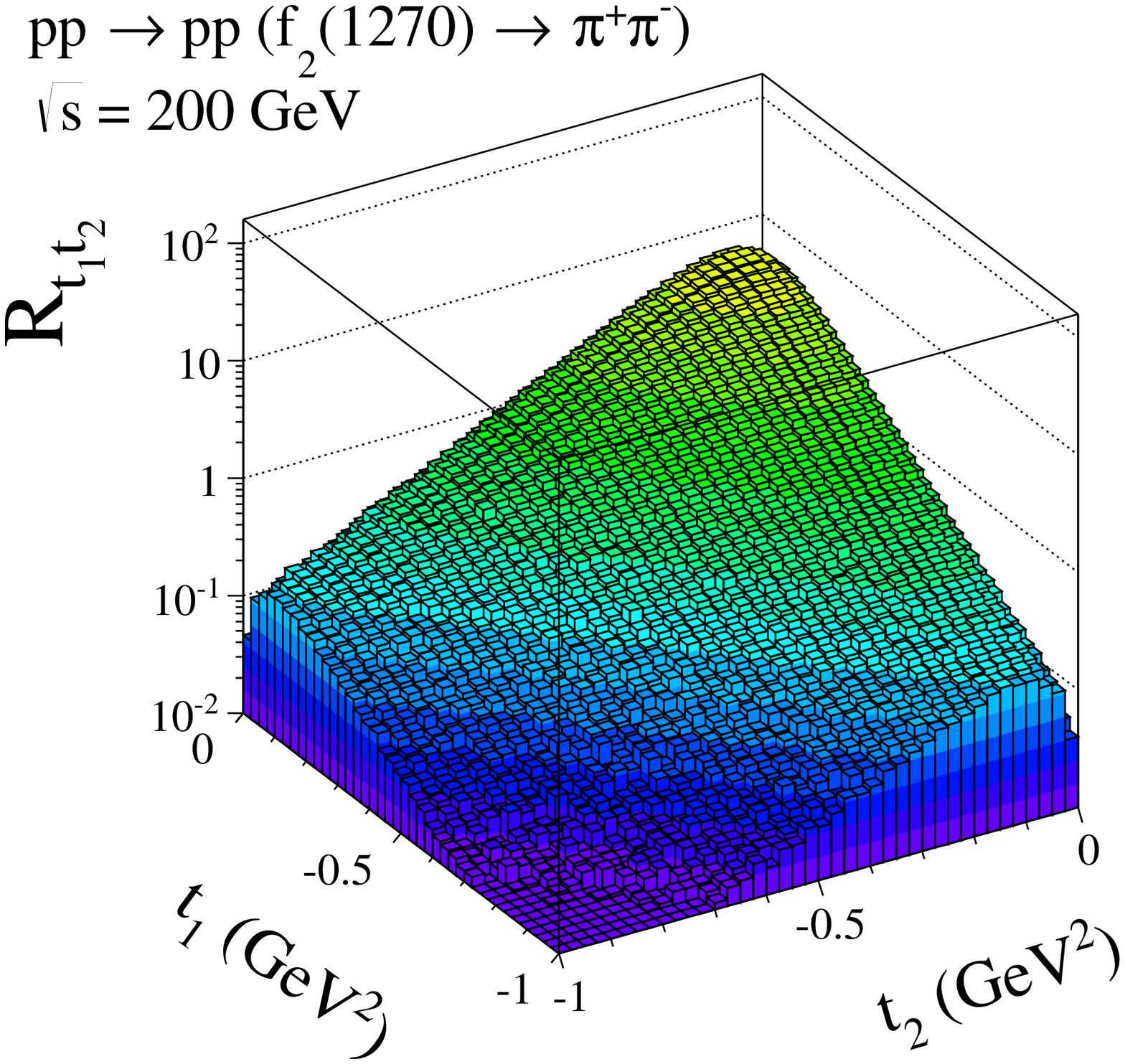}
  \caption{\label{fig:t1t2}
  \small
The distributions in $(t_{1},t_{2})$ space
for the central exclusive production of the $f_{2}(1270)$ meson
via fusion of two tensor pomerons at $\sqrt{s} = 200$~GeV and $|\eta_{\pi}|< 1$.
Plotted is the ratio $R_{t_{1}t_{2}} = \frac{d^{2}\sigma}{dt_{1}dt_{2}} / \int{dt_{1}dt_{2}} \frac{d^{2}\sigma}{dt_{1}dt_{2}}$.
We show as examples the results for the $j=1$ (left panel)
and $j=2$ (right panel) couplings.
No absorption effects were included here.
}
\end{figure}

Different experiments reported results which seem contradictory
\cite{Adamczyk:2014ofa,Aaltonen:2015uva}.
Some of them \cite{Aaltonen:2015uva,Sikora_LowX} observed  
an appearance of the $f_{2}(1270)$ resonance and some not \cite{Adamczyk:2014ofa}.
We think that this fact can be related to different coverage
in $t_1$ and $t_2$ of the different experiments. 
Therefore, before showing any other results we wish to explore the $t_1$ and $t_2$ dependences.
Two examples of the correlation between $t_{1}$ and $t_{2}$ 
for different pomeron-pomeron-$f_{2}$ couplings
are displayed in Fig.~\ref{fig:t1t2}.
The general character of the distributions is rather different.
While for $j=1$ coupling we observe an enhancement of the cross section
when $t_1 \to 0$ or $t_2 \to 0$, in the case of $j=2$ coupling we observe 
a suppression of the cross section when $t_1 \to 0$ or $t_2 \to 0$.

\begin{figure}[!ht]
\includegraphics[width = 0.45\textwidth]{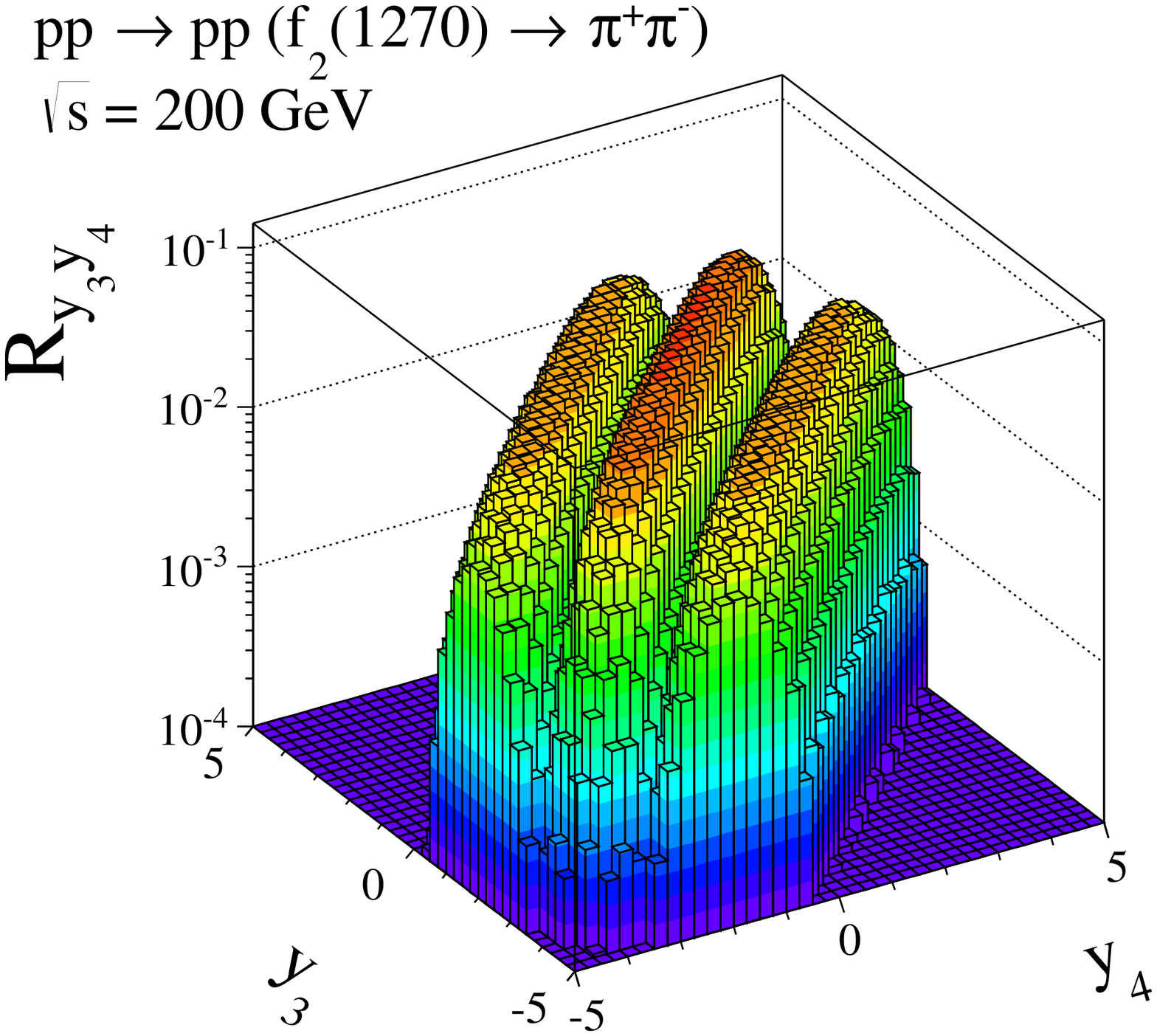}
\includegraphics[width = 0.45\textwidth]{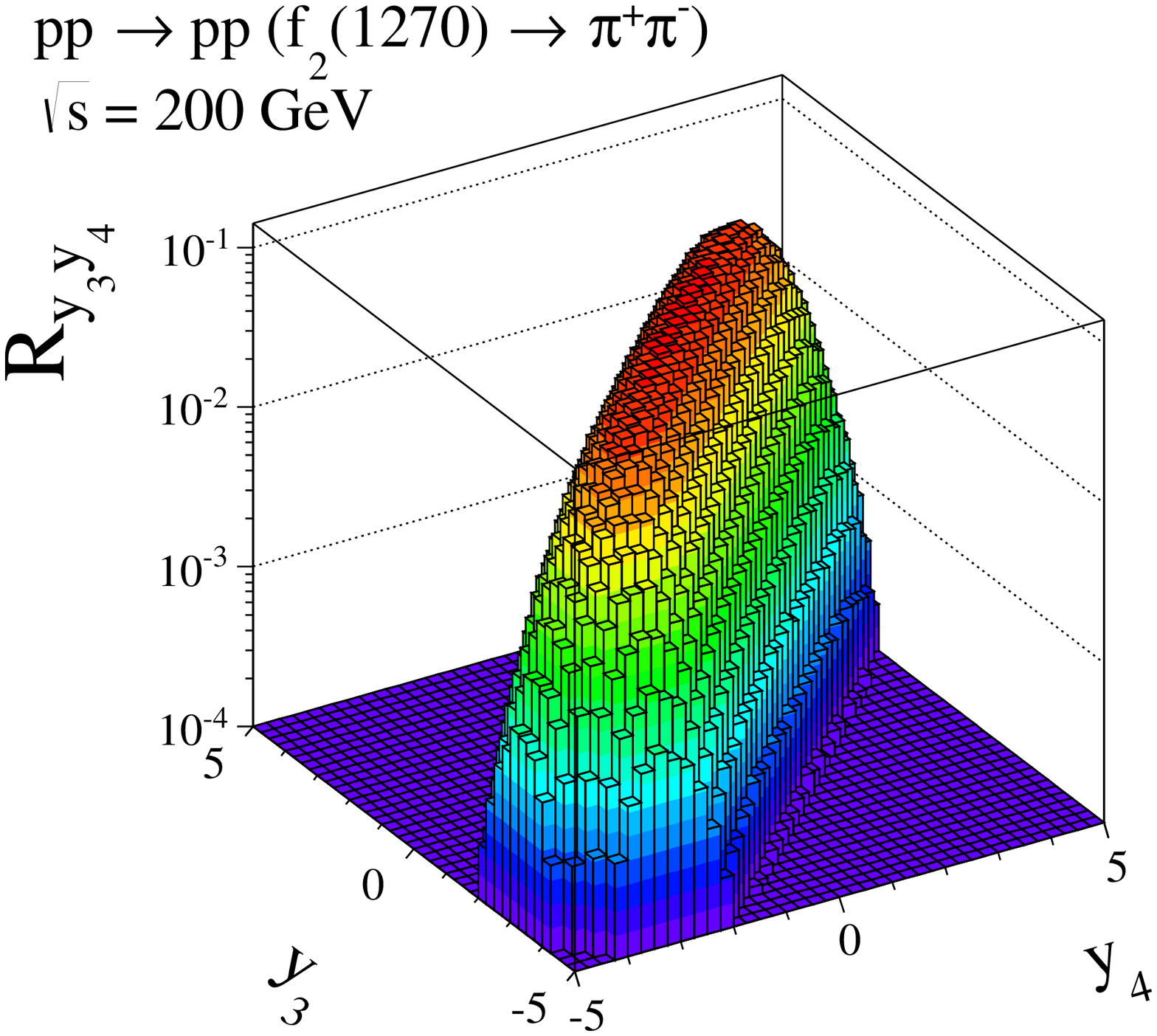}
  \caption{\label{fig:maps_y3y4}
  \small
The distributions in $({\rm y}_{3},{\rm y}_{4})$ space
for the central exclusive production of the $f_{2}(1270)$ meson
via fusion of two tensor pomerons at $\sqrt{s} = 200$~GeV.
Plotted is the ratio $R_{{\rm y}_{3}{\rm y}_{4}} = \frac{d^{2}\sigma}{d{\rm y}_{3}d{\rm y}_{4}} / \int{d{\rm y}_{3}d{\rm y}_{4} \frac{d^{2}\sigma}{d{\rm y}_{3}d{\rm y}_{4}}}$.
We show the results for the $j=1$ (left panel) and $j=2$ (right panel) couplings.
No absorption effects were included here.
}
\end{figure}

\begin{figure}[!ht]
\includegraphics[width = 0.45\textwidth]{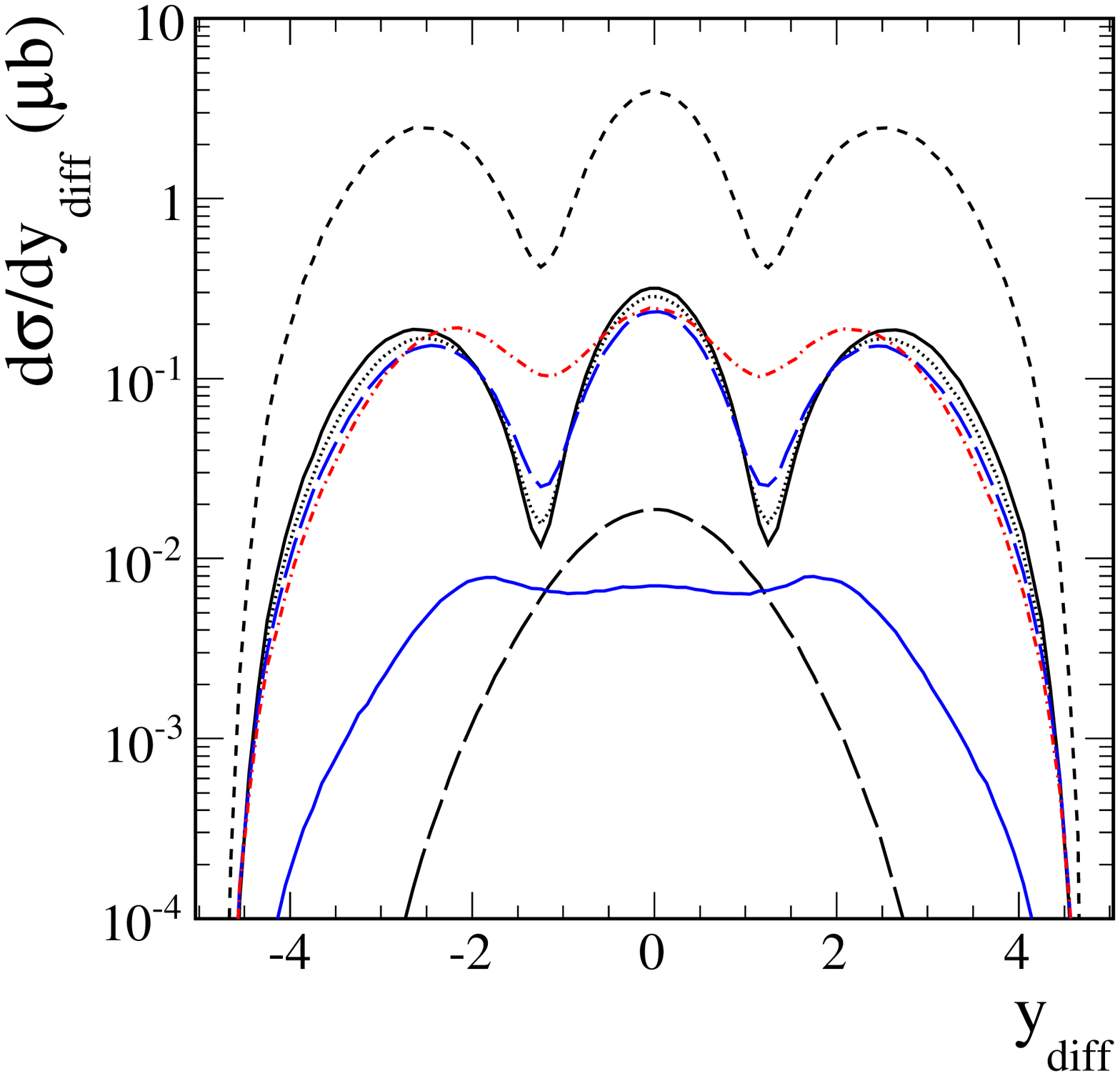}
\includegraphics[width = 0.45\textwidth]{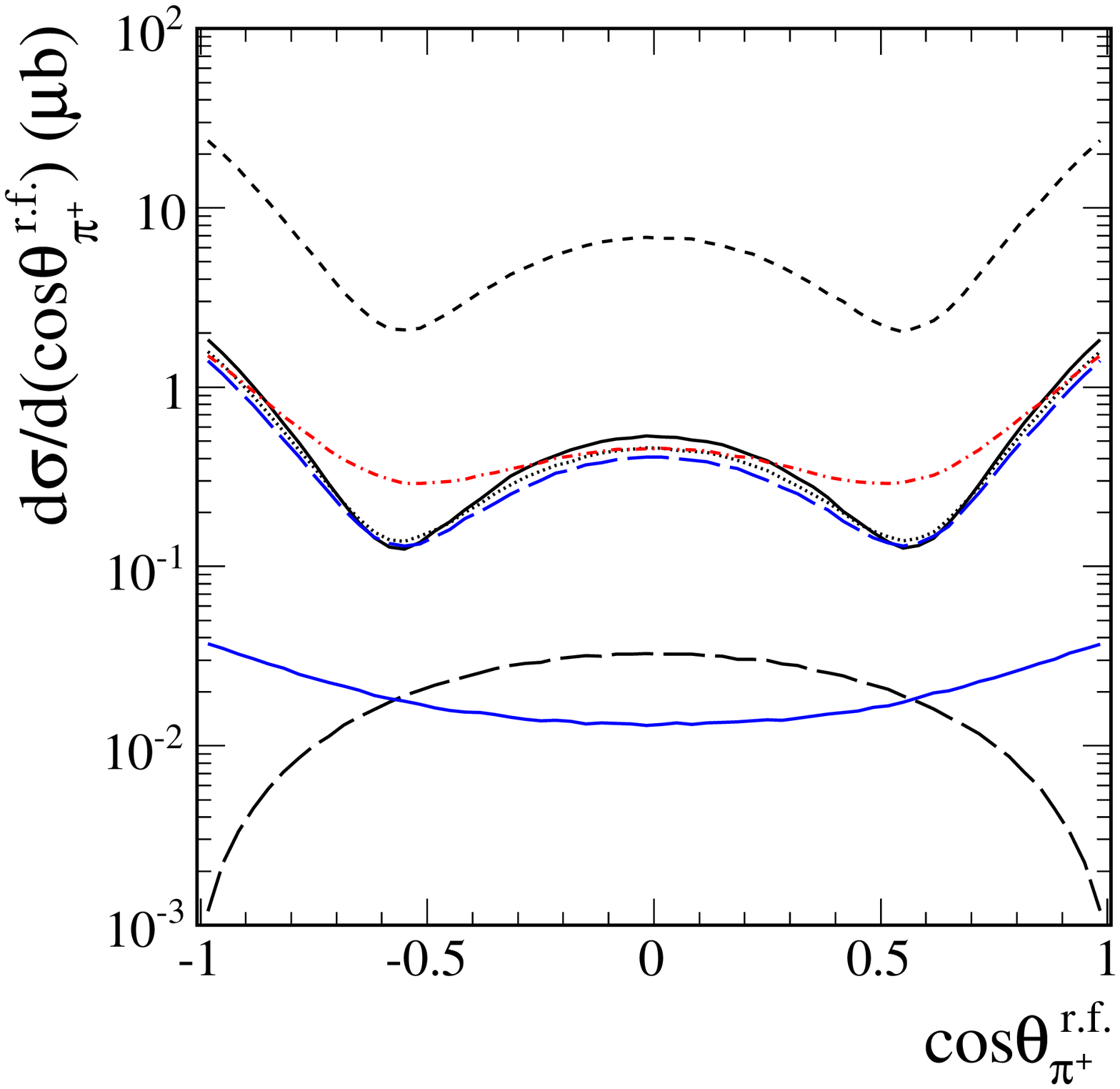}
  \caption{\label{fig:y_diff}
  \small
The distributions in ${\rm y}_{diff} = {\rm y}_{3} - {\rm y}_{4}$
(the left panel) and in $\cos\theta_{\pi^{+}}^{\,\rm{r.f.}}$ (the right panel).
The meaning of the lines is the same as in Fig.~\ref{fig:dsig_dt1} for $\sqrt{s}=200$~GeV
but here the calculation was done
for a broader range of rapidities of both charged pions, $|{\rm y}_{\pi}|< 5$.
No absorption effects were included here.
}
\end{figure}

The correlation in rapidity of the pions is displayed in Fig.~\ref{fig:maps_y3y4}
for two pomeron-pomeron-$f_{2}$ couplings.
A very good one-dimensional observable which can be used for the
comparison of the couplings under discussion could be the differential
cross section $d\sigma/d{\rm y}_{diff}$, 
where ${\rm y}_{diff} = {\rm y}_{3} - {\rm y}_{4}$.
We show the corresponding distribution in Fig.~\ref{fig:y_diff} (the left panel).
In the right panel we show the angular distribution of the $\pi^{+}$ meson,
$\cos\theta_{\pi^{+}}^{\,\rm{r.f.}}$, where $\theta_{\pi^{+}}^{\,\rm{r.f.}}$ 
is the polar angle of the $\pi^{+}$ meson
with respect to the beam axis in the $\pi^{+} \pi^{-}$ rest frame.
One can observe correlations between the left and right panel.
The minima in the left panel correspond to minima in the right panel.
This is related to the kinematical transformation between ${\rm y}_{diff}$ 
and $\cos\theta_{\pi^{+}}^{\,\rm{r.f.}}$.

\begin{figure}[!ht]
\includegraphics[width=0.45\textwidth]{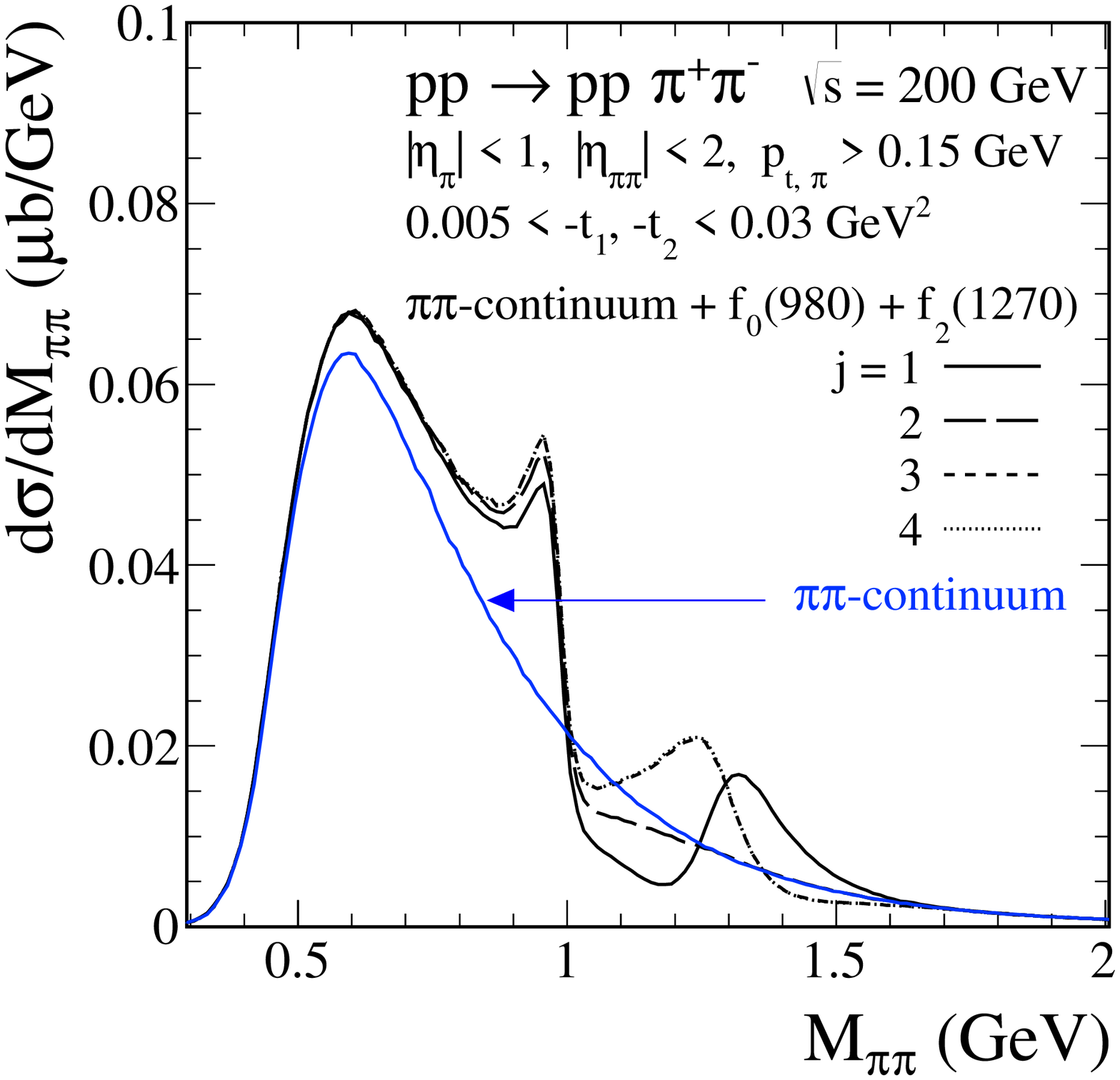}
\includegraphics[width=0.45\textwidth]{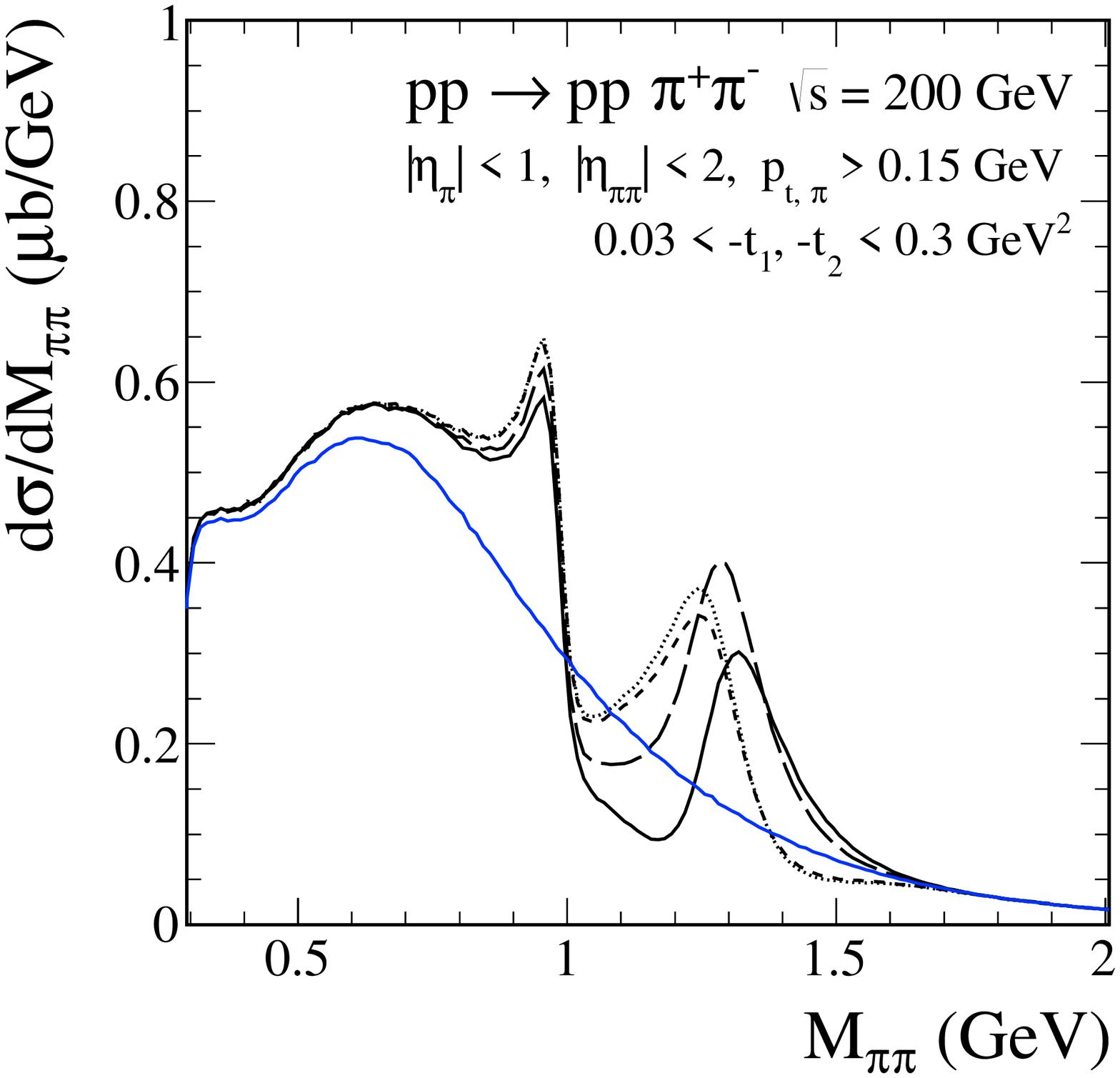}
\includegraphics[width=0.45\textwidth]{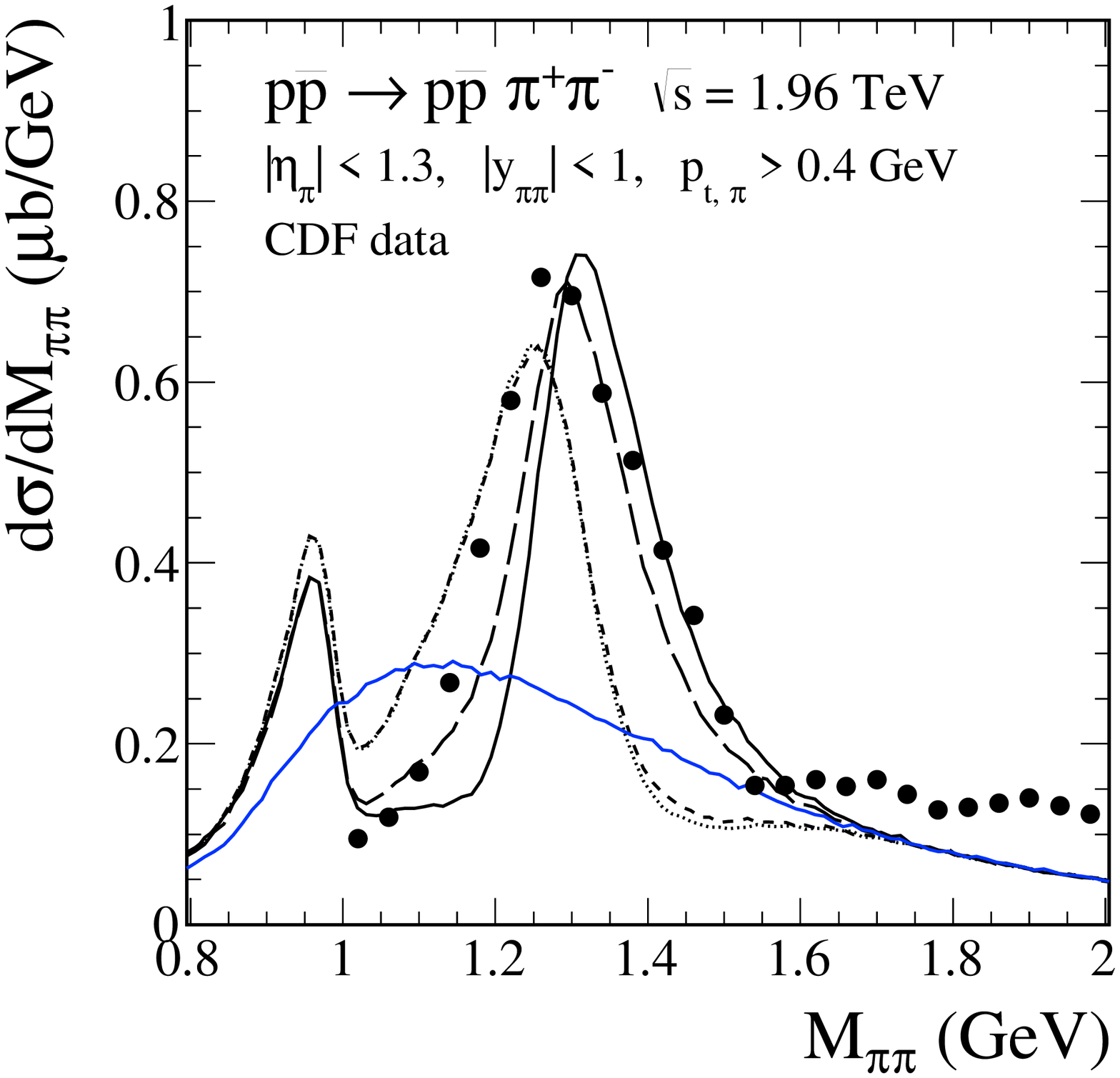}
\includegraphics[width=0.45\textwidth]{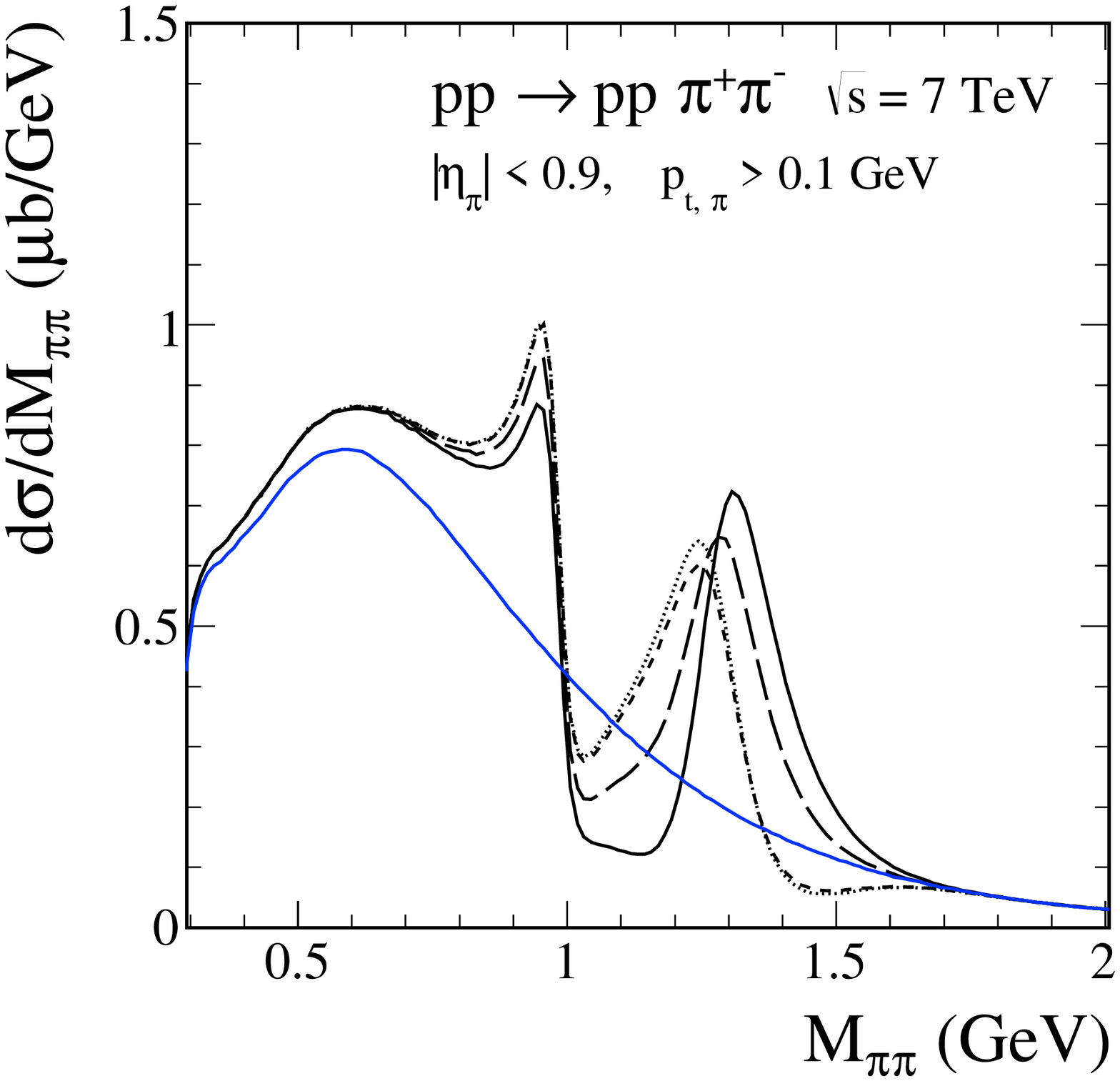}
  \caption{\label{fig:M34}
  \small
Two-pion invariant mass distribution with the relevant 
experimental kinematical cuts specified in the legend.
The results corresponding to the Born calculations
were multiplied for $\sqrt{s}=200$~GeV 
by the gap survival factor $\langle S^{2}\rangle = 0.2$
and by $\langle S^{2}\rangle = 0.1$ for $\sqrt{s}=1.96$ and 7~TeV.
The CDF data from \cite{Aaltonen:2015uva} are shown for comparison.
The blue solid lines represent the non-resonant continuum contribution
obtained for the monopole off-shell pion 
form factors (\ref{off-shell_form_factors_mon}) with $\Lambda_{off,M} = 0.7$~GeV.
The black lines represent a coherent sum of non-resonant continuum, 
$f_{0}(980)$ and $f_{2}(1270)$ resonant terms. 
The individual contributions of different $\Pom \Pom f_{2}$ couplings
$j = 1$ (the solid line), $j = 2$ (the long-dashed line),
$j = 3$ (the dashed line), $j = 4$ (the dotted line) are shown.
The results have been obtained with the coupling constant parameters: 
$g_{\Pom \Pom f_{0}(980)}'=0.2$, $g_{\Pom \Pom f_{0}(980)}''=1.0$,
$g^{(1)}_{\Pom \Pom f_{2}} = 2.0$,
$g^{(2)}_{\Pom \Pom f_{2}} = 9.0$, $g^{(3)}_{\Pom \Pom f_{2}} = 0.5$,
and $g^{(4)}_{\Pom \Pom f_{2}} = 2.0$.
}
\end{figure}

In the present preliminary analysis we wish to understand whether one can approximately describe
the dipion invariant mass distribution observed by different experiments
assuming only one $\Pom \Pom f_{2}$ tensorial coupling. 
The calculations were done at Born level
and the absorption corrections were taken into account
by multiplying the cross section for the corresponding collision energy
by a common factor $\langle S^{2}\rangle$ obtained from \cite{Lebiedowicz:2015eka}.
The two-pion continuum was fixed 
by choosing as a parameter of the form factor for off-shell pion $\Lambda_{off,M} = 0.7$~GeV;
see (\ref{off-shell_form_factors_mon}).
In addition we include the $f_{0}(980)$ contribution where we chose
the $\Pom \Pom f_{0}(980)$ coupling parameters as
$g_{\Pom \Pom f_{0}(980)}'=0.2$ and $g_{\Pom \Pom f_{0}(980)}''=1.0$.
\footnote{
Note, that we take here smaller values of the coupling parameters
than in our previous paper \cite{Lebiedowicz:2013ika} 
because they were fixed there at the WA102 energy where
we expect also large contributions to the cross section from 
the reggeon exchanges.}
For each choice of the $\Pom \Pom f_{2}$ coupling defined by the index $j$ we have adjusted
the corresponding coupling constant to get the same cross section
in the maximum corresponding to the $f_{2}(1270)$ resonance in the CDF data \cite{Aaltonen:2015uva}.
We assume that the peak observed experimentally corresponds mainly to the $f_2$ resonance 
\footnote{In principle there may also
be a contribution from the broad scalar $f_{0}(1370)$.}.
As can be clearly seen from Fig.~\ref{fig:M34} 
different couplings generate different interference patterns. 
We can observe that the $j=2$ coupling gives results close to those observed 
by the CDF Collaboration \cite{Albrow_Project_new,Aaltonen:2015uva}.
In this preliminary study we do not try to fit the existing data 
by mixing different couplings because the CDF data are not fully exclusive
(the outgoing $p$ and $\bar{p}$ were not measured).
Comparing the two upper panels ($\sqrt{s}=200$~GeV)
we see again that for $j=2$ (the long-dashed line)
the $f_{2}$ is practically absent at small $|t_{1,2}|$
but is prominent at large $|t_{1,2}|$.

\begin{figure}[!ht]
\includegraphics[width = 0.45\textwidth]{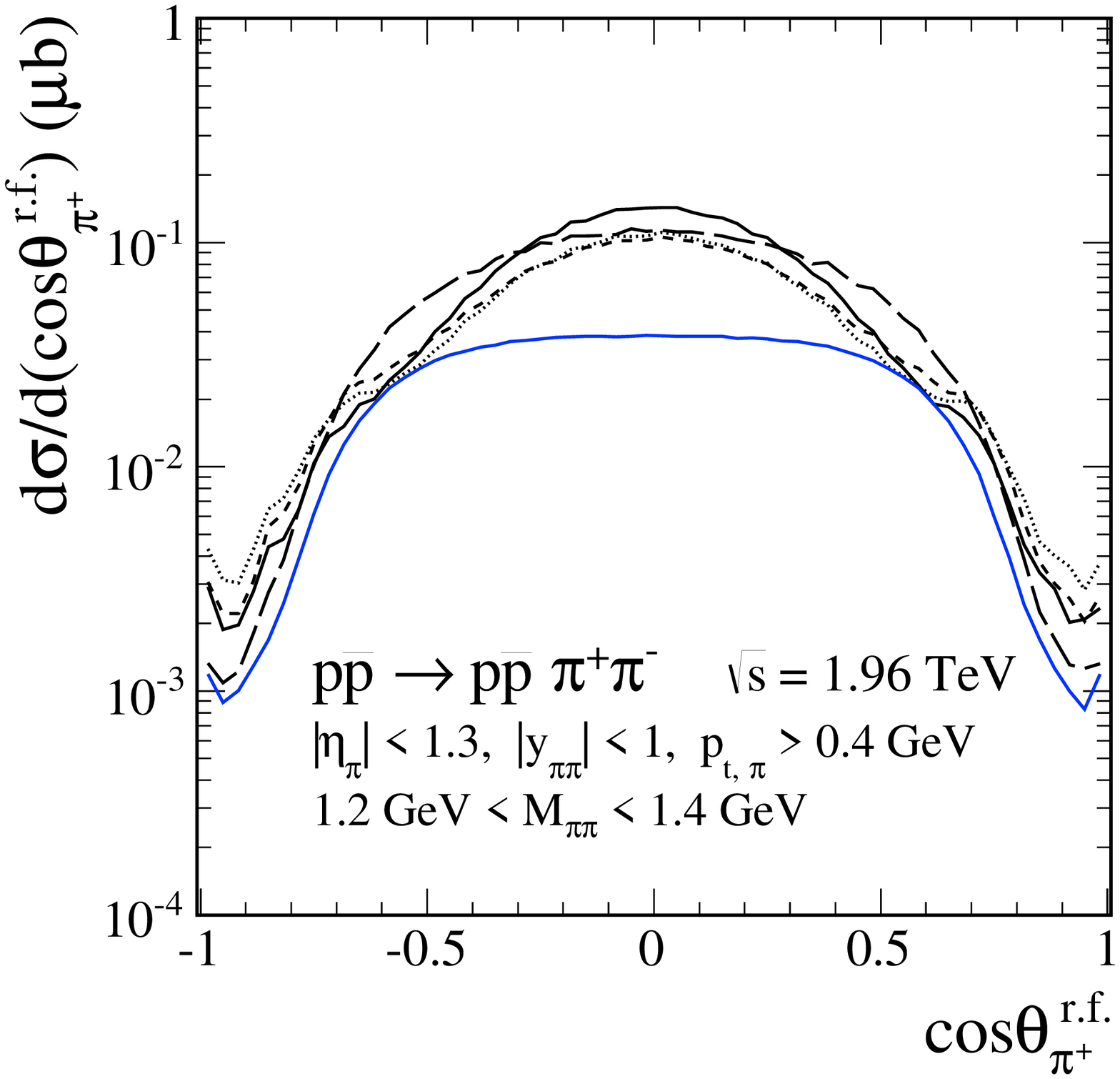}
  \caption{\label{fig:z3rf_cuts}
  \small
Differential cross section $d\sigma/d(\cos\theta_{\pi^{+}}^{\,\rm{r.f.}})$
as a function of the cosine of the polar angle $\theta_{\pi^{+}}^{\,\rm{r.f.}}$
in the $\pi^{+}\pi^{-}$ rest frame for $\sqrt{s}=1.96$~TeV
and in the mass region 1.2~GeV $\leqslant M_{\pi\pi} \leqslant$ 1.4~GeV.
The meaning of the lines is the same as in Fig.~\ref{fig:M34}.
Both signal ($f_{2}(1270)$) and background are included here.
}
\end{figure}

In Fig.~\ref{fig:z3rf_cuts} the $\cos\theta_{\pi^{+}}^{\,\rm{r.f.}}$ distribution
is shown in the $f_{2}$ mass region 1.2~GeV $\leqslant M_{\pi\pi} \leqslant$ 1.4~GeV
with the CDF kinematical cuts.
The limited CDF acceptance, in particular $p_{t,\pi} > 0.4$ GeV,
cause that differences for the different couplings are now less pronounced.
Whether it is possible to pin down the correct couplings may require detailed studies
of the CDF data; see \cite{Albrow_Project_new}.
We expect that these differences should be better visible 
in future LHC experiments.

\begin{figure}[!ht]
\includegraphics[width=0.45\textwidth]{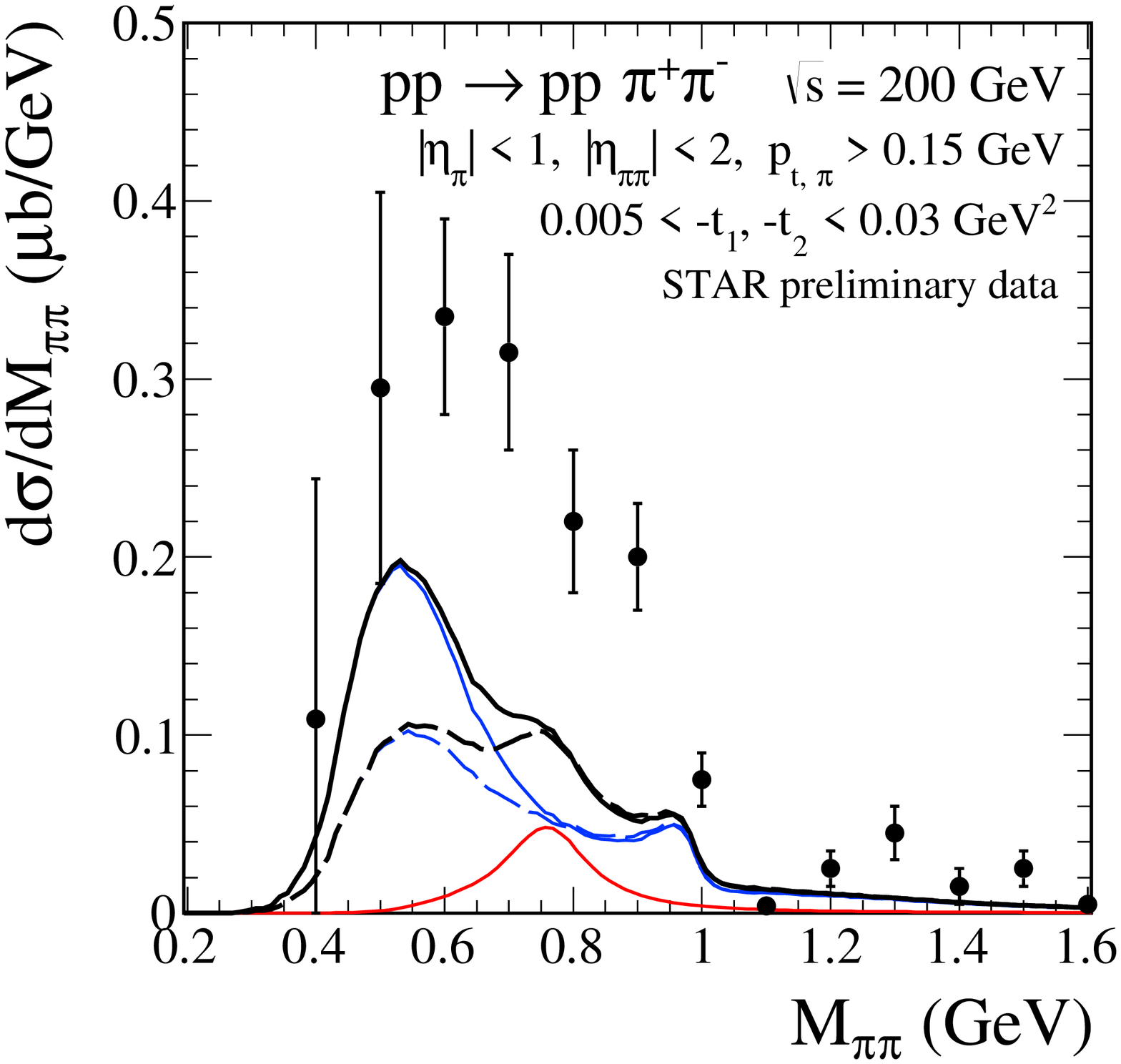}
\includegraphics[width=0.45\textwidth]{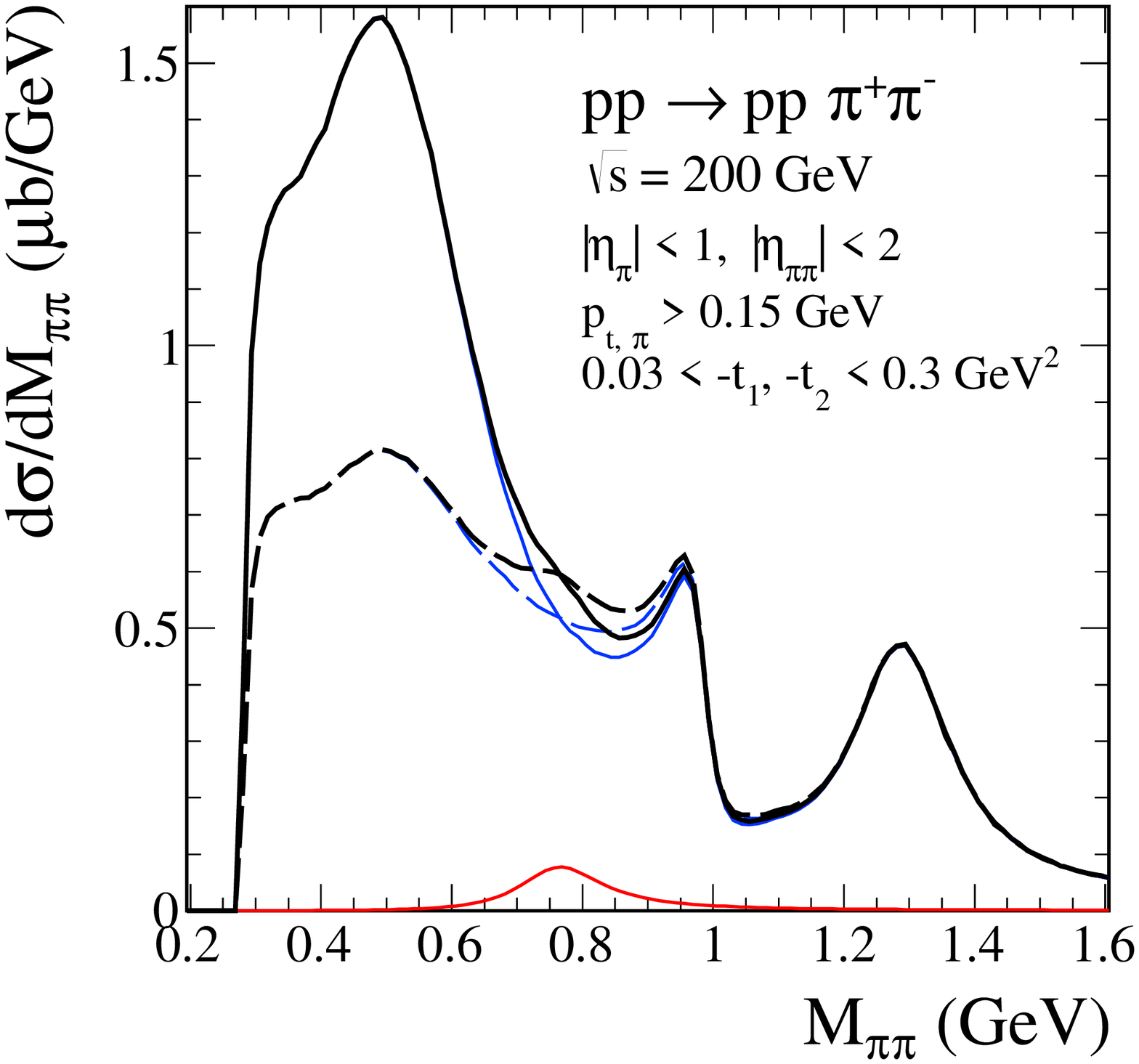}
\includegraphics[width=0.45\textwidth]{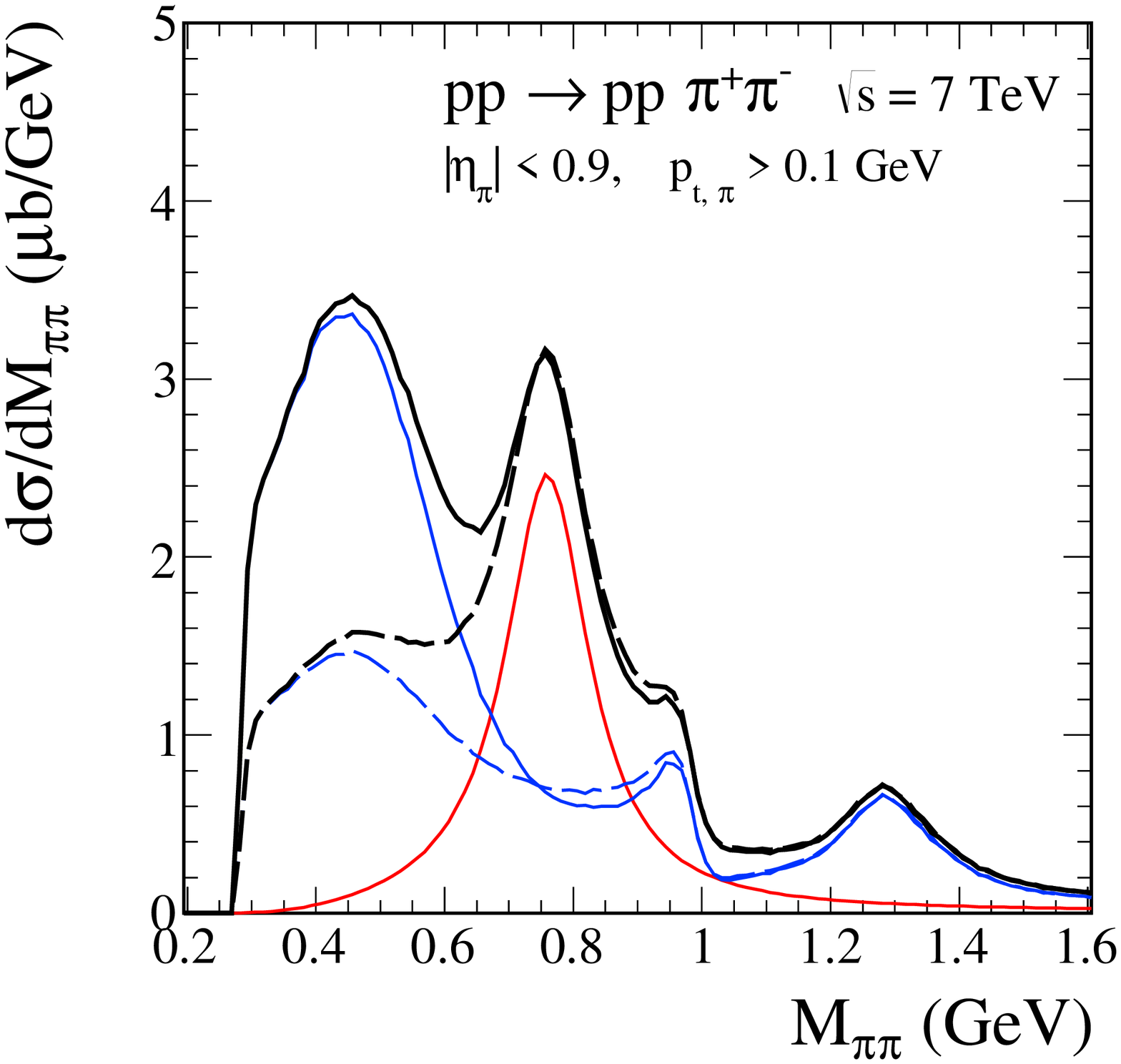}
  \caption{\label{fig:M34_rho}
  \small
Two-pion invariant mass distribution for different experimental kinematical cuts.
The results corresponding to the Born calculations for $\sqrt{s}=200$~GeV 
and $\sqrt{s}=7$~TeV were multiplied 
by the gap survival factors $\langle S^{2}\rangle = 0.2$
and $\langle S^{2}\rangle = 0.1$, respectively.
The STAR \cite{Adamczyk:2014ofa} preliminary data are shown for comparison.
The red solid lines represent results for the photoproduction contribution
with $\langle S^{2}\rangle = 0.9$.
The blue solid and long-dashed lines represent the coherent sum 
of the purely diffractive production terms, that is, 
the continuum, $f_{0}(500)$, $f_{0}(980)$, and $f_{2}(1270)$ contributions.
The complete results for 
$g_{\Pom \Pom f_{0}(500)}'=0.2$ and 0.5 
($g_{\Pom \Pom f_{0}(500)}''=0$ in (\ref{amplitude_f0_pomTpomT_approx}))
correspond to the black long-dashed line and the solid line, respectively.
The other parameters were chosen as
$\Lambda_{off,M} = 0.7$~GeV, $g_{\Pom \Pom f_{0}(980)}'=0.2$, 
$g_{\Pom \Pom f_{0}(980)}''=1.0$, and $g^{(2)}_{\Pom \Pom f_{2}} = 9.0$.
}
\end{figure}

\begin{figure}[!ht]
\includegraphics[width=0.45\textwidth]{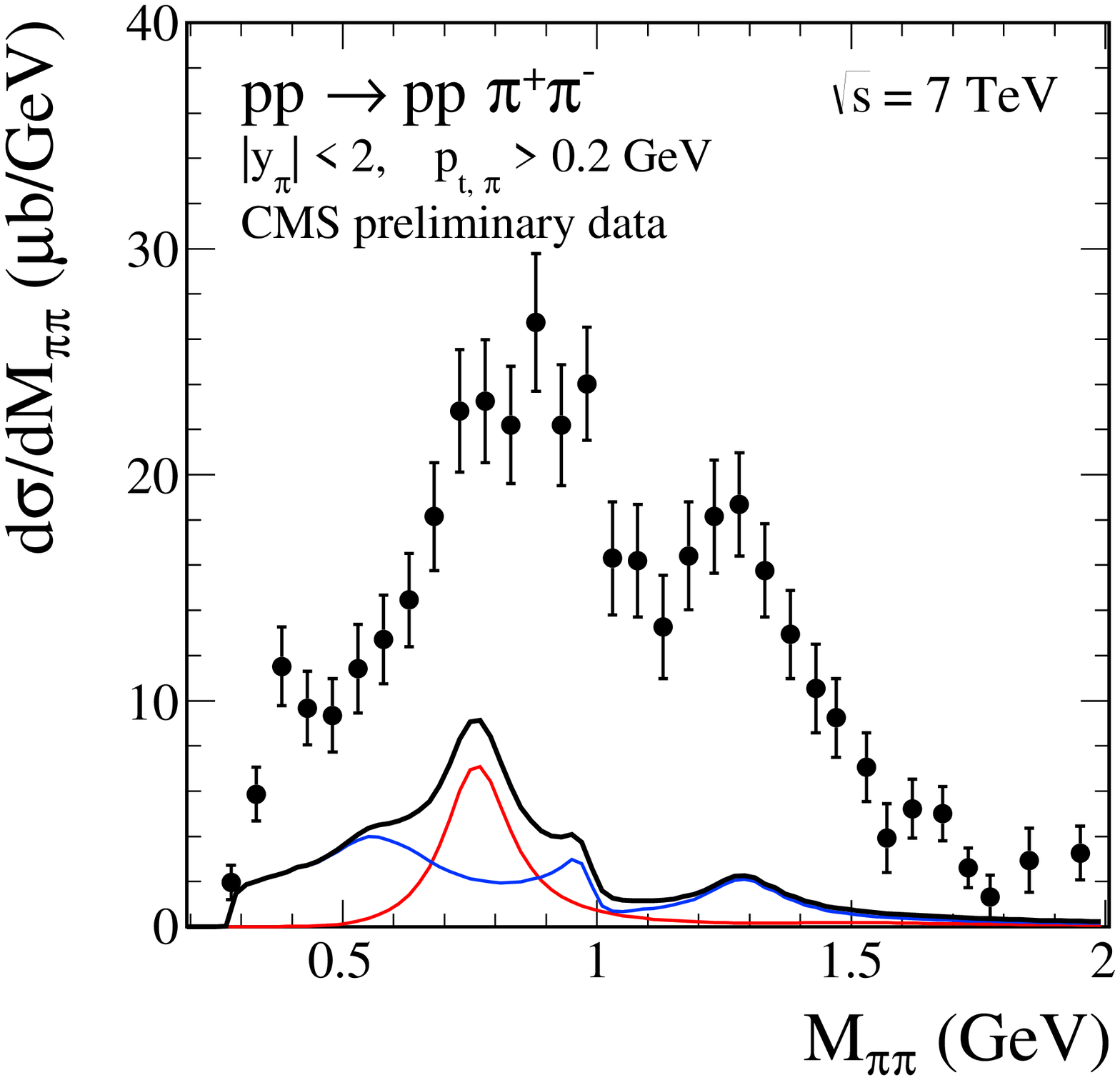}
\includegraphics[width=0.45\textwidth]{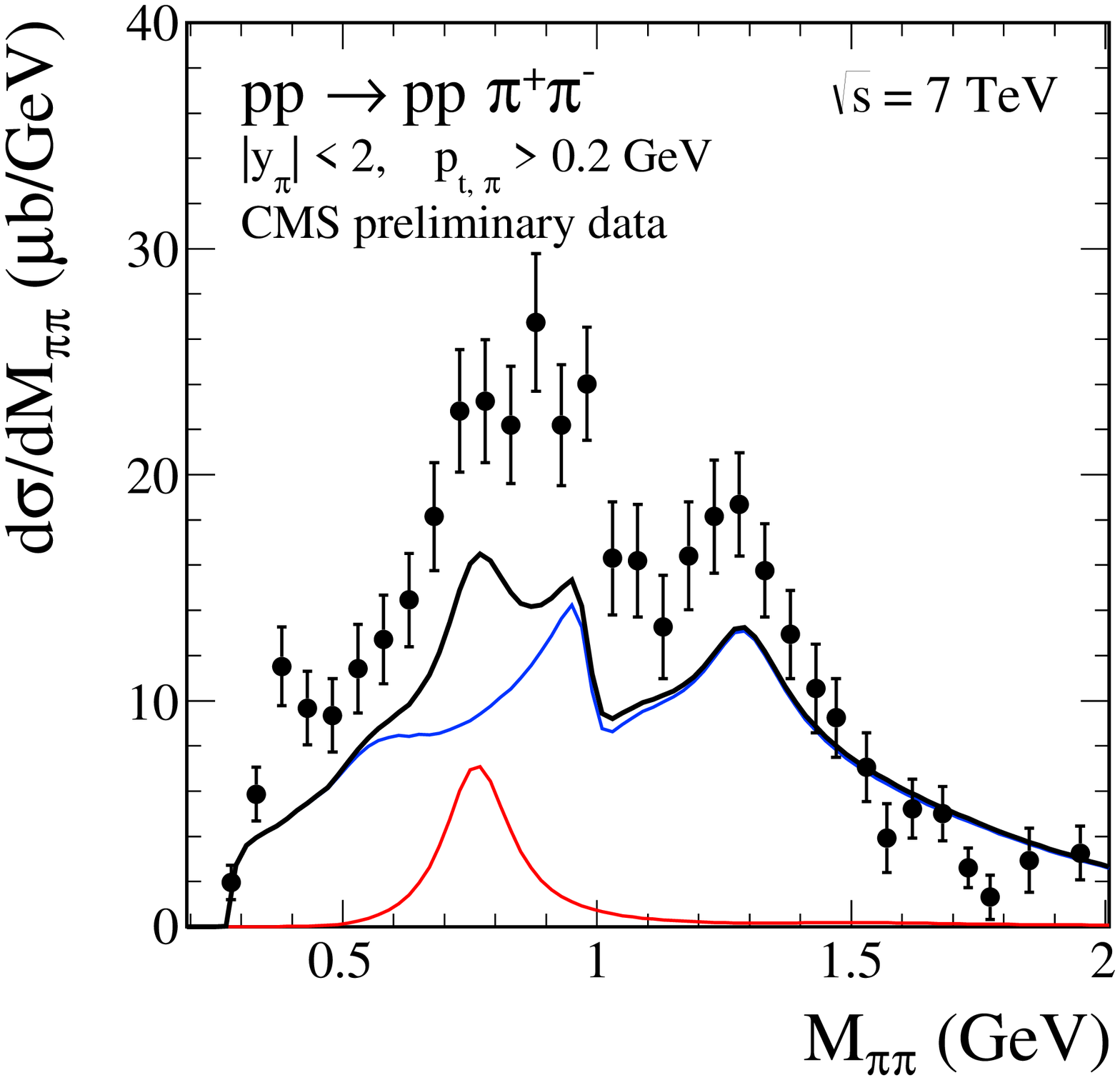}
  \caption{\label{fig:M34_rho_CMS}
  \small
Two-pion invariant mass distribution for the CMS kinematics at $\sqrt{s}=7$~TeV. 
The meaning of the lines is the same as in Fig.~\ref{fig:M34_rho}.
Both photoproduction and purely diffractive contributions are included here.
The complete results correspond to the black solid line.
The CMS preliminary data \cite{CMS:2015diy} are shown for comparison.
In the left panel we show results for the standard parameter set 
as specified in the legend of Fig.~\ref{fig:M34_rho}.
In the right panel we take $\Lambda_{off,E} = 1.6$~GeV in (\ref{off-shell_form_factors_exp}),
$g_{\Pom \Pom f_{0}(500)}'=0.5$, $g_{\Pom \Pom f_{0}(980)}'=0.2$, 
$g_{\Pom \Pom f_{0}(980)}''=1.0$, and $g^{(2)}_{\Pom \Pom f_{2}} = 15.0$.
}
\end{figure}

In Figs.~\ref{fig:M34_rho} and \ref{fig:M34_rho_CMS}
we show the $\pi^{+}\pi^{-}$ invariant mass distribution
for the STAR, ALICE and CMS experiments, respectively.
The experimental data on central exclusive $\pi^{+}\pi^{-}$ production
measured at the energies of the ISR, RHIC, and the LHC collider 
all show a broad continuum 
in the $\pi^{+}\pi^{-}$ invariant mass region of $M_{\pi\pi}<1$~GeV.
This region is experimentally difficult to access due to the missing acceptance
for pion pairs and low pion transverse momentum.
In addition this region of the phase space
may be affected by $\pi\pi$ final state interaction
which may occur in addition to the direct coupling of pomerons
to $f_{0}(500)$ meson considered here
\footnote{The low-energy $\pi \pi$ final state interaction was discussed e.g. in \cite{Pumplin:1976dm,Au:1986vs,Lebiedowicz:2009vt,Lebiedowicz:2009pj}.}.
Therefore, we show here results including 
in addition to the non-resonant $\pi^{+}\pi^{-}$ continuum, 
the $f_{2}(1270)$ and the $f_{0}(980)$,
the contribution from photoproduction, both resonant ($\rho^{0} \to \pi^{+}\pi^{-}$)
and non-resonant (Drell-S\"oding), as well as the $f_{0}(500)$ contribution
\footnote{We have checked numerically that the interference effect
between the two classes of processes,  diffractive and photoproduction, 
is always below 1\%.}.
The complete results for two values of coupling constant, $g_{\Pom \Pom f_{0}(500)}'=0.2$ and 0.5,
correspond to the black long-dashed and solid lines, respectively.
For comparison we show also the contributions of the individual terms separately.
The red solid lines represent the results 
for the $\pi^{+}\pi^{-}$-photoproduction contribution as obtained in \cite{Lebiedowicz:2014bea},
where both the resonant ($\rho(770)$, $\rho(1450)$) 
and the non-resonant (Drell-S\"oding) terms were included.
The blue long-dashed (solid) lines are the results for
the purely diffractive $\pi^{+}\pi^{-}$ production
with $g_{\Pom \Pom f_{0}(500)}'=0.2$ (0.5)
and the other parameters as specified in the legend of Fig.~\ref{fig:M34_rho}.
The absorption effects 
lead to huge damping of the cross section for the purely diffractive term and relatively small
reduction of the cross section for the photoproduction term.
Therefore we expect one could observe the photoproduction term, especially at higher energies.

In Fig.~\ref{fig:M34_rho_CMS} we show very recent results obtained 
by the CMS collaboration. 
This measurement \cite{CMS:2015diy} is not fully exclusive
and the $M_{\pi\pi}$ spectrum contains
therefore contributions associated
with one or both protons undergoing dissociation.
In the left panel we show results
obtained with the parameter set used to ''describe`` 
STAR \cite{Adamczyk:2014ofa} and CDF \cite{Aaltonen:2015uva} data.
At present we cannot decide whether the disagreement is due to
a large dissociation contribution in the CMS data \cite{CMS:2015diy}
or due to an inappropriate parameter set.
Therefore, in the right panel we show results with parameters
better adjusted to the new CMS data.
If we used this set for STAR or CDF measurements
our results there would be above the preliminary STAR data \cite{Adamczyk:2014ofa}
at $M_{\pi\pi} > 1$~GeV
and in complete disagreement with the CDF data
from \cite{Aaltonen:2015uva}, see Figs.~9 and 11.
Only purely central exclusive data expected from
CMS-TOTEM and ATLAS-ALFA will allow to draw definite conclusions.

\begin{figure}[!ht]
\includegraphics[width=0.45\textwidth]{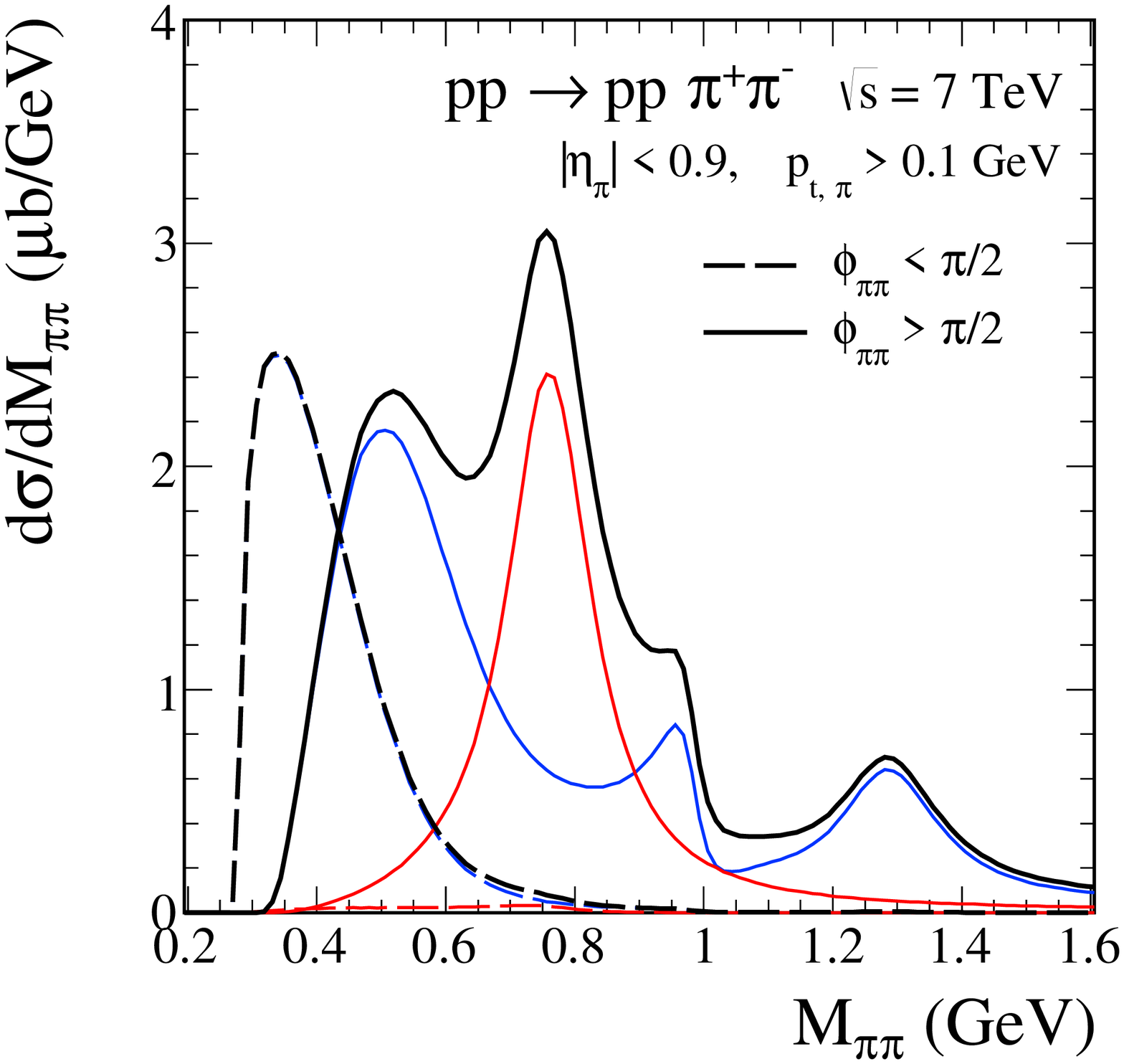}
\includegraphics[width=0.45\textwidth]{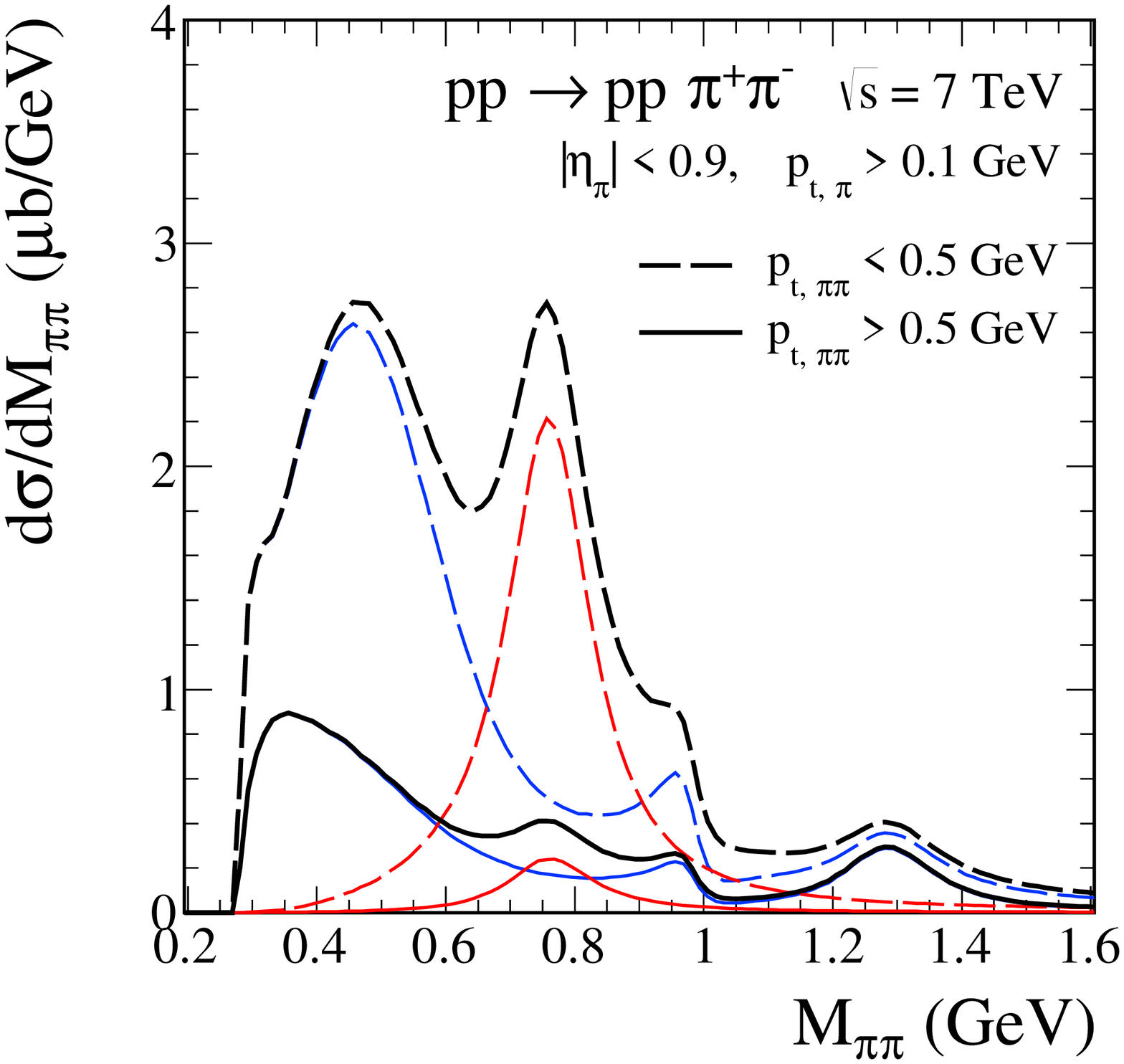}
  \caption{\label{fig:M34_ALICE_cuts}
  \small
Two-pion invariant mass distribution for different experimental kinematical cuts for $\sqrt{s}=7$~TeV.
We show distributions with extra
restrictions on azimuthal angle between the outgoing pions
$\phi_{\pi \pi}$ (the left panel)
and on transverse momentum of the pion pair (the right panel).
The red solid lines represent results for the photoproduction contribution
with $\langle S^{2}\rangle = 0.9$.
The blue lines represent the coherent sum of continuum, $f_{0}(500)$, $f_{0}(980)$ 
and $f_{2}(1270)$ contributions with the same set of parameters as in Fig.~\ref{fig:M34_rho}
(we take here $g_{\Pom \Pom f_{0}(500)}'=0.5$ and $\langle S^{2}\rangle = 0.1$).
The complete results correspond to the black solid and long-dashed lines, respectively.
}
\end{figure}

\begin{figure}[!ht]
\includegraphics[width = 0.48\textwidth]{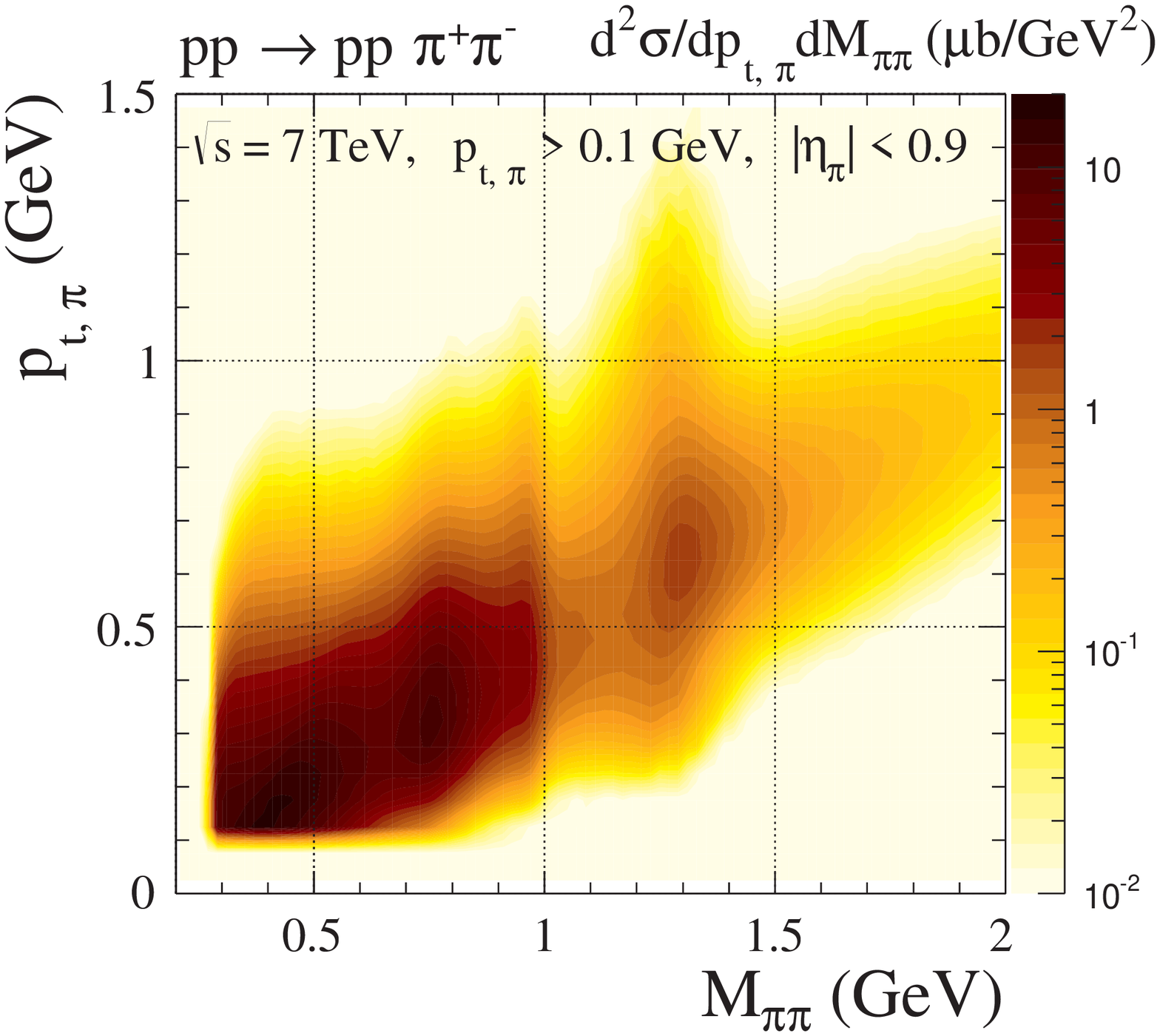}
\includegraphics[width = 0.48\textwidth]{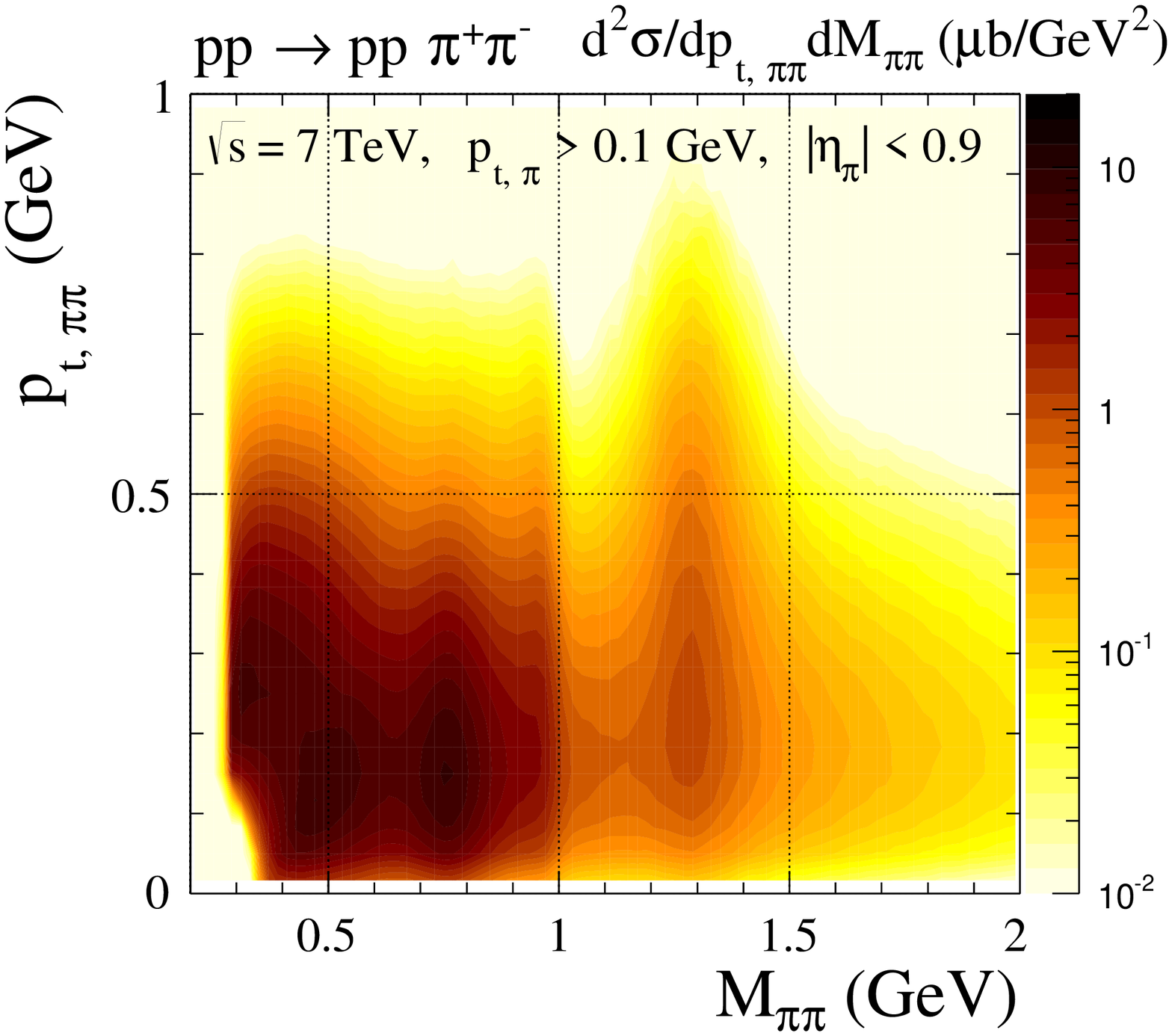}
\includegraphics[width = 0.48\textwidth]{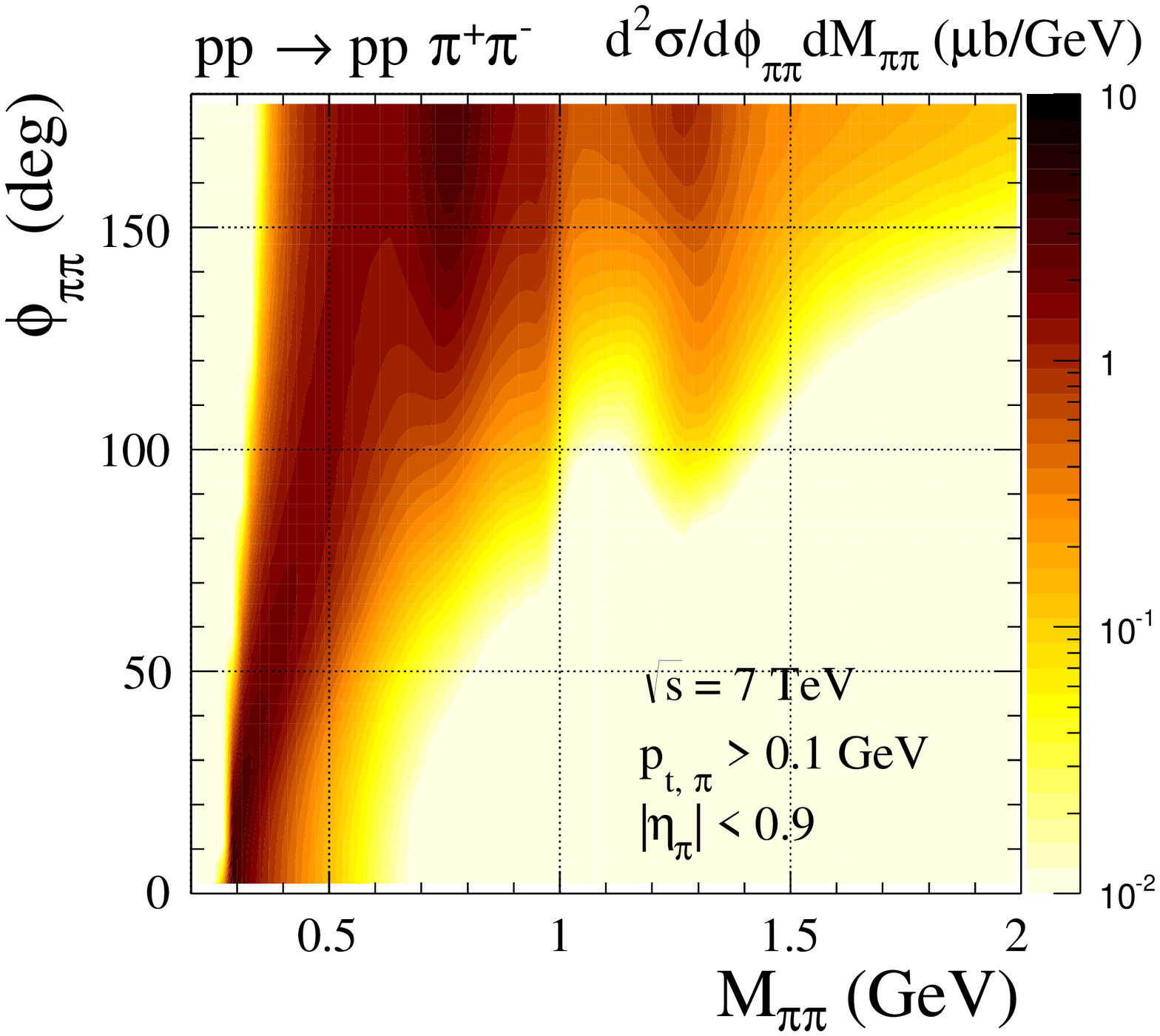}
\includegraphics[width = 0.48\textwidth]{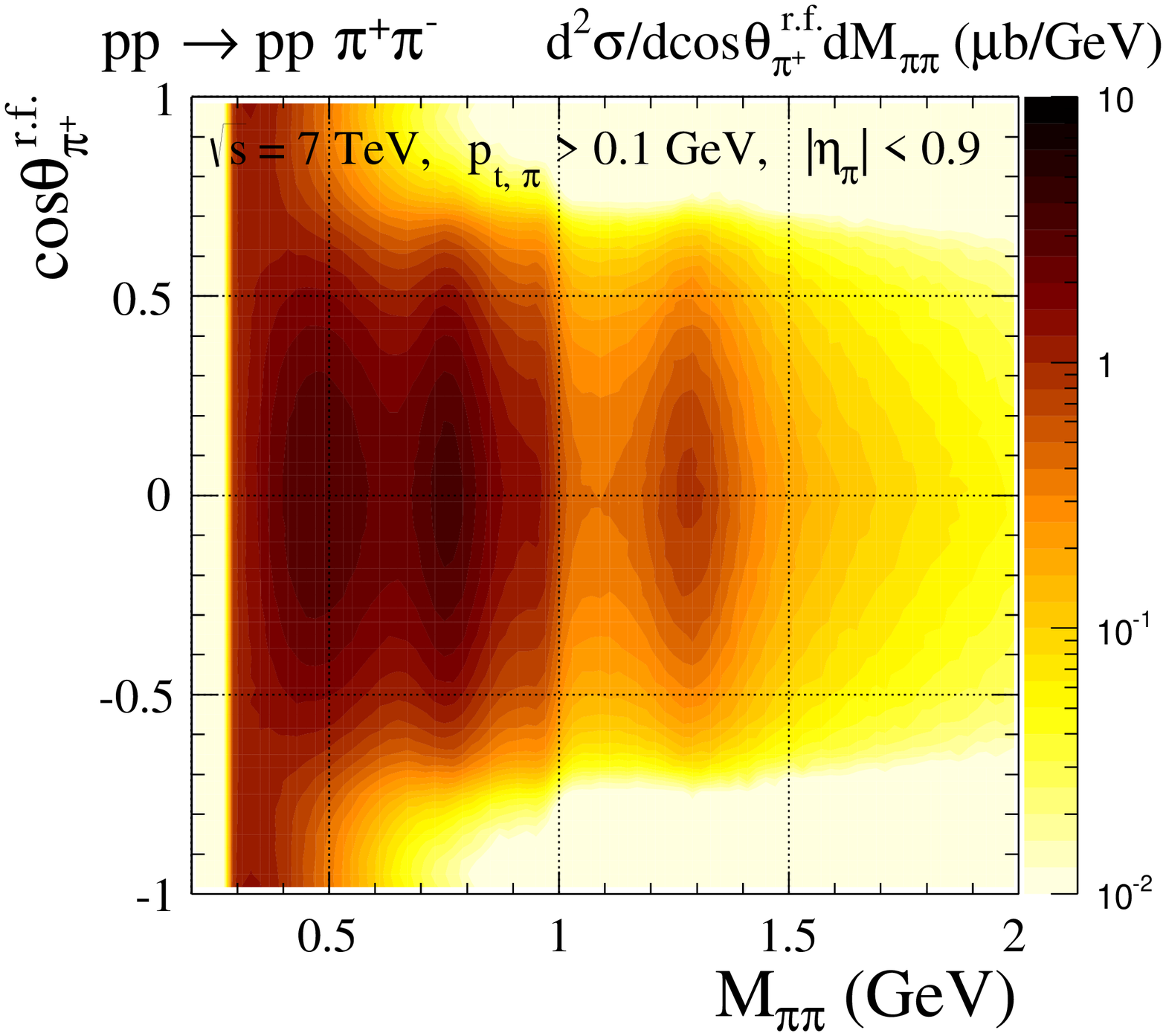}
  \caption{\label{fig:maps_ALICE}
  \small
The distributions for the ALICE kinematics at $\sqrt{s} = 7$~TeV.
Plotted is $\frac{d^{2}\sigma}{dp_{t,\pi}dM_{\pi\pi}}$ (the top left panel),
$\frac{d^{2}\sigma}{dp_{t,\pi\pi}dM_{\pi\pi}}$ (the top right panel),
$\frac{d^{2}\sigma}{d\phi_{\pi\pi}dM_{\pi\pi}}$ (the bottom left panel), and
$\frac{d^{2}\sigma}{d\cos\theta_{\pi^{+}}^{\,r.f.}dM_{\pi\pi}}$ (the bottom right panel).
Both the purely diffractive 
(the non-resonant and resonant $f_{0}(500)$, $f_{0}(980)$, $f_{2}(1270)$ production)
and the photoproduction (the Drell-S\"oding and resonant $\rho(770)$ production) 
processes were included in the calculations.
Here, the model parameters were chosen as in Fig.~\ref{fig:M34_rho}.
Absorption corrections were taken into account effectively 
by the gap survival factors $\langle S^{2}\rangle = 0.1$ and 0.9 
for the fully diffractive and photoproduction contributions, respectively.
}
\end{figure}

The dipion invariant mass spectrum depends on cuts and/or selection conditions.
As an example, we show in Fig.~\ref{fig:M34_ALICE_cuts} the $M_{\pi\pi}$ distribution 
for the ALICE kinematics at $\sqrt{s}=7$~TeV and with
extra restrictions on azimuthal angle between the outgoing pions 
(the left panel) and with restrictions on transverse momentum of the pion pair
(the right panel).
Here we use again only the $j=2$ coupling for $g_{\Pom \Pom f_{2}}$.
In the left panel, the complete results, 
including all interference terms are shown as
black full (for $\phi_{\pi\pi} > \pi/2$) and 
black long-dashed (for $\phi_{\pi\pi} < \pi/2$) lines.
We show the contributions from photoproduction (red line)
and diffractive production (blue line) separately.
In the right panel the red and blue lines have the same meaning
with the full and long-dashed lines corresponding to
$p_{t, \pi \pi} > 0.5$~GeV and $p_{t, \pi \pi} < 0.5$~GeV, respectively.
If we impose a $\phi_{\pi \pi} > \pi/2$ cut, we can see
that the $\rho^{0}$ and $f_{2}$ resonance contributions are strongly enhanced.
Two-dimensional correlations between the variables 
$p_{t, \pi}$, $p_{t, \pi \pi}$, $\phi_{\pi \pi}$, $\cos\theta_{\pi^{+}}^{\,r.f.}$, 
and $M_{\pi\pi}$ are displayed in Fig.~\ref{fig:maps_ALICE} for $\sqrt{s} = 7$~TeV.
We predict complex and interesting patterns which could be checked by the ALICE Collaboration.

\section{Conclusions}

In the present paper we have concentrated on the exclusive production of
the tensor meson $f_2(1270)$ and the dipion continuum 
in central diffractive production via ``fusion'' of two tensor pomerons. 
We have presented for the first time the corresponding amplitudes at Born level.
In the case of a tensor meson and tensor pomerons we have written down all (seven) possible
pomeron-pomeron-$f_{2}$ couplings (vertices) and 
the corresponding amplitudes using
the effective field theoretical approach proposed in \cite{Ewerz:2013kda}.
The corresponding coupling constants in such a model are, however, unknown.
In the future they could be adjusted by comparison with precise experimental data.

Here we have tried to see whether one of the pomeron-pomeron-$f_{2}$ couplings
(tensorial structures) could be sufficient.
Thus we have tried to adjust only one coupling constant 
at the time to the cross section at the resonance maximum. 
The different couplings (tensorial structures) give different results due 
to different interference effects of the resonance and the dipion continuum. 
By assuming dominance of one of the couplings 
we can get only a rough description of the recent CDF and preliminary STAR experimental data.
The model parameters of the optimal coupling ($j=2$) have been roughly adjusted
to recent CDF data and then used for the predictions for the STAR, ALICE, and CMS experiments.
\footnote{A better adjustment of the model parameters
including more than one pomeron-pomeron-$f_{2}$ coupling will be possible with better
experimental data which are expected soon.
Then a corresponding Monte Carlo generator could be constructed.
At present most of the existing Monte Carlo codes \cite{Harland-Lang:2013dia,Kycia:2014hea}
include only the purely diffractive $\pi^{+}\pi^{-}$ continuum 
(for an exception see \cite{Petrov:2007rg}).}
We have also included the scalar $f_{0}(500)$ and $f_{0}(980)$ resonances,
and the vector $\rho(770)$ resonance in a consistent way.
We have shown that the resonance structures in the measured two-pion invariant mass spectra
depend on the cut on proton transverse momenta and/or 
on four-momentum transfer squared $t_{1,2}$ used in an experiments.
The cuts may play then the role of a $\pi \pi$ resonance filter.
We have presented several interesting correlation distributions
which could be checked by the experiments.

To summarize: we have given a consistent treatment of $\pi^{+}\pi^{-}$ continuum
and resonance production in central exclusive $pp$ and $p\bar{p}$ collisions
in an effective field-theoretic approach.
A rich structure emerged which should give experimentalists
interesting challenges to check and explore it.
In this way we shall in particular gain insight
how two pomerons couple to tensor mesons like the $f_{2}(1270)$.
Supposing this to be clarified in future experiments we have then
the big theory challenge to derive such couplings from basic QCD.

\acknowledgments
We are indebted to Mike Albrow, Carlo Ewerz, W{\l}odek Guryn and Lidia G\"orlich 
for useful discussions.
This research was partially supported by 
the MNiSW Grant No. IP2014~025173 (Iuventus Plus),
the Polish National Science Centre Grants No. DEC-2014/15/B/ST2/02528 (OPUS)
and No. DEC-2015/17/D/ST2/03530 (SONATA),
and by the Centre for Innovation and Transfer of Natural Sciences 
and Engineering Knowledge in Rzesz\'ow.

\appendix

\section{$\Pom \Pom f_{2}$ couplings}
\label{section:Tensorial_Couplings}
Here we discuss the couplings $\Pom \Pom f_{2}$.
Consider first the fictitious fusion reaction of two ``tensor pomeron particles''
giving the $f_{2}$ state; see Appendix~A of \cite{Lebiedowicz:2013ika}.
From Table~VI of \cite{Lebiedowicz:2013ika} we find that the following
values of $(l,S)$ can lead to the $f_{2}$ state, a $J^{PC} = 2^{++}$ meson:
$(l,S)$ = $(0,2)$, $(2,0)$, $(2,2)$, $(2,4)$, $(4,2)$, $(4,4)$, $(6,4)$.
Thus, we should be able to construct 
seven independent coupling Lagrangians $\Pom \Pom f_{2}$:
\begin{eqnarray}
{\cal L}'(x) = \sum_{j=1}^{7} {\cal L}'^{(j)}(x)\,.
\label{A1}
\end{eqnarray}
In order to write the corresponding formulae in a compact and convenient form
we find it useful to define the tensor
\begin{eqnarray}
R_{\mu \nu \kappa \lambda} = \frac{1}{2} g_{\mu \kappa} g_{\nu \lambda} 
                           + \frac{1}{2} g_{\mu \lambda} g_{\nu \kappa}
                            -\frac{1}{4} g_{\mu \nu} g_{\kappa \lambda}\,,
\label{A2}
\end{eqnarray}
which fulfils the following relations
\begin{equation}
\begin{split}
& R_{\mu \nu \kappa \lambda} = R_{\nu \mu \kappa \lambda}
= R_{\mu \nu \lambda \kappa} = R_{\kappa \lambda \mu \nu} \,, \quad
R_{\mu \nu \kappa \lambda} g^{\kappa \lambda} = 0\,,\quad
R_{\mu \nu \kappa \lambda} R^{\kappa \lambda}_{\quad \rho \sigma} = R_{\mu \nu \rho \sigma} \,,\\
& R_{\mu \nu \rho_{1} \alpha}\,
R_{\kappa \lambda \sigma_{1}}^{\quad \;\;\, \alpha}\, R_{\rho \sigma}^{\quad \rho_{1} \sigma_{1}}
=
R_{\mu \nu \mu_{1} \nu_{1}}\, R_{\kappa \lambda \alpha_{1} \lambda_{1}}\,
R_{\rho \sigma \rho_{1} \sigma_{1}}\,
g^{\nu_{1} \alpha_{1}}\,g^{\lambda_{1} \rho_{1}}\,g^{\sigma_{1} \mu_{1}}\,.
\end{split}
\label{A3}
\end{equation}
For every tensor $T_{\alpha \beta}$ with $T_{\alpha \beta} = T_{\beta \alpha}$
and $T_{\alpha \beta}\,g^{\alpha \beta}=0$ we have
\begin{equation}
\begin{split}
R_{\kappa \lambda \alpha \beta} T^{\alpha \beta} = T_{\kappa \lambda}\,.
\end{split}
\label{A3a}
\end{equation}

Now we write down the coupling Lagrangians. 
In the following $\Pom_{\mu \nu}(x)$ and $\phi_{\rho \sigma}(x)$ 
are the effective tensor-pomeron and $f_{2}$ field operators, respectively.
We define:
\begin{equation}
\begin{split}
{\cal L}'^{(1)}(x) = M_{0}\, g^{(1)}_{\Pom \Pom f_{2}}\,
\Pom_{\mu_{1} \nu_{1}}(x)\, \Pom_{\alpha_{1} \lambda_{1}}(x)\, \phi_{\rho_{1} \sigma_{1}}(x)
R^{\mu_{1} \nu_{1} \mu_{2} \nu_{2}}\,
R^{\alpha_{1} \lambda_{1} \alpha_{2} \lambda_{2}}\,
R^{\rho_{1} \sigma_{1} \rho_{2} \sigma_{2}}\,
g_{\nu_{2} \alpha_{2}}\,g_{\lambda_{2} \rho_{2}}\,g_{\sigma_{2} \mu_{2}}\,,
\end{split}
\label{A4}
\end{equation}
\begin{equation}
\begin{split}
{\cal L}'^{(2)}(x) = &\frac{1}{M_{0}}\, g^{(2)}_{\Pom \Pom f_{2}}\,
\Bigl( \partial_{\mu}     \Pom_{\kappa \alpha}(x) -
        \partial_{\kappa}  \Pom_{\mu \alpha}(x) \Bigr)
\Bigl( \partial_{\nu}     \Pom_{\lambda \beta}(x) -
        \partial_{\lambda} \Pom_{\nu \beta}(x) \Bigr)
        g^{\alpha \beta}\,g^{\mu \nu}\,R^{\kappa \lambda \rho \sigma}\,\phi_{\rho \sigma}(x)\,,
\end{split}
\label{A5}
\end{equation}
\begin{equation}
\begin{split}
{\cal L}'^{(3)}(x) = &\frac{1}{M_{0}}\, g^{(3)}_{\Pom \Pom f_{2}}\,
\Bigl( \partial_{\mu}     \Pom_{\kappa \alpha}(x) +
        \partial_{\kappa}  \Pom_{\mu \alpha}(x) \Bigr)
\Bigl( \partial_{\nu}     \Pom_{\lambda \beta}(x) +
        \partial_{\lambda} \Pom_{\nu \beta}(x) \Bigr)
        g^{\alpha \beta}\,g^{\mu \nu}\,R^{\kappa \lambda \rho \sigma}\,\phi_{\rho \sigma}(x)\,,
\end{split}
\label{A6}
\end{equation}
\begin{equation}
\begin{split}
{\cal L}'^{(4)}(x) = \frac{1}{M_{0}}\, g^{(4)}_{\Pom \Pom f_{2}}\,
\Bigl( \partial^{\kappa}     \Pom_{\mu \nu}(x) \Bigr)
\Bigl( \partial^{\mu}     \Pom_{\kappa \lambda}(x) \Bigr) \, \phi^{\nu \lambda}(x)\,,
\end{split}
\label{A7}
\end{equation}
\begin{equation}
\begin{split}
{\cal L}'^{(5)}(x) = \frac{1}{M_{0}^{3}}\, g^{(5)}_{\Pom \Pom f_{2}}\,
\left[ \partial_{\kappa} \Bigl( \partial_{\mu} \Pom_{\nu \alpha}(x) -
                                \partial_{\nu} \Pom_{\mu \alpha}(x)\Bigr) \right]
\left[ \partial_{\lambda}\Bigl( \partial^{\mu} \Pom^{\nu \alpha}(x) -
                                \partial^{\nu} \Pom^{\mu \alpha}(x)\Bigr)  \right]
\, \phi^{\kappa \lambda}(x)\,,
\end{split}
\label{A8}
\end{equation}
\begin{equation}
\begin{split}
{\cal L}'^{(6)}(x) = \frac{1}{M_{0}^{3}}\, g^{(6)}_{\Pom \Pom f_{2}}\,
\Bigl( \partial^{\kappa} \partial^{\lambda}    \Pom_{\mu \nu}(x) \Bigr)
\Bigl( \partial^{\mu} \partial_{\rho}    \Pom_{\kappa \lambda}(x) \Bigr)
\, \phi^{\nu \rho}(x)\,,
\end{split}
\label{A9}
\end{equation}
\begin{equation}
\begin{split}
{\cal L}'^{(7)}(x) = \frac{1}{M_{0}^{5}}\, g^{(7)}_{\Pom \Pom f_{2}}\,
\Bigl( \partial^{\rho} \partial^{\kappa} \partial^{\lambda}    \Pom_{\mu \nu}(x) \Bigr)
\Bigl( \partial^{\sigma} \partial^{\mu} \partial^{\nu}    \Pom_{\kappa \lambda}(x) \Bigr)
\, \phi_{\rho \sigma}(x)\,.
\end{split}
\label{A10}
\end{equation}
In (\ref{A4}) to (\ref{A10}) $M_{0} \equiv 1$~GeV and the $g^{(j)}_{\Pom \Pom f_{2}}$
are dimensionless coupling constants.
The values of coupling constants $g^{(j)}_{\Pom \Pom f_{2}}$, where $j = 1, ..., 7$,
as of nonperturbative origin, are not known
and are not easy to be found from first principles.

\begin{figure}
\includegraphics[width=5.cm]{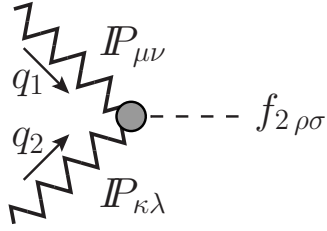}    
  \caption{\label{fig:pompomf2}
  \small Generic diagram for the $\Pom \Pom f_{2}$ vertices
  (\ref{A11}) - (\ref{A17}) with momentum and Lorentz-indices assignments.
}
\end{figure}
The vertices as obtained from (\ref{A4}) to (\ref{A10})
with the momentum and Lorentz-indices assignments shown in Fig.~\ref{fig:pompomf2}
are as follows:
\begin{equation}
\begin{split}
i\Gamma_{\mu \nu,\kappa \lambda,\rho \sigma}^{(\Pom \Pom f_{2})(1)}=
2 i \,g^{(1)}_{\Pom \Pom f_{2}} M_{0}\, 
R_{\mu \nu \mu_{1} \nu_{1}}\,
R_{\kappa \lambda \alpha_{1} \lambda_{1}}\,
R_{\rho \sigma \rho_{1} \sigma_{1}}\,
g^{\nu_{1} \alpha_{1}}\,g^{\lambda_{1} \rho_{1}}\,g^{\sigma_{1} \mu_{1}}\,,
\end{split}
\label{A11}
\end{equation}
\begin{equation}
\begin{split}
i\Gamma_{\mu \nu,\kappa \lambda,\rho \sigma}^{(\Pom \Pom f_{2})(2)} (q_{1},q_{2})=
-\frac{2i}{M_{0}} \,g^{(2)}_{\Pom \Pom f_{2}} \, 
\Bigl( &(q_{1} \cdot q_{2})\, R_{\mu \nu \rho_{1} \alpha}\, R_{\kappa \lambda \sigma_{1}}^{\quad \;\;\, \alpha}
- q_{1 \rho_{1}} \,q_{2}^{\mu_{1}} \,
  R_{\mu \nu \mu_{1} \alpha}\, R_{\kappa \lambda \sigma_{1}}^{\quad \;\;\, \alpha}\\
&- q_{1}^{\mu_{1}} \,q_{2 \sigma_{1}} \,
  R_{\mu \nu \rho_{1} \alpha}\, R_{\kappa \lambda \mu_{1}}^{\quad \;\;\, \alpha}
+ q_{1 \rho_{1}} \,q_{2 \sigma_{1}} \, R_{\mu \nu \kappa \lambda}
\Bigr) R_{\rho \sigma}^{\quad \rho_{1} \sigma_{1}}\,, 
\end{split}
\label{A12}
\end{equation}
\begin{equation}
\begin{split}
i\Gamma_{\mu \nu,\kappa \lambda,\rho \sigma}^{(\Pom \Pom f_{2})(3)} (q_{1},q_{2})=
-\frac{2i}{M_{0}} \,g^{(3)}_{\Pom \Pom f_{2}} \, 
\Bigl( &(q_{1} \cdot q_{2})\, R_{\mu \nu \rho_{1} \alpha}\, R_{\kappa \lambda \sigma_{1}}^{\quad \;\;\, \alpha}
+ q_{1 \rho_{1}} \,q_{2}^{\mu_{1}} \,
  R_{\mu \nu \mu_{1} \alpha}\, R_{\kappa \lambda \sigma_{1}}^{\quad \;\;\, \alpha}\\
&+ q_{1}^{\mu_{1}} \,q_{2 \sigma_{1}} \,
  R_{\mu \nu \rho_{1} \alpha}\, R_{\kappa \lambda \mu_{1}}^{\quad \;\;\, \alpha}
+ q_{1 \rho_{1}} \,q_{2 \sigma_{1}} \, R_{\mu \nu \kappa \lambda}
\Bigr) R_{\rho \sigma}^{\quad \rho_{1} \sigma_{1}}\,, 
\end{split}
\label{A13}
\end{equation}
\begin{equation}
\begin{split}
i\Gamma_{\mu \nu,\kappa \lambda,\rho \sigma}^{(\Pom \Pom f_{2})(4)} (q_{1},q_{2})=
-\frac{i}{M_{0}} \,g^{(4)}_{\Pom \Pom f_{2}} \, 
\Bigl( 
q_{1}^{\alpha_{1}} \,q_{2}^{\mu_{1}} \, 
R_{\mu \nu \mu_{1} \nu_{1}}\, R_{\kappa \lambda \alpha_{1} \lambda_{1}} + 
q_{2}^{\alpha_{1}} \,q_{1}^{\mu_{1}} \, 
R_{\mu \nu \alpha_{1} \lambda_{1}}\, R_{\kappa \lambda \mu_{1} \nu_{1}}
\Bigr) R^{\nu_{1} \lambda_{1}}_{\quad \; \; \; \rho \sigma}\,, 
\end{split}
\label{A14}
\end{equation}
\begin{equation}
\begin{split}
i\Gamma_{\mu \nu,\kappa \lambda,\rho \sigma}^{(\Pom \Pom f_{2})(5)} (q_{1},q_{2})=
-\frac{2i}{M_{0}^{3}} \,g^{(5)}_{\Pom \Pom f_{2}} \, 
\Bigl( &
q_{1}^{\mu_{1}} \,q_{2}^{\nu_{1}} \, 
R_{\mu \nu \nu_{1} \alpha}\, R_{\kappa \lambda \mu_{1}}^{\quad \;\;\, \alpha} + 
q_{1}^{\nu_{1}} \,q_{2}^{\mu_{1}} \, 
R_{\mu \nu \mu_{1} \alpha}\, R_{\kappa \lambda \nu_{1}}^{\quad \;\;\, \alpha} \\
&-2 (q_{1} \cdot q_{2})\, R_{\mu \nu \kappa \lambda}
\Bigr) 
q_{1 \alpha_{1}} \,q_{2 \lambda_{1}} \, 
R^{\alpha_{1} \lambda_{1}}_{\quad \; \; \; \rho \sigma}\,, 
\end{split}
\label{A15}
\end{equation}
\begin{equation}
\begin{split}
i\Gamma_{\mu \nu,\kappa \lambda,\rho \sigma}^{(\Pom \Pom f_{2})(6)} (q_{1},q_{2})=
\frac{i}{M_{0}^{3}} \,g^{(6)}_{\Pom \Pom f_{2}} \, 
\Bigl( &
q_{1}^{\alpha_{1}} \,q_{1}^{\lambda_{1}} \, q_{2}^{\mu_{1}} \, q_{2 \rho_{1}} \, 
R_{\mu \nu \mu_{1} \nu_{1}}\, R_{\kappa \lambda \alpha_{1} \lambda_{1}} \\
&+ 
q_{2}^{\alpha_{1}} \,q_{2}^{\lambda_{1}} \, q_{1}^{\mu_{1}} \, q_{1 \rho_{1}} \, 
R_{\mu \nu \alpha_{1} \lambda_{1}}\, R_{\kappa \lambda \mu_{1} \nu_{1}}
\Bigr) 
R^{\nu_{1} \rho_{1}}_{\quad \; \; \; \rho \sigma}\,, 
\end{split}
\label{A16}
\end{equation}
\begin{equation}
\begin{split}
i\Gamma_{\mu \nu,\kappa \lambda,\rho \sigma}^{(\Pom \Pom f_{2})(7)} (q_{1},q_{2})=
-\frac{2i}{M_{0}^{5}} \,g^{(7)}_{\Pom \Pom f_{2}} \, 
q_{1}^{\rho_{1}} \, q_{1}^{\alpha_{1}} \, q_{1}^{\lambda_{1}} \, 
q_{2}^{\sigma_{1}} \, q_{2}^{\mu_{1}} \, q_{2}^{\nu_{1}} \, 
R_{\mu \nu \mu_{1} \nu_{1}}\, R_{\kappa \lambda \alpha_{1} \lambda_{1}}\,
R_{\rho \sigma \rho_{1} \sigma_{1}}\,.
\end{split}
\label{A17}
\end{equation}
From (\ref{A3}) and (\ref{A11}) - (\ref{A17})
we have
\begin{equation}
\begin{split}
g^{\mu \nu} \, \Gamma_{\mu \nu,\kappa \lambda,\rho \sigma}^{(\Pom \Pom f_{2})(j)}(q_{1},q_{2}) =0\,, \quad
g^{\kappa \lambda} \, \Gamma_{\mu \nu,\kappa \lambda,\rho \sigma}^{(\Pom \Pom f_{2})(j)}(q_{1},q_{2})=0\,, \quad
g^{\rho \sigma} \, \Gamma_{\mu \nu,\kappa \lambda,\rho \sigma}^{(\Pom \Pom f_{2})(j)}(q_{1},q_{2})=0\,.
\end{split}
\label{A18}
\end{equation}
The expressions (\ref{A11}) - (\ref{A17}) represent our bare vertices which
we use in (\ref{vertex_pompomT}) multiplied by a form factor.

Investigating the contributions of the vertices (\ref{A11}) to (\ref{A17})
to the fictitious reaction of
two ``real tensor pomerons'' annihilating to the $f_{2}$ meson
we find that we can associate the couplings $j = 1, ..., 7$
with the following $(l,S)$ values
$(0,2)$, $(2,0)-(2,2)$, $(2,0)+(2,2)$, $(2,4)$, $(4,2)$, $(4,4)$, $(6,4)$, respectively.

\section{Asymmetries due to interference of different charge-conjugation exchanges}
\label{section:Asymmetries}
Here we return to asymmetries generated by interference
of $(C_{1},C_{2}) = (1,1)$, that is, our purely
diffractive continuum and resonance terms and the
$(C_{1},C_{2}) = (1,-1) + (-1,1)$ terms from photoproduction;
see section~\ref{sec:continuum}.
In the $\pi^{+}\pi^{-}$ rest frame 
we choose the Collins-Soper basis \cite{Collins:1977iv} with unit vectors
\begin{equation}
\begin{split}
& \bea = \frac{\bhpa+\bhpb}{|\bhpa+\bhpb|}\,,\\
& \beb = \frac{\bhpa\times\bhpb}{|\bhpa\times\bhpb|}\,,\\
& \bec = \frac{\bhpa-\bhpb}{|\bhpa-\bhpb|}\,,
\end{split}
\label{B1}
\end{equation}
where $\bhpa = \bpa / |\bpa|$ and $\bhpb = \bpb / |\bpb|$.
We define in this frame the unit vector
\begin{equation}
\begin{split}
& \bhk = \frac{\bpip-\bpim}{|\bpip-\bpim|} =
\left( \begin{array}{c}
\sin\chi \sin\psi \\
\cos\chi \\
\sin\chi \cos\psi \\
\end{array} \right)\,,\\
& 
0 \leqslant \chi \leqslant \pi, \quad
0 \leqslant \psi < 2\pi \,.
\end{split}
\label{B2}
\end{equation}

We are interested in the distribution of the vector $\bhk$.
From parity invariance, which holds for strong
and electromagnetic processes,
we get that this distribution must be symmetric under
$\bhk \cdot \beb \to -\bhk \cdot \beb$.
That is, parity requires symmetry under the replacement
\begin{equation}
\begin{split}
\cos\chi &\to -\cos\chi\\
\psi &\to \psi \,.
\end{split}
\label{B3}
\end{equation}
If only $(C_{1},C_{2}) = (1,1) + (-1,-1)$ or 
$(C_{1},C_{2}) = (1,-1) + (-1,1)$ amplitudes contribute 
the distribution must be symmetric under $\bhk \to -\bhk$.
Thus, an asymmetry under $\bhk \to -\bhk$ signals
an interference of the $(C_{1},C_{2}) = (1,1) + (-1,-1)$
and $(C_{1},C_{2}) = (1,-1) + (-1,1)$ amplitudes.
For $\cos\chi$ and $\psi$ the replacement $\bhk \to -\bhk$ means
\begin{equation}
\begin{split}
\cos\chi &\to -\cos\chi\\
\psi &\to \pi + \psi \,.
\end{split}
\label{B4}
\end{equation}
%

\nocite{}

{
\begin{small}
\bibliography{refs}
\end{small}
}

\end{document}